\documentclass[
twocolumn,
 amsmath,amssymb,
 aps,
 prx,
 superscriptaddress
]{revtex4-2}

\usepackage{dcolumn}
\usepackage[usenames,dvipsnames,svgnames,table]{xcolor}
\usepackage{float}
\usepackage{graphicx,amsmath,amsfonts,bm,mathtools}
\usepackage[colorlinks=true, citecolor=blue, linkcolor=blue, urlcolor=blue]{hyperref}
\usepackage{braket}
\usepackage{dsfont}
\usepackage{mathrsfs}
\usepackage[normalem]{ulem}
\usepackage{comment}

\newcommand{\narrow}{\mkern-1mu}
\newcommand{\svec}[2]{%
  \smash{\scalebox{.8}{$(\narrow #1\,#2\narrow)$}}%
}
\newcommand{\tvec}[2]{%
  \smash{\scalebox{.55}{$(\narrow #1\,#2\narrow)$}}%
}

\newcommand{\smplus}{\scriptscriptstyle+\narrow}
\newcommand{\smminus}{\scriptscriptstyle-\narrow}

\begin{document}

\title{Higher-Dimensional Information Lattice: Quantum State Characterization through Inclusion-Exclusion Local Information}

\author{Ian Matthias Flór}
\affiliation{Department of Physics, KTH Royal Institute of Technology, Stockholm, 106 91 Sweden}

\author{Claudia Artiaco}
\affiliation{Department of Physics, KTH Royal Institute of Technology, Stockholm, 106 91 Sweden}
\affiliation{Institut f\"ur Theoretische Physik, Universit\"at zu K\"oln, Z\"ulpicher Stra{\ss}e 77, D-50937 Cologne, Germany}

\author{Thomas Klein Kvorning}
\affiliation{Department of Physics, KTH Royal Institute of Technology, Stockholm, 106 91 Sweden}

\author{Jens H. Bardarson}
\affiliation{Department of Physics, KTH Royal Institute of Technology, Stockholm, 106 91 Sweden}

%\date{\today}

\begin{abstract}
    Developing a local description of quantum information is a central problem in quantum many-body physics: the goal is to identify how correlations are organized across positions and scales without choosing specific observables.
    Here we introduce a higher-dimensional information lattice: an inclusion-exclusion construction that decomposes the von Neumann information of reduced density matrices into local contributions labeled by subsystem position and scale.
    In dimensions higher than one, the inclusion-exclusion construction involves overlapping subsystems that can form loops, allowing information associated with one subsystem to be inferred nonlocally from the density matrices of the others.
    The sign of the inclusion-exclusion local information allows such overlap redundancies to be resolved in position and scale: positive values mark positions and scales with a net information gain, while negative values occur where redundant contributions dominate over new ones.
    We implement the construction explicitly in two dimensions and apply it to many-body ground states with qualitatively different entanglement structures.
    For localized states, such as disordered-localized states or gapped ground states of local Hamiltonians, the inclusion-exclusion information decays exponentially along different directions, defining direction-dependent information length scales.
    In critical states with a Fermi surface, this decay becomes algebraic, with dominant directions tied to the velocity of the gapless modes.
    In the toric code, topological order manifests as a negative information value at the largest scale of the lattice, that we interpret as a global information redundancy between the smallest loop-enclosing overlapping subsystems.
    We also isolate the universal information carried by edge states and resolve where in position and scale the signatures of non-Abelian fusion channels appear.
    Our results establish the higher-dimensional information lattice as an observable-independent framework for resolving universal correlation structures through the spatial and scale distribution of quantum information.
\end{abstract}

\maketitle

\section{Introduction}

A central way of characterizing a quantum many-body state is through the structure of its correlations.
In conventional many-body physics, this structure is often accessed through correlation functions: gapped states exhibit exponentially decaying correlations, critical states exhibit algebraic decay, and higher-dimensional systems can display strong directional dependence tied to anisotropy and gapless modes.
Correlation functions thus provide a local and scale-dependent language for characterizing quantum many-body states. 
However, they depend on a choice of observables.
Quantum information quantities, by contrast, are independent of any particular local probe, and can therefore expose universal properties of quantum states without reference to microscopic details~\cite{zeng2019quantum,laflorencie2016quantum,eisert2010colloquium}.
This has made quantum information theory central to condensed-matter physics~\cite{zeng2019quantum,laflorencie2016quantum,eisert2010colloquium}, quantum computing~\cite{terhal2015quantum,bennett2000quantum}, and holographic approaches to quantum gravity~\cite{ryu2006holographic,vanraamsdonk2010building,swingle2012entanglement}.
The challenge is to develop an information-theoretic analogue of the local picture provided by correlation functions: a description that is independent of the choice of observables, but still resolves how universal information is distributed across position, scale, shape, and direction.

The von Neumann entropy is the paradigmatic observable-independent diagnostic of quantum correlations.
In one-dimensional (1D) systems, it is constant in subsystem size in ground states of local gapped Hamiltonians, while it grows logarithmically at criticality~\cite{srednicki1993entropy,vidal2003entanglement,calabrese2004entanglement,wolf2008area}.
In midspectrum states of ergodic Hamiltonians it instead obeys a volume law~\cite{page1993average,vidmar2017entanglement,bianchi2022volume}.
These scaling laws underlie much of the modern understanding of quantum many-body states, including the success of tensor-network methods, which exploit the 1D area law to efficiently represent ground states of local gapped Hamiltonians~\cite{white1992density,ostlund1995thermodynamic,schollwock2011density}.
Distinguishing state universality classes from the scaling of the subsystem von Neumann entropy is often less direct in higher dimensions than in 1D.
In low-entangled states, the von Neumann entropy is dominated by area-law contributions, while universal features are often contained in subleading geometry-dependent terms at quantum critical points~\cite{casini2007universal,metlitski2009entanglement,fradkin2006entanglement,melko2013entanglement}, in a multiplicative logarithm for states with a Fermi surface~\cite{gioev2006entanglement,swingle2010entanglement,ding2012entanglement,wolf2006violation,pretko2017nodal}, or as a constant offset in topologically ordered phases~\cite{kitaev2006topological,levin2006detecting}.
Thus, the subsystem von Neumann entropy is a powerful diagnostic of quantum correlations.
However, it does not by itself specify where the relevant information is located, at which scale it appears, or along which spatial direction it is organized.

A useful way to extract more detailed structure from subsystem entropies is to consider suitable linear combinations of von Neumann entropies for different subsystems.
This idea is already central to several important quantum-information diagnostics.
The mutual information cancels contributions internal to each subsystem and quantifies correlations between two subsystems~\cite{wilde2013quantum}.
Its decay in area-law ground states can be used to define information based length scales~\cite{wolf2008area}.
The topological entanglement entropy cancels area-law terms to reveal long-range topological order~\cite{kitaev2006topological,levin2006detecting}.
The conditional mutual information measures the information shared between two subsystems that is not already contained in an intermediate subsystem, and is closely tied to reconstruction of quantum states~\cite{wilde2013quantum,winter2004,fawzi2015quantum}.
Recently, the scale dependence of conditional mutual information has also been used to define information-theoretic correlation lengths, such as the Markov length in mixed states~\cite{sang2025stability,sang2025mixed, zhang2024nonlocal, lee2024universal, zhang2025conditional, kuwahara2025clustering} and in parallel the correlation decay length in the 1D information lattice~\cite{artiaco2025universal}.
These examples show that linear combinations of entropies can carry content beyond the individual subsystem entropies entering them: by canceling dominant contributions, they reveal structures that are not visible in any single subsystem entropy alone.
At the same time, the latter constructions are tied to quasi-one-dimensional geometries or to specific partitions.
What is missing is a systematic way to assign this information locally in genuinely higher-dimensional systems.

Several approaches have addressed related aspects of this problem by seeking local or spatially resolved descriptions of quantum correlations.
These include multipartite entanglement measures~\cite{doherty2005detecting,szalay2015multipartite,walter2016multipartite}, quantum Fisher information~\cite{pezze2009entanglement,hyllus2012fisher,toth2012multipartite,strobel2014fisher,hauke2016measuring}, entanglement contours~\cite{chen2014entanglement,coser2017contour,rolph2022local}, entanglement-link representations~\cite{singha2020entanglement,roy2021link,santalla2023entanglement}, and local pictures of entanglement dynamics such as the quasiparticle picture~\cite{calabrese2005evolution,calabrese2016introduction,calabrese2020entanglement}, the membrane picture~\cite{nahum2017quantum,mezei2018membrane}, and the entanglement tsunami~\cite{liu2014entanglement}.
In parallel, the information lattice introduced for 1D open-boundary chains provides a fully local decomposition of the von Neumann information by assigning to each interval the contribution that is not contained in intervals of smaller scale or different position~\cite{klein2022time,artiaco2025universal}.
This structure has led to information-based universality classes of quantum states~\cite{artiaco2025universal,bilinskaya2025witnessing}, continuity equations for information dynamics~\cite{bauer2025local,artiaco2025out,barata2025hadronic,nava2026information} and efficient time-evolution algorithms~\cite{klein2022time,artiaco2024efficient,harkins2025nanoscale}. 
The computational strategy of retaining relevant information while discarding less relevant correlations also appears in several other approaches to simulating many-body dynamics, such as the density-matrix truncation method~\cite{white2018quantum,ye2020emergent,ye2022universal,peng2023exploiting,yi2024comparing,yang2025beyond}, dissipation-assisted operator evolution~\cite{rakovszky2022dissipation,keyserlingk2022operator,yoo2023open,lloyd2024ballistic,kuo2024energy}, approaches that trade entanglement for mixtures~\cite{surace2019simulating,frias2024converting}, cluster truncated Wigner approaches~\cite{wurtz2018cluster}, and other discussions of the entanglement barrier~\cite{pastori2019disentangling,rams2020breaking}.
Together with diagnostics based on conditional mutual information in mixed states, these developments organize information through conditional dependence along a single ordered direction.
This makes them essentially quasi-one-dimensional.
Higher-dimensional quantum states require a genuinely higher-dimensional information structure.
Subsystems admit various shapes and topologies, and the interval-based ordering that constrains many 1D entanglement constructions is no longer available~\cite{roy2021link, miao2025convergence, wong2025entanglement, zhang20252d, wen2020entanglement, rottoli2025entanglement}.
The increased geometric complexity is familiar in computational approaches: projected entangled pair states are a conceptually direct generalization of matrix product states, but their contraction and optimization are substantially more demanding and lack a simple canonical form~\cite{Verstraete2008,Cirac2021}.
This geometric richness is tied to physical phenomena absent in 1D, including extended Fermi surfaces, topologically ordered phases~\cite{wen1990topological}, and the possibility of braiding Abelian or non-Abelian anyons~\cite{wilczek1982quantum,arovas1984fractional,nayak2008nonabelian}.
In such systems, one would like to know how correlations are organized across position and scale.
This motivates an information-theoretic framework that turns such position- and scale-dependence into local contributions assigned to subsystems.

The main obstruction to assigning information uniquely to position and scale in higher dimensions is that subsystems overlap in a fundamentally different way from intervals in an open 1D chain.
For open 1D intervals, smaller intervals are naturally nested inside larger ones and linearly ordered along the physical chain, making the local decomposition unique.
In higher dimensions, and already in periodic 1D geometries, overlapping subsystems can form loop-like structures.
The same information can then be accessible through several distinct overlapping subsystems.
We refer to this as \emph{overlap redundancy}.
Overlap redundancies are not specific to quantum mechanics; related phenomena already appear in classical multivariate information theory~\cite{mcgill1954multivariate,ting1962amount,watanabe1960information}.
They show that, unlike for intervals in an open 1D chain, information shared among overlapping subsystems in higher dimensions cannot always be assigned uniquely to a single position and scale.
Any attempt to assign information locally to position and scale in higher dimensions must therefore account for overlap redundancies.

In this work, we introduce a higher-dimensional information lattice: an inclusion-exclusion decomposition of the von Neumann information of subsystem reduced density matrices into local contributions labeled by subsystem position and scale.
These local contributions sum to the total von Neumann information and can be positive or negative.
Positive values identify a net information gain at the corresponding position and scale, whereas negative values indicate overlap redundancy, where information is shared among smaller overlapping subsystems.
The sign structure therefore resolves how information is locally gained or redundantly shared across the subsystem lattice.
In this sense, the framework gives an operator-independent, spatially resolved counterpart to correlation-function diagnostics: it resolves how information is organized across position and scale without choosing local observables. 

We implement the higher-dimensional information lattice explicitly in two dimensions using rectangular subsystems and apply it to representative many-body ground states with qualitatively different entanglement structures.
In gapped states, the inclusion-exclusion information decays exponentially along different directions, defining direction-dependent information decay lengths.
In critical states with a Fermi surface, this decay becomes algebraic, with dominant directions tied to the modes on the Fermi surface.
For chiral systems with boundaries, the framework isolates the universal contribution of edge states.
For the toric code, topological order appears as a negative information contribution at the largest scales, reflecting redundancy among local constraints at short scales.
With non-Abelian defects, the information lattice resolves the position and scale at which fusion-channel information is encoded.
These results demonstrate that the higher-dimensional information lattice provides a general framework for resolving the organization of correlations across position and scale.

The paper is organized as follows.
In Sec.~\ref{sec:generalization}, we introduce the higher-dimensional construction and discuss the role of overlap redundancy.
In Sec.~\ref{sec:2d_information_lattice}, we specialize the framework to rectangular subsystems in two dimensions.
We then apply it to localized and critical free-fermion states in Sec.~\ref{sec:gapped_critical}, to systems with chiral edge modes in Sec.~\ref{sec:edge_mode}, and to topologically ordered states with and without non-Abelian defects in Sec.~\ref{sec:tc}.
Finally, in Sec.~\ref{sec:conclusion}, we summarize our results and discuss future directions.

\section{Generalizing the information lattice to higher dimensions} \label{sec:generalization}

\begin{figure}
    \centering
    \includegraphics[width=\linewidth]{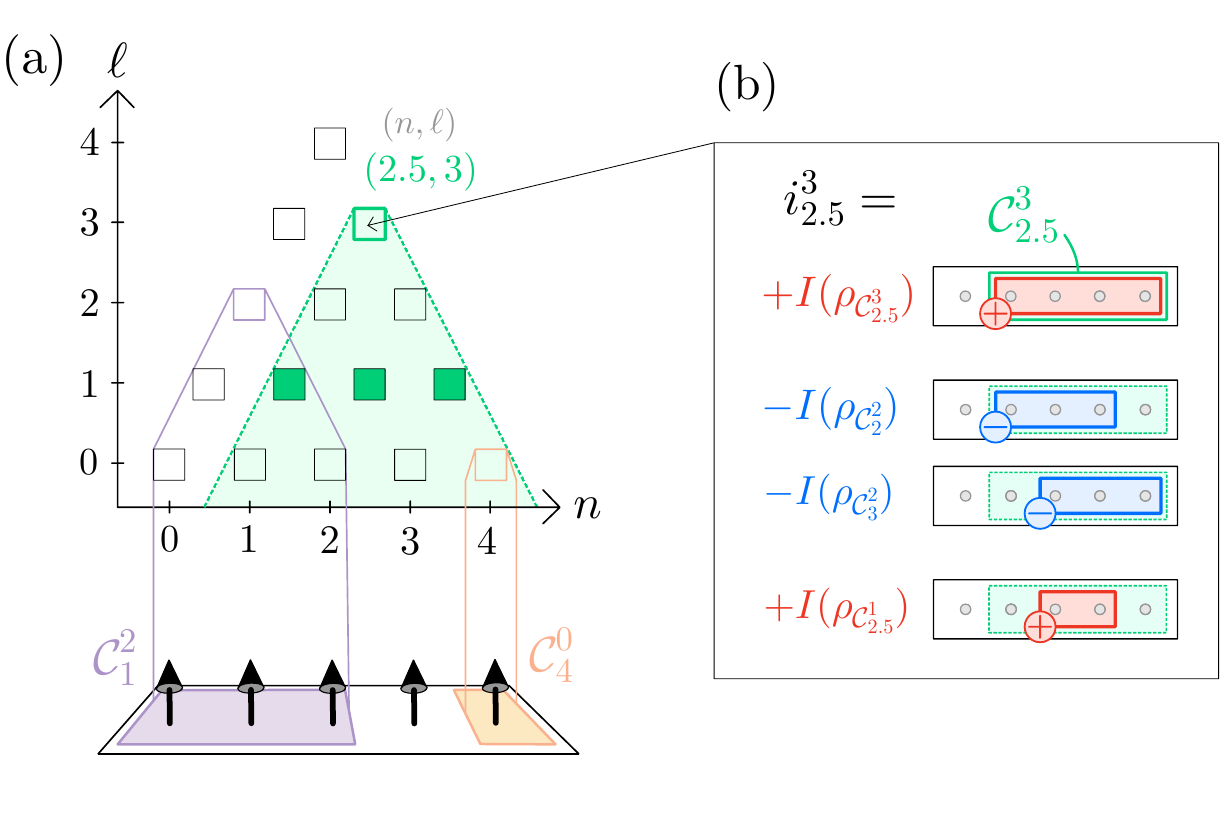}
    \caption{
        On a 1D chain of five physical sites [indicated by black arrows at the bottom of (a)], we consider the subsystems $\mathcal{C}_n^\ell$ consisting of $\ell+1$ consecutive sites centered at position $n$ on the chain.
        Each subsystem is labeled by the indices $(n,\ell)$, which in (a) are represented as square vertices and arranged hierarchically in a triangular structure.
        Examples are shown for $\mathcal{C}_1^2$ in lilac and $\mathcal{C}_4^0$ in yellow.
        The local information $i_n^\ell$ is calculated for each vertex $(n,\ell)$ by an addition-subtraction of the subsystem von Neumann information [see Eq.~\eqref{eq:local_information}] as illustrated in (b) for $i_{2.5}^3$.
        The shaded green area in (b) represents the subsystem $\mathcal{C}_{2.5}^3$.
        The shaded green area in (a) includes all the vertices associated with subsystems contained within $\mathcal{C}_{2.5}^3$.
        The total information in $\mathcal{C}_{2.5}^3$ is the sum of all the local information $i_n^\ell$ associated with the vertices within the shaded green area in (a) [see Eq.~\eqref{eq:decomposition}].
        The filled green vertices label all the subsystems of scale 1 which are contained in the larger subsystem $\mathcal{C}_{2.5}^3$.
        The total information at scale $\ell=1$ in $\mathcal{C}_{2.5}^3$ is the sum of all local information $i_n^\ell$ on the filled green vertices [see Eq.\eqref{eq:informationonscaleinaregion}].
}
    \label{fig:info1d}
\end{figure}

The von Neumann information (or total information) of a state $\rho$ is
\begin{align}
  I(\rho) = \log_2(\dim\rho) - S(\rho)
  \label{eq:von_neumann_info}
\end{align}
where $\dim{\rho}$ is the Hilbert space dimension of the state and $S(\rho)=-\mathrm{Tr}[\rho\log_2\rho]$ is the von Neumann entropy.
$I(\rho)$ is the average number of bits that can be inferred from $\rho$ about the distribution of measurement outcomes~\footnote{
	For example, if there is an observable whose repeated measurements on a single-qubit state $\rho$ always yield the same outcome, then the measurement answers one yes/no question with certainty, resulting in one bit of information (meaning $\rho$ is pure).
	In contrast, if for every observable the outcomes of repeated measurements are fully random, they provide no information (meaning $\rho$ is maximally mixed).
    $I(\rho)$ is the average number of independent yes/no questions that can be answered from $\rho$ about the outcomes of measurements.
}.
Analogously, the von Neumann information of the reduced density matrix $\rho_{\mathcal{C}}=\mathrm{Tr}_{\bar{\mathcal{C}}}\rho$, defined by
\begin{align}
  I(\rho_\mathcal{C}) = \log_2(\dim\rho_\mathcal{C}) - S(\rho_\mathcal{C}),
  \label{eq:von_neumann_info_reduced}
\end{align}
is the total information in the subsystem $\mathcal{C}$, that is, the number of bits that can be inferred from $\rho_\mathcal{C}$ about measurements supported on $\mathcal{C}$.
The total information $I(\rho_\mathcal{C})$ in a subsystem $\mathcal{C}$ also accounts for the information already present in its subsystems $\mathcal{C}' \subset \mathcal{C}$.

\subsection{The 1D information lattice \label{sec:1d_information_lattice}}

Before generalizing the information lattice to higher-dimensional systems, we briefly review its construction in the 1D case.
While the total information of a subsystem includes that of all its parts, in a 1D chain the \textit{local information} quantifies the information that is exclusive to a given subsystem, namely that it cannot be found in any subsystem at a lower scale or different location.
The \textit{information lattice} organizes the local information of each subsystem according to its spatial location and its length scale.

To achieve a decomposition of the information into local contributions from different subsystems, one considers the set of all subsystems $\{\mathcal{C}_n^\ell\}$ of $\ell+1$ neighboring sites centered at position $n$ on the one-dimensional chain.
Each subsystem is uniquely identified by a label $(n,\ell)$.
The set of labels can be placed on a triangular structure illustrated in Fig.~\ref{fig:info1d}(a), where each vertex (square marker) is associated with the position $n$ along the physical chain (horizontal axis) and the scale $\ell$ (vertical axis).
This defines a hierarchical arrangement, where the labels associated with larger subsystems are placed at higher levels.
The information lattice decomposes the von Neumann information of a subsystem $\mathcal{C}_n^\ell$ into a sum of individual contributions $i^{\ell^\prime}_{n^\prime}$ for each subsystem $\mathcal{C}_{n'}^{\ell'}\subset\mathcal{C}_n^\ell$ through
\begin{align}
\label{eq:decomposition}
	I(\rho_{\mathcal{C}_{n}^\ell})
	&= \!\!\!\!
   \sum_{\{ (n',\ell') \mid \mathcal{C}^{\ell'}_{n'} \subseteq \mathcal{C}^{\ell}_{n} \}}
   \!\!\!\!
   i^{\ell^\prime}_{n^\prime}.
\end{align}
For example, $I(\rho_{\smash{\mathcal{C}_{2.5}^3}})$ is the sum of the values $i^{\ell'}_{n'}$ on all subsystems covered by the shaded green area in Fig.~\ref{fig:info1d}(a).
Imposing that the above condition be satisfied for all subsystems results in $i_n^\ell$ taking the form
\begin{align}
    i^\ell_n &= I_n^\ell - I^{\ell-1}_{n-1/2} - I^{\ell-1}_{n+1/2} + I^{\ell-2}_n,
    \label{eq:local_information}
\end{align}
where $I_n^\ell$ is a shorthand for $I(\rho_{\mathcal{C}_{n}^\ell})$.
This subtraction of von Neumann information values is represented schematically in Fig.~\ref{fig:info1d}(b).
The expression of $i^\ell_n$ in Eq.~\eqref{eq:local_information} corresponds to the quantum conditional mutual information~\cite{wilde2013quantum}; $i^\ell_n$ singles out the portion of information that is exclusive to scale $\ell$ and position $n$.
The quantum conditional mutual information is greater than or equal to zero by the strong subadditivity of von Neumann entropy~\cite{lieb1973proof}.
If $i_n^\ell = 0$, the outcome distribution of an optimal measurement on $\mathcal C_n^\ell$ is already fixed by the reduced density matrices of its two smaller subsystems $\rho_{\smash{\mathcal C_{n\pm1/2}^{\ell-1}}}$~\cite{winter2004}: no new bits of information are gained by performing measurements on the larger subsystem $\mathcal{C}_n^\ell$.
If $i_n^\ell > 0$, the state $\rho_{\mathcal C_n^\ell}$ contributes exactly $i_n^\ell$ additional bits beyond those contained in its smaller subsystems.
Thus, $i_n^\ell$ is a local quantity in position and scale that singles out the portion of information unique to $\mathcal{C}_n^\ell$~\cite{klein2022time,artiaco2024efficient,harkins2025nanoscale,artiaco2025universal,bauer2025local}.

By summing the local information $i_n^\ell$ over the positional indices $n$, we obtain information per scale quantities~\cite{artiaco2025universal}.
The information at scale $\ell'$ inside a subsystem $\mathcal{C}_n^\ell$ is
\begin{equation}
    \label{eq:informationonscaleinaregion}
    I(\ell^\prime)\big|_{\smash{\mathcal{C}_n^\ell}}
    = \sum_{\{n^\prime \mid \mathcal{C}^{\ell^\prime}_{n^\prime} \subseteq \mathcal{C}^\ell_n\}} i^{\ell^\prime}_{n^\prime}.
\end{equation}
For example, the information in $\mathcal{C}_{2.5}^3$ at scale $\ell'=1$ is the sum of the local information over the filled green vertices in Fig.~\ref{fig:info1d}(a).
This captures the exclusive information at scale $\ell'$ inside $\mathcal C_{n}^{\ell}$ since it tells us how many extra bits can be inferred about outcomes at scale $\ell'$ from all density matrices at scale $\ell'$, compared with what can be learned from only those at scale $\ell' - 1$.

\subsection{Inclusion-exclusion local information\label{sec:ie_procedure}}

The starting point of our generalization of the 1D information lattice to higher dimensions is the decomposition of the information \eqref{eq:decomposition} on a selection of subsystems.
The perspective of the information decomposition allows us to recast the problem of defining local information for each individual subsystem into that of choosing a set of subsystems ordered in position and scale appropriately.
The most direct generalization is to consider the set of all connected subsystems which differ from each other by a single site, as in the 1D case. 
However, the 1D information lattice benefits from the fact that an intersection of any two subsystems is labeled by its own vertex on the lattice.
This is a necessary condition to achieve a correct scale-assignment of information~\footnote{Consider a four-site open-boundary 1D chain with sites labeled $a,b,c$ and $d$. If $|\psi\rangle=|\!\uparrow\rangle\otimes\frac{\ket{\uparrow\downarrow}-\ket{\downarrow\uparrow}}{\sqrt{2}}\otimes\ket{\uparrow}$ is the state, but we assume the subsystem $bc$ is discarded from the lattice, then the decomposition enforces that $i_{abc}=I(\rho_{abc})-I(\rho_{ab})-I(\rho_c)$. This would assign the $2$ bits of information associated to the singlet on $bc$ to $abc$ instead. However, it would also assign those bits somewhere else on the lattice, namely to $bcd$. This structural ambiguity cannot occur if all subsystem intersections each have their own vertex on the lattice.}
, which we therefore also impose on our set of subsystems in higher dimensions.
In higher dimensions, the intersection of subsystems generically yields disconnected subsystems, unless the subsystems are convex (contain every shortest lattice path between any two of their sites).
This leaves us with two possibilities.
The first is to consider a lattice whose vertices also label disconnected subsystems.
However, this causes the number of vertices to grow exponentially with system size and adds a complication in the purpose of a scale-resolved description of information.
The second option is to impose that the subsystems are convex.
We define the higher-dimensional information lattice by considering a set of convex subsystems.
We denote those subsystems as $\mathcal{C}^{\boldsymbol{\ell}}_{\boldsymbol{n}}$ with
\begin{itemize}
    \item a multiscale label ${\boldsymbol{\ell}}$, which encodes the spatial extent of a subsystem along independent axes;
    \item a positional label ${\boldsymbol{n}}$, which locates the subsystem within the system.
\end{itemize}
When clear from context, we will refer to $\boldsymbol{\ell}$ as the scale.
The subsystems in the set $\{\mathcal{C}^{\boldsymbol{\ell}}_{\boldsymbol{n}}\}$ are partially ordered by all $n_i\in\boldsymbol{n}$ and $\ell_i\in\boldsymbol{\ell}$, where increasing ${\boldsymbol{\ell}}$ corresponds to enlarging subsystems in one or more dimensions.
The most natural choice of subsystems depends on the geometry of the physical system.
For instance, on a 2D square lattice it is natural to consider rectangular subsystems specified by position $(n_x, n_y)$ and scale $(\ell_x, \ell_y)$ (see next section), whereas in a radial geometry one may instead use coordinates $(n_r, n_\theta)$ and scales $(\ell_r, \ell_\theta)$.
We impose the decomposition property~\eqref{eq:decomposition} on this set of subsystems.
The decomposition~\eqref{eq:decomposition} represents a system of equations where $i_{\boldsymbol{n}'}^{\boldsymbol{\ell}'}$ are the unknowns.
Its solution is obtained by so-called Möbius inversion~\cite{rota1964foundations, stanley2011enumerative} and is given by the expression
\begin{align}
    i^{\boldsymbol{\ell}}_{\boldsymbol{n}} &= I(\rho_{\mathcal{C}^{\boldsymbol{\ell}}_{\boldsymbol{n}}})
    - \sum_{i \in \mathfrak{I}^{\boldsymbol{\ell}}_{\boldsymbol{n}}} I(\rho_{\mathcal{A}_i})
    + \sum_{i<j \in \mathfrak{I}^{\boldsymbol{\ell}}_{\boldsymbol{n}}} I(\rho_{\mathcal{A}_i \cap \mathcal{A}_j}) \nonumber \\
    & \hspace{1cm}- \sum_{i<j<k \in \mathfrak{I}^{\boldsymbol{\ell}}_{\boldsymbol{n}}} I(\rho_{\mathcal{A}_i \cap \mathcal{A}_j \cap \mathcal{A}_k})
    + \cdots,
    \label{eq:inclusion_exclusion}
\end{align}
where $\mathfrak{I}^{\boldsymbol{\ell}}_{\boldsymbol{n}}$ is the set of indices of subsystems $\mathcal{A}_i\subset\mathcal{C}^{\boldsymbol{\ell}}_{\boldsymbol{n}}$ obtained by removing one layer of sites (i.e., decreasing one of the components of $\boldsymbol \ell$ by one) along any coordinate direction.
The second term subtracts the information from pairwise intersections of the subsystems $\mathcal{A}_\textit{i}$.
Subsequent terms alternately add and subtract the information from the higher-order intersections between three or more subsystems $\mathcal{A}_\textit{i}$, which correspond to progressively smaller subsystems in $\mathcal{C}^{\boldsymbol{\ell}}_{\boldsymbol{n}}$.
We refer to~\eqref{eq:inclusion_exclusion} as the \textit{inclusion-exclusion local information}, and when clear from context as local information for short.
We require the set of subsystems to be closed under intersections; that is, for any two subsystems $\mathcal{C}^{\boldsymbol{\ell}_1}_{\boldsymbol{n}_1}$ and $\mathcal{C}^{\boldsymbol{\ell}_2}_{\boldsymbol{n}_2}$ in the set, their intersection $\mathcal{C}^{\boldsymbol{\ell}_1}_{\boldsymbol{n}_1} \cap \mathcal{C}^{\boldsymbol{\ell}_2}_{\boldsymbol{n}_2}$ must also belong to the set.

\subsection{Overlap redundancies in higher dimensions \label{sec:obstacles}}

In the 1D information lattice, the subsystems $\{\mathcal{C}_n^{\ell}\}$ are not only closed under intersections but they also form a set chain at each scale $\ell$: the set $\{\mathcal{C}^{{\ell}}_{{n}}\}_{\text{all}\ {n}}$ satisfies $(\bigcup_{m<n}\mathcal{C}^\ell_m)\cap \mathcal{C}^\ell_n = \mathcal{C}^\ell_{n-1}\cap \mathcal{C}_n^\ell$.
In that case, $i_n^\ell$ in Eq.~\eqref{eq:local_information} can be expressed as the quantum conditional mutual information
\begin{align}
    I(A:C|B) = I(\rho_{ABC}) - I(\rho_{AB}) - I(\rho_{BC}) + I(\rho_{B}) \label{eq:cmi}
\end{align}
between $A$ and $C$ conditioned on $B$.
Indeed, comparing with Eq.~\eqref{eq:local_information}, $ABC$ denotes $\mathcal{C}_n^\ell$, and the subsystems $A=\mathcal{C}^0_{n-\ell/2},B=\mathcal{C}^{\ell-2}_{n}$ and $C=\mathcal{C}^0_{n+\ell/2}$ are disjoint.
Eq.~\eqref{eq:cmi} is non-negative due to the strong sub-additivity of the von Neumann entanglement entropy~\cite{lieb1973proof}.
In geometries other than the open 1D chain, the set of subsystems $\{\mathcal{C}_{\boldsymbol{n}}^{\boldsymbol{\ell}}\}$ in Eq.~\eqref{eq:inclusion_exclusion} closed under intersections typically does not form a set chain.
For example, when the subsystems at one scale form a loop $\{AB,BC,CA\}$ the inclusion-exclusion information takes the form
\begin{align}
    I(A:B:C) &= I(\rho_{ABC}) - I(\rho_{AB}) - I(\rho_{BC}) - I(\rho_{AC}) \nonumber \\
    &\hspace{1cm} + I(\rho_{A}) + I(\rho_{B}) + I(\rho_{C}).\label{eq:interaction_information}
\end{align}
Eq.~\eqref{eq:interaction_information} is a possible multipartite generalization of the conditional mutual information, known as the \textit{interaction information}, among the disjoint subsystems $A,B,C$~\cite{mcgill1954multivariate,watanabe1960information, ting1962amount,yeung1991new,bell2003co,williams2010nonnegative,rosas2019quantifying, horodecki2009quantum}.
Unlike Eq.~\eqref{eq:cmi}, $I(A:B:C)$ may be negative.
This already occurs in classical probability theory, where it corresponds to the situation where conditioning on one variable reduces the dependence between the others, expressed as $I(A:C|\emptyset) > I(A:C|B)$ with $\emptyset$ the empty set~
\footnote{
    If we have the events $A$: ``it is cloudy'', $B$: ``it is raining'', and $C$: ``the ground is wet'', the pairwise distributions $p(A,B)$, $p(B,C)$, and $p(A,C)$ all describe the shared likelihood of rain, clouds and wet ground: knowing the distribution of one gives us some amount of predictive power over the others.
    Indeed, if we do not know $B$, then $p(A)$ gives us some predictive power over $C$, so $I(A:C|\emptyset) > 0$.
    But if we do know $B$ (i.e., we know it is raining), then $p(A)$ provides less predictive power over $C$, since once we know it is raining, knowing whether it is cloudy becomes redundant for predicting whether the ground is wet.
    Thus $I(A:C|B) < I(A:C|\emptyset)$ and, correspondingly, $I(A:B:C)=I(A:C|B)-I(A:C|\emptyset)<0$, signaling the redundancy.
    On the other hand, if $A,B$ are uniform Bernoulli distributed random variables and $C=A\oplus B$ is the mod(2) addition, then we have that each marginal density $p(A)$, $p(B)$, and $p(C)$ describes fully random outcomes and therefore carries no predictive power about the others.
    Similarly $p(A,B),p(B,C),p(C,A)$ also give no predictive power over measurement outcomes.
    In this case, only by knowing $p(A,B,C)$ we can fully predict what one of the random variables will output, so in this case $I(A:B:C)=1$ bit $>0$ gives new information compared to what the marginal densities provide.
    See Ref.~\cite{jakulin2003quantifying} for a detailed discussion.
}.
Analogous to the loop $\{AB,BC,CA\}$, the subsystems in the set $\{\mathcal{C}^{\boldsymbol{\ell}}_{{\boldsymbol{n}}}\}_{\text{all }\boldsymbol{n}}$ at a given multiscale $\boldsymbol{\ell}$ in \eqref{eq:inclusion_exclusion} overlap in space but do not form a set chain, which results in $i_{\boldsymbol{n}}^{\boldsymbol{\ell}}$ being a signed quantity.

\begin{figure}
    \includegraphics[width=.9\linewidth]{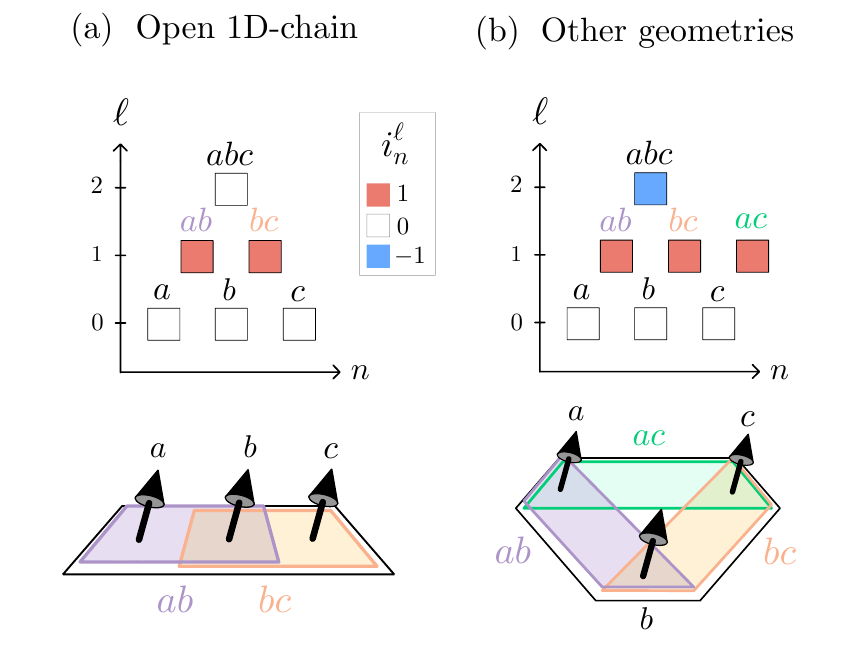}
    \caption{
        (a) 1D chain of three qubits denoted $a,b,c$ (bottom) in the state $\rho_{abc}=\tfrac12(|\!\!\uparrow\uparrow\uparrow\rangle\!\langle\uparrow\uparrow\uparrow\!\!| + |\!\!\downarrow\downarrow\downarrow\rangle\!\langle\downarrow\downarrow\downarrow\!\!|)$.
        The 1D information lattice (top) is represented for this state, where for clarity the names of the subsystems are shown next to the vertices that label them (as $a$, $ab$, and so on).
        It shows that local information is exclusively located at scale $\ell=1$ in subsystems $ab$ and $bc$ that have 1 bit each, and the total information in the state is 2 bits given by the sum of those two contributions.
        (b) Consider the same state $\rho_{abc}$ for a system in a triangular arrangement (bottom).
        There are three two-site subsystem, the new one being $ac$ (green).
        (Top) On the information lattice based on the inclusion-exclusion information \eqref{eq:inclusion_exclusion} at scale $\ell=1$ each subsystem contributes 1 bit of information.
        However, 1 out of these 3 bits is redundant; the 2 bits of information are contained in any pair of $\rho_{ab},\rho_{bc},\rho_{ac}$.
        Using Eq.~\eqref{eq:inclusion_exclusion}, the overcounting of this redundant information at scale $\ell=1$ is compensated by a negative value at scale $\ell=2$, the scale at which all overlapping subsystems are contained and the dependence in their correlations is resolved.
    }
    \label{fig:loops}
\end{figure}

We illustrate the origin of negative inclusion-exclusion information through a simple three-qubit example.
We consider the state $\rho_{abc}=\tfrac12(\ket{\uparrow\uparrow\uparrow}\!\!\bra{\uparrow\uparrow\uparrow}+\ket{\downarrow\downarrow\downarrow}\!\!\bra{\downarrow\downarrow\downarrow})$ on three qubits labeled $a,b,c$ in an open 1D chain.
The qubits and their respective positions are shown in Fig.~\ref{fig:loops}(a) (bottom).
The total information of the state is $I(\rho_{abc}) = 2$ bits.
In Fig.~\ref{fig:loops}(a), those two bits are decomposed on the 1D information lattice by evaluating \eqref{eq:local_information}.
We see that at $\ell=0$ the single site subsystems contribute no information $i^0_{a}=i^0_{b}=i^0_{c}=0$ (their density matrices are maximally mixed).
The $\ell=1$ subsystems each contribute $i^1_{ab}=i^1_{bc}=1$ bit, so summing all information at scale $\ell=1$ according to \eqref{eq:informationonscaleinaregion} already gives the total information of 2 bits in the state.
Correspondingly, there are $i^2_{abc}=0$ new bits at the largest scale.
The 1D information lattice unambiguously shows that all the information about the state $\rho_{abc}$ is contained in $ab$ and $bc$, as we cannot better predict the distribution of measurement outcomes on $abc$ from $\rho_{abc}$ than can already be inferred from the density matrices $\rho_{ab}$ and $\rho_{bc}$.

We now consider the same state on three qubits $a,b,c$, but in a triangular arrangement as shown in Fig.~\ref{fig:loops}(b) (bottom) (for instance, on a periodic 1D-chain or a higher-dimensional lattice) and examine the information that contributes to certain scales.
At $\ell=1$ in Fig.~\ref{fig:loops}(b) (top), each of the three two-site subsystems provides 1 bit of information.
This time, summing their individual contributions following \eqref{eq:informationonscaleinaregion} gives 3 bits, which exceeds the total information of the state.

The over-counting of information at a certain scale manifests a relation between the reduced density matrices that we call an \textit{overlap redundancy}. 
In the triangular three-site example, the measurement statistics on $ac$ cannot be inferred from $\rho_a$ and $\rho_c$ alone.
This tells us that, \textit{a priori}, knowledge of the larger density matrix $\rho_{ac}$ is required to infer the measurement statistics on $ac$.
It turns out that, in this case, the full measurement statistics on $ac$ can also be inferred collectively from $\rho_{ab}$ and $\rho_{bc}$, even though neither is supported on all of $ac$.
Geometrically, the subsystems $ab$ and $bc$ form an alternative path between $a$ and $c$, denoted $\mathcal{P}=\{ab,bc\}$, which closes a loop; such an alternative path is absent for intervals of the open 1D chain.
Whether the reduced density matrices along this path actually suffice to nonlocally infer any statistics on $ac$ is a property of the state.

The existence of an overlap redundancy must be separated from whether and how a given inclusion-exclusion information captures it.
In the three site triangular example in Fig.~\ref{fig:loops}, the inclusion-exclusion information decomposes $I(\rho)$ on the set $\{\mathcal{C}_{n}^\ell\}=\{a,b,c,ab,bc,ac,abc\}$ and hence produces the interaction information
\begin{align}
    i^{2}_{abc} & = I(a:b:c) = -1 \text{ bit}\label{eq:inclusion_exclusion_3party}
\end{align}
at scale $\ell=2$.
This negative value is shown in Fig.~\ref{fig:loops}(b) (top) at scale $\ell=2$: the subsystem $abc$ is the smallest subsystem that resolves the overlap redundancy between the density matrices $\rho_{ab},\rho_{bc}$ and $\rho_{ac}$.
The decomposition property \eqref{eq:decomposition} gives negative values a particularly simple interpretation: the excess of information at smaller scales due to an overlap redundancy must be compensated at a larger encompassing scale so that the total information
in every subsystem is recovered.
Note that if genuinely new correlations also become accessible in $abc$ at the same scale, their positive contribution can partially or fully compensate the negative contribution associated with the redundancy.
The inclusion-exclusion information therefore reflects the net balance of these contributions.
A strictly negative value of Eq.~\eqref{eq:inclusion_exclusion_3party} is a definite signature of an overlap redundancy between lower-scale density matrices.

The three-site example in Fig.~\ref{fig:loops}(b) reflects the general case, in which we decompose the information on a given set of subsystems
$\{\mathcal{C}_{\boldsymbol{n}}^{\boldsymbol{\ell}}\}$ in Eq.~\eqref{eq:decomposition}.
On some subsystem $\mathcal{C}_{\boldsymbol{n}'}^{\boldsymbol{\ell}'}$ in that set, consider some measurement statistics that cannot be inferred from the reduced density matrices of its proper subsystems.
An \textit{overlap redundancy} occurs when these statistics can nevertheless be inferred collectively from the reduced density matrices associated with a subset
$\mathcal{P}\subset\{\mathcal{C}_{\boldsymbol{n}}^{\boldsymbol{\ell}}\}$,
none of whose subsystems contains $\mathcal{C}_{\boldsymbol{n}'}^{\boldsymbol{\ell}'}$.
The subsystems in $\mathcal{P}$ are connected through spatial overlaps, so they form a network within the set $\{\mathcal{C}_{\boldsymbol{n}}^{\boldsymbol{\ell}}\}$.
We refer to such subsytems connected via overlaps as a \textit{path} in $\{\mathcal{C}_{\boldsymbol{n}}^{\boldsymbol{\ell}}\}$, using the term in a generalized sense.
An overlap redundancy means that the relevant measurement statistics on $\mathcal{C}_{\boldsymbol{n}'}^{\boldsymbol{\ell}'}$ can be accessed through such an alternative path $\mathcal{P}$.
The existence of such a redundancy is therefore a relation among the reduced density matrices in the chosen set of subsystems $\{\mathcal{C}_{\boldsymbol{n}}^{\boldsymbol{\ell}}\}$.
Which redundancies can be resolved depends on that choice.
The inclusion-exclusion information separately describes how these redundancies enter the decomposition of the total information on $\{\mathcal{C}_{\boldsymbol{n}}^{\boldsymbol{\ell}}\}$.
When an overlap redundancy among lower-scale reduced density matrices becomes resolved at a position $\boldsymbol{n}$ and scale $\boldsymbol{\ell}$, it contributes negatively to
$i_{\boldsymbol{n}}^{\boldsymbol{\ell}}$, while genuinely new information becoming accessible there contributes positively.
Like in the three-site example, $i_{\boldsymbol{n}}^{\boldsymbol{\ell}}$ is a net diagnostic of the balance between new and redundant information: positive values indicate that new information dominates, whereas negative values indicate that overlap redundancy dominates.

\section{The 2D Information Lattice using Rectangular Subsystems \label{sec:2d_information_lattice}}

In the remainder, we evaluate the inclusion-exclusion information \eqref{eq:inclusion_exclusion} on rectangular subsystems on two-dimensional lattices to illustrate the key features of our method.
However, the construction itself can be applied in arbitrary dimensions and one is free to pick whichever set of subsystems closed under intersections.

\subsection{Rectangular inclusion-exclusion information}

\begin{figure}
    \centering
    \includegraphics[width=\linewidth]{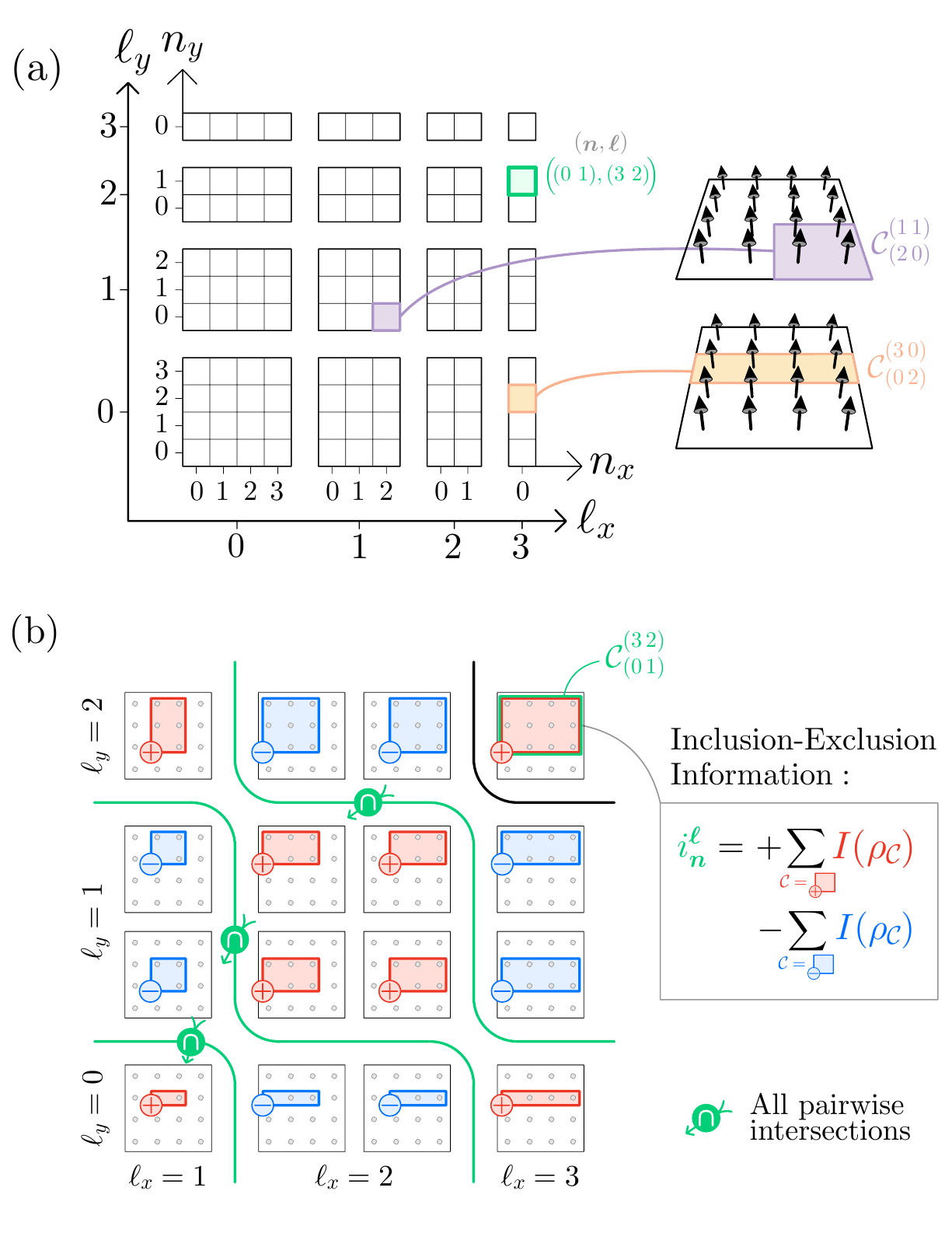}
    \caption{
    	(a) On a system of $4\times4$ physical sites (black arrows), each rectangular subsystem is labeled by its scale $\boldsymbol{\ell} = (\ell_x\ \ell_y)$ and the coordinate of its bottom-left corner $\boldsymbol{n} = (n_x\ n_y)$, denoted as $\mathcal{C}_{\tvec{n_x}{n_y}}^{\tvec{\ell_x}{\ell_y}}$ (for example $\mathcal{C}_{\tvec{0}{2}}^{\tvec{1}{1}}$ in lilac and $\mathcal{C}_{\tvec{0}{2}}^{\tvec{3}{0}}$ in yellow).
        Each subsystem label is represented by a square marker on a lattice, where they are ordered across scales and positions as shown on the left.
        (b) For every subsystem, we compute its inclusion-exclusion information $i_{\boldsymbol{n}}^{\boldsymbol{\ell}}$ in Eq.~\eqref{eq:rectangular_info}.
        The structure of the inclusion-exclusion equation is illustrated for the subsystem $\mathcal{C}_{\tvec{0}{1}}^{\tvec{3}{2}}$.
        We first take the total information in $\mathcal{C}_{\tvec{0}{1}}^{\tvec{3}{2}}$ and then subtract (blue) the information of the subsystems at scales $(3\ 1)$ and $(2\ 2)$ [first and second lines in \eqref{eq:rectangular_info}].
        The information from all pairwise subtractions (indicated by crossing the green curve) must be re-added (red) [third line in \eqref{eq:rectangular_info}].
        Higher-order intersections are re-subtracted and re-added until only one set is left.
        The green line separates levels of the inclusion-exclusion order of intersections: the subsystems to the left are intersections of those to the right (marked by the direction of the green arrow).
    }
\label{fig:inclusion_exclusion}
\end{figure}

We consider a system of $N_x\times N_y$ physical sites arranged on a square lattice as shown for a $4\times 4$ system in Fig.~\ref{fig:inclusion_exclusion}(a) (right).
Instead of connected intervals in 1D, we now consider rectangular subsystems $\mathcal{C}_{\boldsymbol{n}}^{\boldsymbol{\ell}}$ in 2D which are labeled by the coordinate of their bottom-left corner $\boldsymbol{n}=(n_x,n_y)$ and their scale $\boldsymbol{\ell}=(\ell_x,\ell_y)$, where a scale $\ell_x$ corresponds to a subsystems that extends over $\ell_x+1$ physical sites along the $x$ direction and $\ell_y+1$ physical sites along the $y$ direction.
We represent each pair $(\boldsymbol{n},\boldsymbol{\ell})$ by a vertex (square marker) in Fig.~\ref{fig:inclusion_exclusion}(a) labeling the subsystems $\mathcal{C}_{\boldsymbol{n}}^{\boldsymbol{\ell}}$.
This mapping between vertices and subsystems is illustrated for the subsystems $\mathcal{C}^{\tvec{1}{1}}_{\tvec{2}{0}}$ and $\mathcal{C}^{\tvec{3}{0}}_{\tvec{0}{2}}$ in Fig.~\ref{fig:inclusion_exclusion}(a).
The inclusion-exclusion information \eqref{eq:inclusion_exclusion} can be evaluated for the rectangular subsystems, which explicitly gives
\begin{align}
    i_{\boldsymbol{n}}^{\boldsymbol{\ell}} &= I^{\boldsymbol{\ell}}_{\boldsymbol{n}} \nonumber \\[4pt]
    &\;
    -\Big(
        I^{\boldsymbol{\ell} \smminus \tvec{1}{0}}_{\boldsymbol{n}}
        + I^{\boldsymbol{\ell} \smminus \tvec{1}{0}}_{\boldsymbol{n} \smplus \tvec{1}{0}}
        + I^{\boldsymbol{\ell} \smminus \tvec{0}{1}}_{\boldsymbol{n}}
        + I^{\boldsymbol{\ell} \smminus \tvec{0}{1}}_{\boldsymbol{n} \smplus \tvec{0}{1}}
    \Big) \nonumber \\[6pt]
    &\;+
    \Big(
        I^{\boldsymbol{\ell} \smminus \tvec{1}{1}}_{\boldsymbol{n}}
        \!+\! I^{\boldsymbol{\ell} \smminus \tvec{1}{1}}_{\boldsymbol{n} \smplus \tvec{1}{0}}
        \!+\! I^{\boldsymbol{\ell} \smminus \tvec{1}{1}}_{\boldsymbol{n} \smplus \tvec{0}{1}}
        \!+\! I^{\boldsymbol{\ell} \smminus \tvec{1}{1}}_{\boldsymbol{n} \smplus \tvec{1}{1}}
        \!+\! I^{\boldsymbol{\ell} \smminus \tvec{2}{0}}_{\boldsymbol{n} \smplus \tvec{1}{0}}
        \!+\! I^{\boldsymbol{\ell} \smminus \tvec{0}{2}}_{\boldsymbol{n} \smplus \tvec{0}{1}}
    \Big) \nonumber \\[4pt]
    &\;-
    \Big(
        I^{\boldsymbol{\ell} \smminus \tvec{2}{1}}_{\boldsymbol{n} \smplus \tvec{1}{0}}
        + I^{\boldsymbol{\ell} \smminus \tvec{2}{1}}_{\boldsymbol{n} \smplus \tvec{1}{1}}
        + I^{\boldsymbol{\ell} \smminus \tvec{1}{2}}_{\boldsymbol{n} \smplus \tvec{0}{1}}
        + I^{\boldsymbol{\ell} \smminus \tvec{1}{2}}_{\boldsymbol{n} \smplus \tvec{1}{1}}
    \Big) \nonumber \\[4pt]
    &\;
    +I^{\boldsymbol{\ell} \smminus \tvec{2}{2}}_{\boldsymbol{n} \smplus \tvec{1}{1}} \label{eq:rectangular_info}
\end{align}
for each pair $(\boldsymbol{n},\boldsymbol{\ell})$, where $I^{\boldsymbol{\ell}}_{\boldsymbol{n}}$ is the von Neumann information of the reduced density matrix on the subsystem $\mathcal{C}^{\boldsymbol{\ell}}_{\boldsymbol{n}}$.
The evaluation of \eqref{eq:rectangular_info} is represented visually in Fig.~\ref{fig:inclusion_exclusion}(b) for the vertex $(\boldsymbol{n},\boldsymbol{\ell})=(\svec{0}{1},\svec{3}{2})$: the first line in \eqref{eq:rectangular_info} is the total information in $\mathcal{C}_{\tvec{0}{1}}^{\tvec{3}{2}}$.
The second line of \eqref{eq:rectangular_info} subtracts the information of the next smaller subsystems [i.e., those at scales $(\ell_x-1,\ell_y)$ and $(\ell_x,\ell_y-1)$].
Then, the information of all possible pairwise intersections [separated by a green line and an intersection symbol in Fig.~\ref{fig:inclusion_exclusion}(b)] is re-added in the third line.
The last two lines in \eqref{eq:rectangular_info} re-subtract and re-add the information of higher order intersections.
In Eq.~\eqref{eq:rectangular_info}, it is implicitly assumed that a subsystem with indices $\ell_x$ or $\ell_y$ smaller than zero is empty, with zero information.

It is useful to read Eq.~\eqref{eq:rectangular_info} not only as an inclusion-exclusion formula, but also as a successive local subtraction along the two coordinate directions.
To make this explicit, define the directional difference operator
\begin{align}
\Delta_x I_{\boldsymbol{n}}^{\boldsymbol{\ell}}
=
I_{\boldsymbol{n}}^{\boldsymbol{\ell}}
- I^{\boldsymbol{\ell}\smminus\tvec{1}{0}}_{\boldsymbol{n}}
- I^{\boldsymbol{\ell}\smminus\tvec{1}{0}}_{\boldsymbol{n}\smplus\tvec{1}{0}}
+ I^{\boldsymbol{\ell}\smminus\tvec{2}{0}}_{\boldsymbol{n}\smplus\tvec{1}{0}},
\label{eq:delta_x_rectangular}
\end{align}
and analogously $\Delta_y$ by replacing $(1\ 0)$ with $(0\ 1)$ and $(2\ 0)$ with $(0\ 2)$.
Then Eq.~\eqref{eq:rectangular_info} can be written compactly as
\begin{align}
i_{\boldsymbol{n}}^{\boldsymbol{\ell}}
=
\Delta_y \Delta_x I_{\boldsymbol{n}}^{\boldsymbol{\ell}}
=
\Delta_x \Delta_y I_{\boldsymbol{n}}^{\boldsymbol{\ell}}.
\label{eq:rectangular_delta}
\end{align}
The first subtraction, $\Delta_x I_{\boldsymbol{n}}^{\boldsymbol{\ell}}$, compares the information inferred from the density matrix on $\mathcal{C}^{\boldsymbol{\ell}}_{\boldsymbol{n}}$ with what can already be inferred from the two density matrices obtained by reducing the subsystem by one layer along the $x$ direction.
It therefore isolates the net information that becomes accessible when the subsystem is enlarged along $x$, while keeping its position and scale fixed along $y$.
Applying $\Delta_y$ then asks whether this contribution is itself new at the present position and scale along the transverse direction, or whether it can already be inferred from smaller overlapping subsystems obtained by reducing the subsystem along $y$.
In this sense, the rectangular inclusion-exclusion information resolves the information gained along one direction by its position and scale along the other direction.
A positive value indicates a net gain of information assigned to $\mathcal{C}^{\boldsymbol{\ell}}_{\boldsymbol{n}}$, while a negative value indicates that the transverse subtraction removes information that was redundantly accessed through overlapping smaller subsystems.

\begin{figure}
    \centering
    \includegraphics[width=\linewidth]{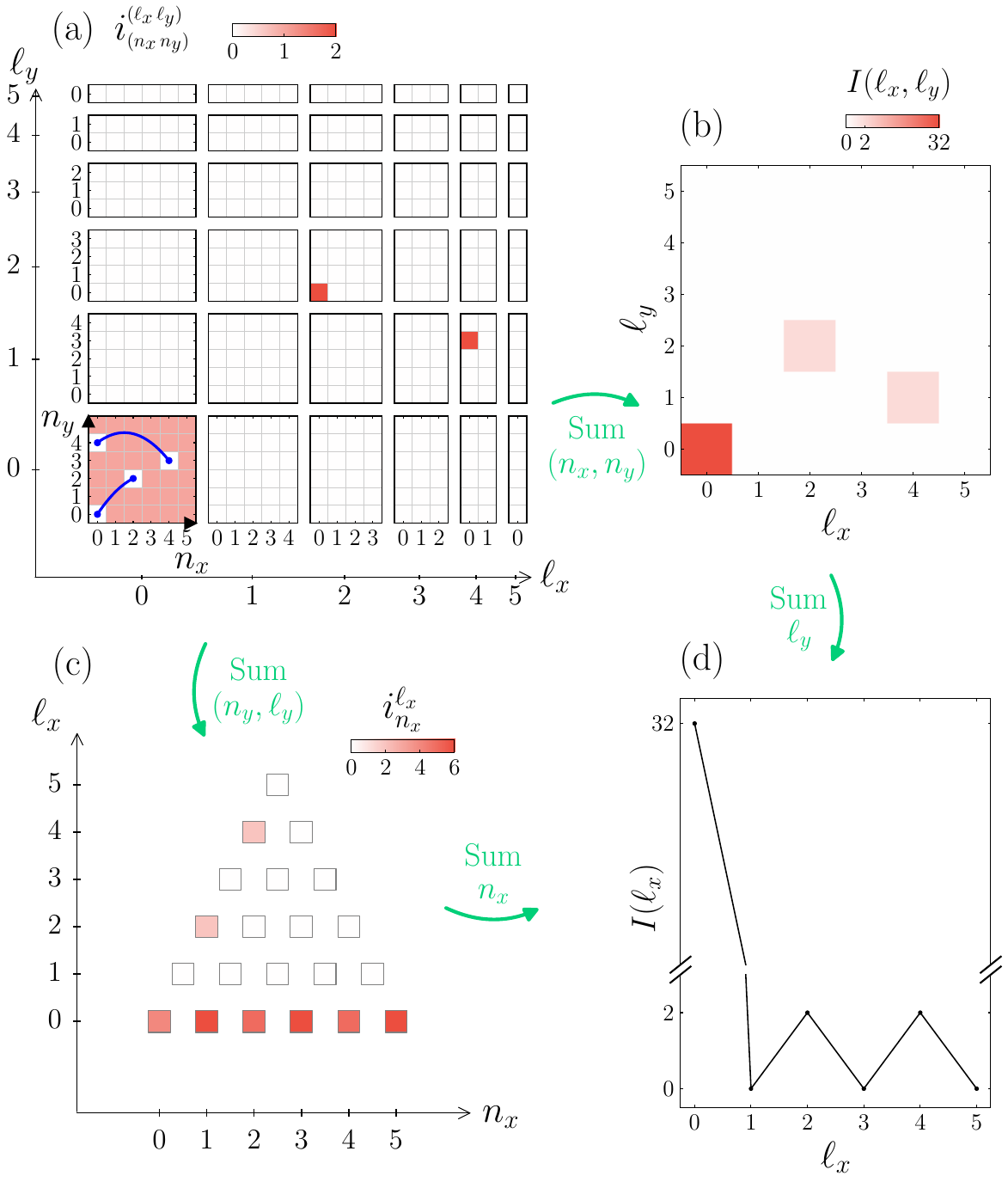}
    \caption{
        (a) We decompose the information of the state \eqref{eq:singlets}, consisting of all sites in a spin-up configuration except for four sites forming two singlets (illustrated by the blue lines), by evaluating the inclusion-exclusion information in Eq.~\eqref{eq:rectangular_info}.
        For each vertex $(\svec{n_x}{n_y},\svec{\ell_x}{\ell_y})$ (square marker), the inclusion-exclusion information $i_{\tvec{n_x}{n_y}}^{\tvec{\ell_x}{\ell_y}}$ is represented by a color.
        The information of the singlets is captured at the smallest scales that fully enclose them (at $(\svec{0}{0},\svec{2}{2})$ and $(\svec{0}{3},\svec{4}{1})$ respectively).
        (b) Summing out $(n_x\ n_y)$ leaves the total scale resolved information of the state $I(\ell_x, \ell_y)$: 32 bits are at $(0\ 0)$ while 2 bits are found at $(2\ 2)$ and 2 others at $(4\ 1)$.
        (c) Summing out an axis $(n_y\ \ell_y)$ leaves a quasi-1D information lattice over scales $(n_x\ \ell_x)$, where the information of the state is only resolved along the $x$ direction.
        (d) Summing (b) over $\ell_y$ or (c) over $n_x$ gives the quasi-1D information per scale along $\ell_x$.
	}
    \label{fig:singlets}
\end{figure}

We illustrate the 2D information lattice with a simple state on a lattice of $6\times6$ sites, in which all sites are in the spin-up configuration, except for two entangled pairs
\begin{equation}
    |{\Psi}_\mathrm{singlets}\rangle = \Bigl(\!
      \bigotimes_{\substack{\tvec{n_x}{n_y}\notin
        \{\tvec{0}{0},\\ \tvec{2}{2}, \tvec{0}{4},\tvec{4}{3}\}}}
          |\!\uparrow\rangle_{\tvec{x}{y}}
     \!\Bigr)
     \!\otimes\!
     |\Psi_{-}\rangle_{\tvec{0}{0},\tvec{2}{2}}
     \!\otimes\!
     |\Psi_{-}\rangle_{\tvec{0}{4},\tvec{4}{3}}
    \label{eq:singlets}
\end{equation}
where $|\Psi_{-}\rangle_{\tvec{n_x}{n_y},\tvec{n'_x}{n'_y}} = \frac{1}{\sqrt{2}}(|\!\!\uparrow\rangle_{\tvec{n_x}{n_y}}|\!\!\downarrow\rangle_{\tvec{n'_x}{n'_y}} - |\!\!\downarrow\rangle_{\tvec{n_x}{n_y}}|\!\!\uparrow\rangle_{\tvec{n'_x}{n'_y}})$ denotes a singlet between the sites positioned at $(n_x,n_y)$ and $(n'_x,n'_y)$ [these positions are connected by blue lines in Fig.~\ref{fig:singlets}(a)].
A single-site subsystem carries 1 bit of information when the site is unentangled (its density matrix is pure) but 0 bits when it belongs to a singlet (the density matrix is maximally mixed).
This is what the distribution of information shows in the lowest scale $\boldsymbol{\ell}=(0,0)$ of the information lattice in Fig.~\ref{fig:singlets}(a).
Each entangled pair contributes 2 bits of information at the smallest scale that encompasses it entirely: in this case $|\Psi_{-}\rangle_{\tvec{0}{0},\tvec{2}{2}}$ is fully enclosed in the subsystem $\mathcal{C}_{\tvec{0}{0}}^{\tvec{2}{2}}$ and $|\Psi_{-}\rangle_{\tvec{0}{4},\tvec{4}{3}}$ in $\mathcal{C}_{\tvec{0}{3}}^{\tvec{4}{1}}$.
Operationally, 1 bit is, e.g., associated with the measurement of $\hat X\hat X$ on a given singlet, which yields the outcome $-1$ with certainty, while the other bit is associated with the measurement of $\hat Y\hat Y$, which yields $+1$, for the Pauli operators $\hat X$ and $\hat Y$.

\subsection{Reduced information quantities}

In analogy with the information per scale Eq.~\eqref{eq:informationonscaleinaregion} for 1D, we define reduced information quantities in 2D by partial summations over indices.
We define the information at scale $\boldsymbol{\ell}'$ inside of a subsystem $\mathcal{C}_{\boldsymbol{n}}^{\boldsymbol{\ell}}$ as
\begin{equation}
	I(\boldsymbol{\ell}')\big|_{\mathcal{C}_{\boldsymbol{n}}^{\boldsymbol{\ell}}} = \sum_{\{\boldsymbol{n}^\prime \mid \mathcal{C}^{\boldsymbol{\ell}^\prime}_{\boldsymbol{n}^\prime} \subseteq \mathcal{C}^{\boldsymbol{\ell}}_{\boldsymbol{n}}\}} i^{\boldsymbol{\ell}'}_{\boldsymbol{n}'},
	\label{eq:informationonscaleinaregion_2d}
\end{equation}
which we abbreviate as $I(\ell_x,\ell_y)$ or $I(\boldsymbol{\ell})$ when $\mathcal{C}_{\boldsymbol{n}}^{\boldsymbol{\ell}}$ is the whole system.
$I(\ell_x,\ell_y)$ is illustrated in Fig.~\ref{fig:singlets}(b) for the singlets state in Eq.~\eqref{eq:singlets}: it shows that most information is in the single site scales while two bits appear at two intermediate scales.
We can also consider partial summations over an entire axis, which leaves us with a quasi-1D information lattice along the remaining axis.
For example, summing over the $y$ axis inside a subsystem $\mathcal{C}_{\boldsymbol{n}}^{\boldsymbol{\ell}}$
\begin{equation}
	i_{n_x'}^{\ell_x'}\Big|_{\smash{\mathcal{C}_{\boldsymbol{n}}^{\boldsymbol{\ell}}}} =
   \sum_{\{ \tvec{n_y'}{\ell_y'} | \mathcal{C}^{\boldsymbol{\ell}^\prime}_{\boldsymbol{n}^\prime} \subseteq \mathcal{C}^{\boldsymbol{\ell}}_{\boldsymbol{n}}\}} i^{\boldsymbol{\ell}'}_{\boldsymbol{n}'},
   \label{eq:quasi-1d-local}
\end{equation}
is the quasi-1D local information inside $\mathcal{C}_{\boldsymbol{n}}^{\boldsymbol{\ell}}$, which we abbreviate as $i_{\smash{n_{x}}}^{\smash{\ell_{x}}}$ when $\mathcal{C}_{\boldsymbol{n}}^{\boldsymbol{\ell}}$ is the whole system.
Analogously to the 1D local information \eqref{eq:local_information}, the quasi-1D local information \eqref{eq:quasi-1d-local} can be conveniently placed on a triangular structure with vertices labeled by $(n_x',\ell_x')$; this is illustrated in Fig.~\ref{fig:singlets}(c) for $i_{\smash{n_{x}}}^{\smash{\ell_{x}}}$.
The quasi-1D information lattice tells us that 2 bits become accessible at position and scale $(n_x, \ell_x)=(0, 2)$ and also at $(n_x, \ell_x)=(0, 4)$, but leaves out the resolution along the $y$ direction.
Finally, we define the quasi-1D information at scale $\ell_x'$ inside $\mathcal{C}_{\boldsymbol{n}}^{\boldsymbol{\ell}}$ by summing over the position $n_x'$ in Eq.~\eqref{eq:quasi-1d-local}, or equivalently by summing over the scale $\ell_y'$ in Eq.~\eqref{eq:informationonscaleinaregion_2d},
\begin{equation}
	I(\ell_x')\big|_{\smash{\mathcal{C}_{\boldsymbol{n}}^{\boldsymbol{\ell}}}} = \sum_{n_x'} i_{n_x'}^{\ell_x'}\Big|_{\smash{\mathcal{C}_{\boldsymbol{n}}^{\boldsymbol{\ell}}}}
    \label{eq:quasi-1d-local-scale},
\end{equation}
which we abbreviate as $I(\ell_x)$ when $\mathcal{C}_{\boldsymbol{n}}^{\boldsymbol{\ell}}$ is the whole system.
This tells us the scale in the $x$ direction at which information became accessible.

\subsection{General quasi-1D constructions}

\begin{figure}
    \centering
    \includegraphics[width=\linewidth]{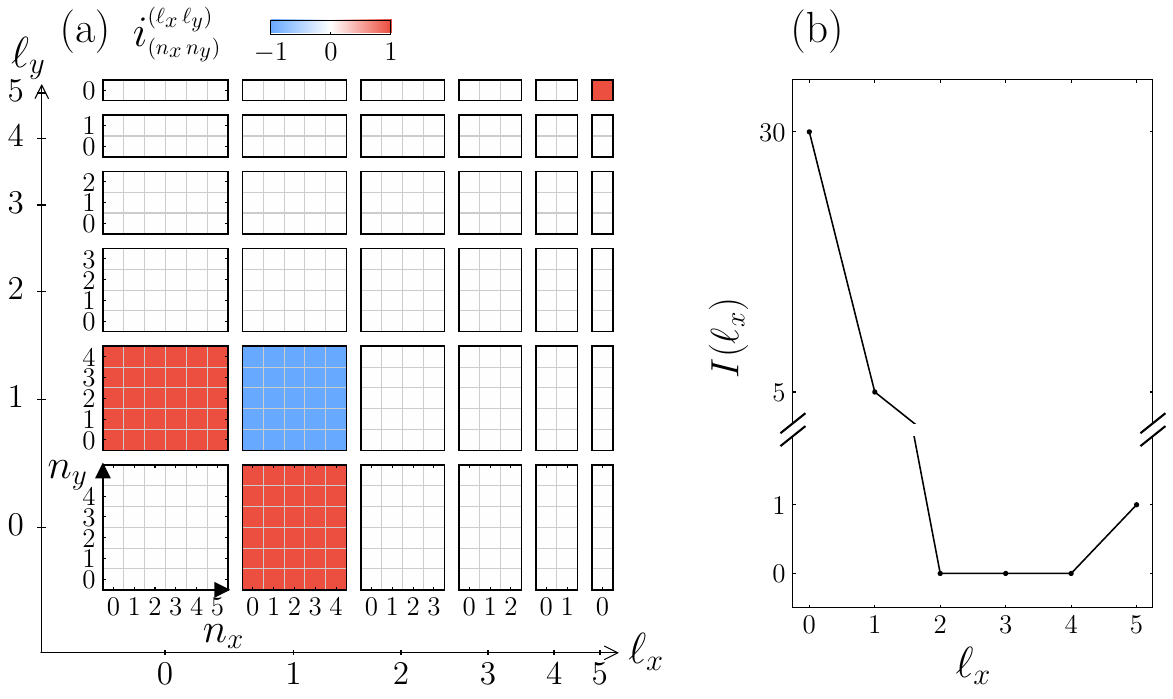}
    \caption{(a) 2D information lattice of the cat state defined in Eq.~\eqref{eq:cat_state}, where the reading of the lattice is identical to Fig.~\ref{fig:singlets}(a).
        There are now negative values at scale $\boldsymbol{\ell}=(1\ 1)$, which compensate for the over-counting of information at scales $\boldsymbol{\ell}=(1\ 0)$ and $(0\ 1)$.
        This signals that the overlapping subsystems at those lower scales have dependent density matrices, what we call an overlap redundancy.
		(b) The quasi-1D information per scale along the $x$ direction defined in \eqref{eq:quasi-1d-local-scale} resolves the information along the $x$ direction, and therefore it is insensitive to loop redundancies.
        Both (a) and (b) show a bit at the largest scale: in cat states a bit of information cannot be inferred by any local measurement and requires access to the entire system.
    }
    \label{fig:ghz}
\end{figure}

The quasi-1D information quantities defined in Eqs.~\eqref{eq:quasi-1d-local} and \eqref{eq:quasi-1d-local-scale} are nonnegative.
This can be seen by substituting Eq.~\eqref{eq:rectangular_info} in \eqref{eq:quasi-1d-local}, which simplifies into
\begin{align}
	i_{n_x'}^{\ell_x'}\big|_{\smash{\mathcal{C}_{\boldsymbol{n}}^{\boldsymbol{\ell}}}}
    & = I^{\boldsymbol{\ell}}_{\boldsymbol{n}}
   - I^{\boldsymbol{\ell} - \tvec{1}{0}}_{\boldsymbol{n}}
   - I^{\boldsymbol{\ell} - \tvec{1}{0}}_{\boldsymbol{n} + \tvec{1}{0}}
   + I^{\boldsymbol{\ell} - \tvec{2}{0}}_{\boldsymbol{n} + \tvec{1}{0}}.
    \label{eq:quasi-1d-expanded}
\end{align}
This has an identical form to the 1D local information \eqref{eq:local_information}.
Here, the $y$ direction has been coarse grained: the physical sites along $y$ are grouped together to form a new elementary site with enlarged Hilbert space; the quasi-1D local information then captures correlations only along $x$.
To illustrate this we use the cat state
\begin{equation}
    |{\Psi_\text{cat}}\rangle = \frac{1}{\sqrt{2}} \left( |{\uparrow\uparrow\cdots\uparrow}\rangle + e^{i\phi}|{\downarrow\downarrow\cdots\downarrow}\rangle \right)
    \label{eq:cat_state}
\end{equation}
placed on $6\times 6$ qubits with relative angle $\phi=0$.
The reduced density matrices of $|{\Psi_\text{cat}}\rangle$ have the form $\tfrac12(|\!\!\uparrow\uparrow\!...\rangle\!\langle\uparrow\uparrow\!...|+|\!\!\downarrow\downarrow\!...\rangle\!\langle \downarrow\downarrow\!...|)$ on each subsystem.
The negative information values at scale $\boldsymbol{\ell}=(1\ 1)$ in Fig.~\ref{fig:ghz}(a) show that there is an information redundancy.
Indeed, the density matrices at scales $\boldsymbol{\ell}=(1\ 0)$ and $\boldsymbol{\ell}=(0\ 1)$ are not independent (they are precisely those discussed in Sec.~\ref{sec:obstacles}).
This contrasts with the quasi-1D information per scale in the $x$ direction \eqref{eq:quasi-1d-local-scale} shown in Fig.~\ref{fig:ghz}(b).
As in 1D, \eqref{eq:quasi-1d-expanded} quantifies how much more information can be inferred about the distribution of measurement outcomes on $\mathcal{C}_{\boldsymbol{n}}^{\boldsymbol{\ell}}$ from $\rho_{\smash{\mathcal C_{\boldsymbol{n}}^{\boldsymbol{\ell}}}}$ than what can be inferred from the two smaller density matrices $\{\rho_{\smash{\mathcal C_{\boldsymbol{n}}^{\boldsymbol{\ell}-\tvec{1}{0}}}},\rho_{\smash{\mathcal C_{\boldsymbol{n}+\tvec{1}{0}}^{\boldsymbol{\ell}-\tvec{1}{0}}}}\}$ on the subsystems that are reduced by a single unit along the $x$ direction.
The bit of information at the largest scale is present in both cases, as the phase $\phi$ can only be learned globally and is invisible to any local measurement.

\begin{figure}
    \centering
    \includegraphics[width=.9\linewidth]{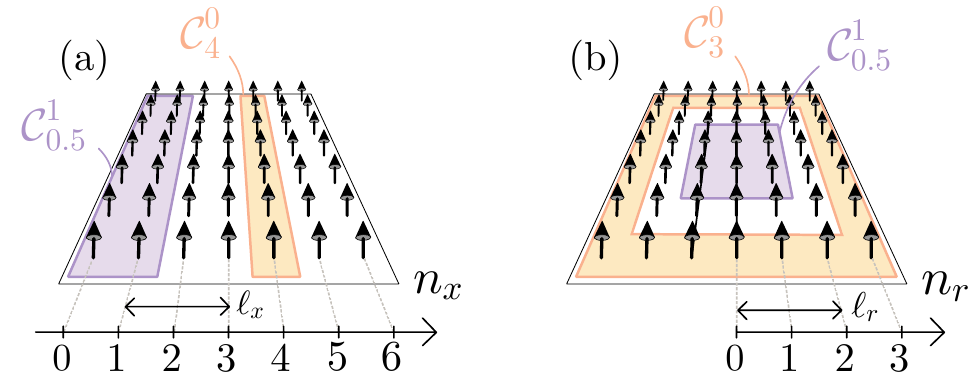}
    \caption{In the information lattice, a higher-dimensional system can be treated as effectively one-dimensional by evaluating the local information [Eq.~\eqref{eq:local_information}] along a chosen 1D direction (specified by position $n$ and scale $\ell$) while coarse graining the other directions into a single site with an enlarged Hilbert space: this yields a \textit{quasi-1D information lattice}.
    (a) illustrates this construction for subsystems $\mathcal{C}^{\ell_x}_{n_x}$ at different horizontal positions $n_x$ and scales $\ell_x$ (e.g., $\mathcal{C}_{0.5}^1$ in lilac and $\mathcal{C}_{4}^0$ in orange), with the $y$ direction coarse grained.
    This produces the quasi-1D local information $i_{n_x}^{\ell_x}$ from Eq.~\eqref{eq:quasi-1d-expanded}.
    (b) shows an analogous construction where Eq.~\eqref{eq:local_information} resolves correlations along the radial direction, treating concentric square rings $\mathcal{C}^{\ell_r}_{n_r}$ (radius $n_r$, thickness $\ell_r$) as the sites of a 1D chain.
    For instance, $\mathcal{C}_{0}^{3}$ (orange) is a ring centered at square radius $n_r=3$ with thickness $\ell_r=1$, whereas $\mathcal{C}_{0.5}^{1}$ (lilac) is a ring at $n_r=0.5$ whose thickness $\ell_r=1$ makes it coincide with a filled square at the origin.
    }
    \label{fig:quasi1d}
\end{figure}

We note that quasi-1D generalizations can also be constructed by directly selecting subsystems ordered along a one-dimensional position $n$ and scale $\ell$ (that is, by considering set chains in higher-dimensions), and evaluating the local information [Eq.~\eqref{eq:local_information}] as in 1D.
Fig.~\ref{fig:quasi1d} illustrates two such constructions: one obtained by coarse graining the $y$ direction, and another by coarse graining concentric square rings around the system center.

\section{Characterizing Quantum States through the 2D Information Lattice\label{sec:inclusion_exclusion}}

Quantum states of one-dimensional chains are universally characterized by their distribution of local information~\cite{artiaco2025universal}.
Building on this, we show how the inclusion-exclusion information \eqref{eq:inclusion_exclusion} specialized to rectangular subsystems \eqref{eq:rectangular_info} can be used to characterize quantum states in two-dimensional systems.

\subsection{Localized and critical ground states of the 2D Anderson model} \label{sec:gapped_critical}

We consider the ground state $\ket{\Psi_\mathrm{AM}}$ of the disordered Anderson model (AM) \cite{anderson1958absence, abrahams201050} on a two-dimensional square lattice of size $N_x\times N_y$ defined by the Hamiltonian
\begin{equation}
    \hat{H}_\mathrm{AM} = -\sum_{i} (t_x\hat{c}_i^\dagger \hat{c}_{i+\hat x} + t_y\hat{c}_i^\dagger \hat{c}_{i+\hat y} + \text{H.c.}) + \sum_i \epsilon_i \hat{c}_i^\dagger \hat{c}_i,
    \label{eq:anderson_hamiltonian}
\end{equation}
where $\hat{c}_i^\dagger$ ($\hat{c}_i$) creates (annihilates) an electron at site $i$, and $\epsilon_i$ are independent random onsite potentials uniformly distributed in $[-W/2,W/2]$ with disorder strength $W$.
The unit vectors $\hat x$ and $\hat y$ connect each site $i$ to its nearest neighbors in the horizontal and vertical directions with hopping amplitudes $t_x$ and $t_y$ that may differ to allow for anisotropy.
We work at half filling. 
Since $\hat{H}_\mathrm{AM}$ is noninteracting, the von Neumann information of the subsystem density matrices of $\ket{\Psi_\mathrm{AM}}$ can be efficiently calculated from the single-particle eigenstates~\cite{peschel2003calculation}.
At zero disorder ($W=0$), the eigenstates are extended and the ground state is a gapless critical state.
In one dimension and in two dimensions for symmetry classes such as the orthogonal and unitary classes, any nonzero uncorrelated disorder $W>0$ produces exponentially localized states~\cite{abrahams1979scaling,evers2008anderson}.
By contrast, in three dimensions a localization-delocalization transition occurs at a finite disorder strength even for uncorrelated disorder, whereas in two dimensions this happens only in special symmetry classes such as the symplectic class~\cite{evers2008anderson}.
Here we consider the ground state in both the localized phase ($W\neq 0$) and the clean case ($W=0$) of the 2D Anderson model in the orthogonal class.

\begin{figure}
    \centering
    \includegraphics[width=\linewidth]{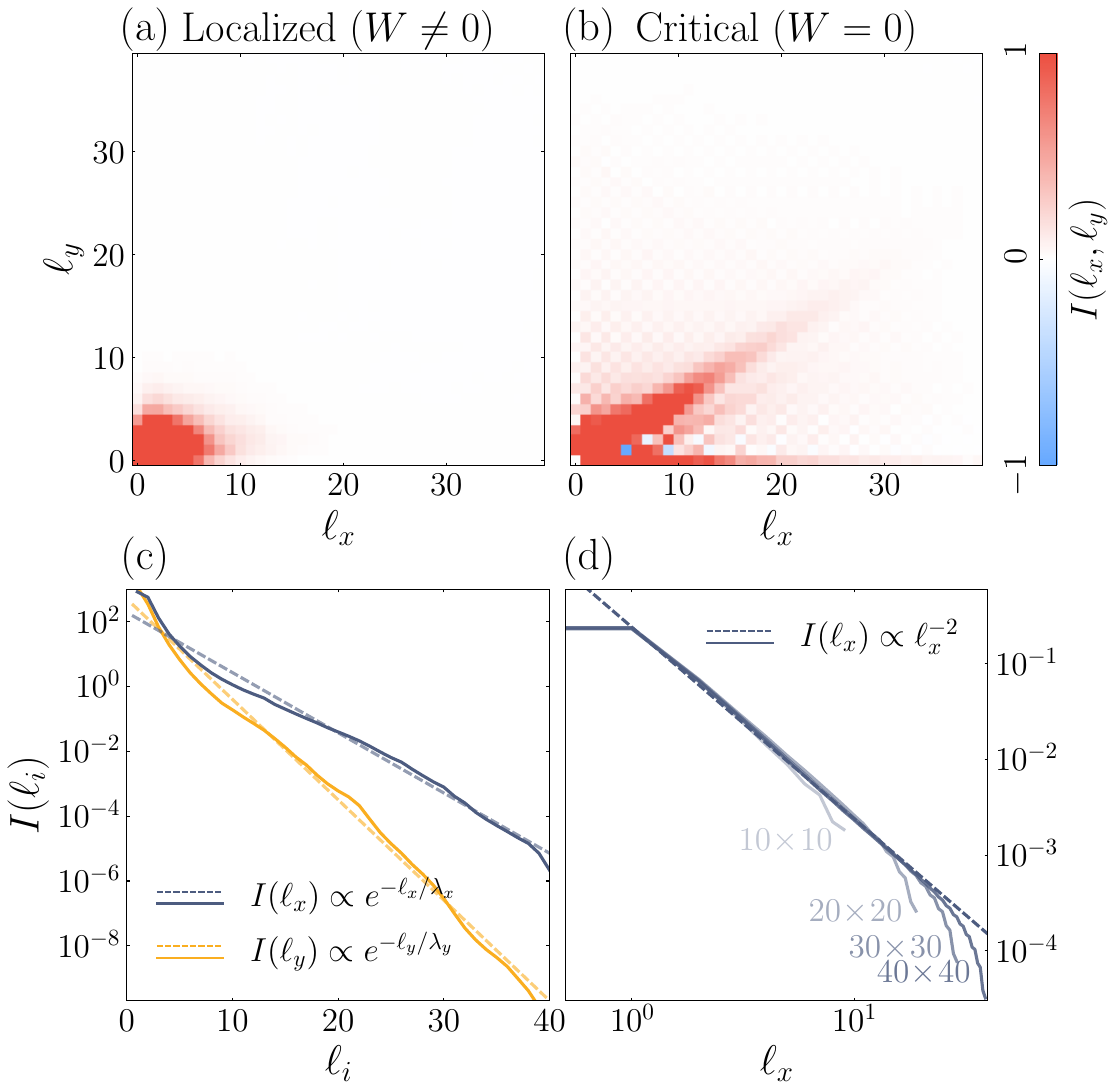}
    \caption{
    Inclusion-exclusion information \eqref{eq:rectangular_info} of the ground state of the Anderson model on a $40\times40$ lattice with hopping amplitudes $t_x=1.5$ and $t_y=1$ [see Eq.~\eqref{eq:anderson_hamiltonian}].
    (Left) For a single disorder realization at $W=10$, panel (a) shows the information per multiscale, concentrated at short scales.
    The anisotropy results in a larger spread of information along the $\ell_x$ direction.
    Panel (c) displays the quasi-1D information per scale $I(\ell_\mu)$ along $x$ and $y$ (solid lines, gray and orange respectively). Exponential fits \eqref{eq:exp_fit} (dashed lines) yield the correlation decay lengths $\lambda_x$ and $\lambda_y$ that characterize the 2D localized phase.
    (Right) For $W=0$, the ground state is critical. (d) The quasi-1D information per scale (solid lines, for multiple system sizes) follows the 1D scale-invariant form $I(\ell_x)\propto \ell_x^{-2}$ (a similar behavior holds along $\ell_y$, not shown), with the fitted power law indicated by a dashed line.
    In panel (b), the information per multiscale is oriented along a dominant direction, which we refer to as the information propagation direction.
    }
    \label{fig:fig4}
\end{figure}

In 1D systems, localized states are characterized by an exponential decay of the total information per scale~\eqref{eq:informationonscaleinaregion}, from which a \textit{correlation decay length} $\lambda$ can be defined as an intrinsic measure of localization~\cite{artiaco2025universal}.
In two dimensions, we may analogously define the correlation decay lengths $\lambda_\mu$ for $\mu\in\{x,y\}$ through the exponential fit of the total information per scale along each quasi-1D direction.
Explicitly,
\begin{equation}
    \ln I(\ell_\mu) = -\ell_\mu/\lambda_\mu + \mathrm{const} \label{eq:exp_fit}.
\end{equation}
A complementary approach consists in defining such decay lengths directly on the inclusion-exclusion information at different positions.
To illustrate this we consider the ground state $\ket{\Psi_\mathrm{AM}}$ for a single disorder realization of the Hamiltonian \eqref{eq:anderson_hamiltonian} with anisotropic hopping $t_x=1.5$ and $t_y=1$, and $W=10$ on a lattice of $N_x\times N_y=40\times40$ sites.
The information per multiscale is illustrated in Fig.~\ref{fig:fig4}(a), showing an elliptically shaped profile at short scales, characteristic of localized states.
The corresponding quasi-1D information per scale is shown in Fig.~\ref{fig:fig4}(c), together with least-squares exponential fits to extract $\lambda_\mu$ along both directions.
The correlation decay lengths offer a system-agnostic characterization of localization that can be extracted from the density matrix and they also apply in interacting states.

We now turn to the gapless critical ground state in the clean case ($W=0$).
Unlike in 1D, where criticality is fully captured by the scale invariance of local information~\cite{artiaco2025universal}, higher-dimensional critical states exhibit substantially richer behavior~\cite{eisert2010colloquium}.
To examine the critical behavior, we first adopt a quasi-1D reduction: we treat the $y$-axis as an internal degree of freedom and analyze scale invariance along $\hat x$ (and vice versa).
In this reduced description, the critical ground state is scale invariant in the sense of Ref.~\cite{artiaco2025universal}, which implies
\begin{equation}
    I(\ell_\mu)\propto \ell_\mu^{-2}.
\end{equation}
This power law is represented in Fig.~\ref{fig:fig4}(d), where it is only represented for $I(\ell_x)$ for visibility.

To go beyond the quasi-1D reduction, we examine the information per multiscale in Fig.~\ref{fig:fig4}(b).
We find that the information extends preferentially along a specific scale direction.
We define the \textit{information propagation direction} $\langle\hat{\boldsymbol{v}}\rangle$ through a least-squares fit as the direction of maximum variance~\cite{joliffe2002pca}
\begin{align}
    \langle\hat{\boldsymbol{v}}\rangle = \underset{||\boldsymbol \alpha||=1}{\mathrm{arg\ max}} \ \, \boldsymbol{\alpha}^{\!\top}\Sigma\, \boldsymbol{\alpha} \label{eq:propagation_direction}
\end{align}
where $\Sigma=\sum_{\boldsymbol{\ell}} I(\boldsymbol{\ell}) \boldsymbol{\ell}\boldsymbol{\ell}^{\!\top} / \sum_{\boldsymbol{\ell}} I(\boldsymbol{\ell})$ is the $2\times2$ information per multiscale weighted covariance matrix.
$\langle\hat{\boldsymbol{v}}\rangle$ is the normalized eigenvector of $\Sigma$ with the largest eigenvalue: the unit vector along which a scale increase yields the largest information gain.
The inclusion-exclusion information $i_{\boldsymbol{n}}^{\boldsymbol{\ell}}$ decays with a power of $-2$ for values of $\boldsymbol{\ell}$ along $\langle\hat{\boldsymbol{v}} \rangle$ (not shown).
In general, the information per multiscale may extend along several directions.
The fit Eq.~\eqref{eq:propagation_direction} identifies the dominant among these directions by construction, providing a single effective information propagation direction.
In our case, this dominant direction is clearly visible in Fig.~\ref{fig:fig4}(b) as the bright, anisotropic feature extending along a direction close to the diagonal.
The information per multiscale displays a weaker contribution to the information along $\ell_x$ for subsystems with $\ell_y = 0$.
This contribution originates from the horizontal 1D chains embedded in the 2D system, including those in the bulk.
In the fully anisotropic limit of $t_y=0$, the 2D system reduces to decoupled critical horizontal 1D chains.
In this case, the inclusion-exclusion information at scale $\ell_y=0$ (horizontal chains) decays with a power law $\propto \ell_x^{-2}$, while at scale $\ell_x=0$ (vertical chains) there is no information gain along the $\ell_y$ direction.
This scaling behaviour is observed in the weakly anisotropic case considered here as well (scaling $\propto \ell_x^{-2}$ not shown).
We leave a full characterization of why this scaling persists beyond this limit to future work.

\begin{figure}
    \centering
    \includegraphics[width=\linewidth]{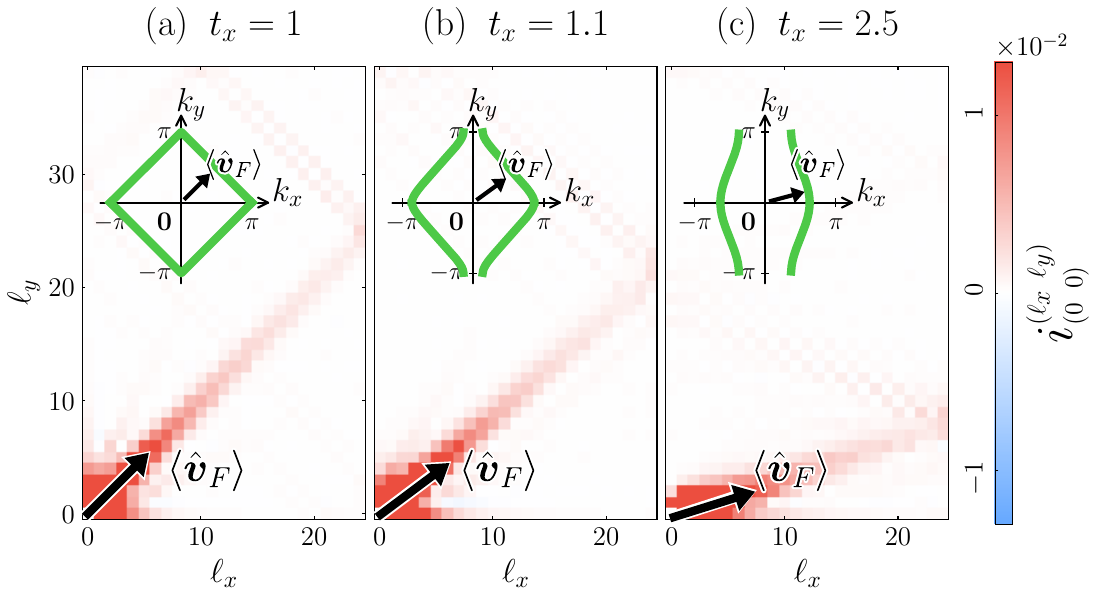}
    \caption{
        Inclusion-exclusion information $i_{\tvec{0}{0}}^{\tvec{\ell_x}{\ell_y}}$ (colormap) of a subsystem positioned at the bottom-left corner of the system, as its scale $(\ell_x,\ell_y)$ increases, in the ground state of the anisotropic Anderson model \eqref{eq:anderson_hamiltonian} at half-filling and at criticality $W=0$ on a lattice of $N_x\times N_y=25\times40$ sites.
        It is shown for (a) $t_x=1$, (b) $t_x=1.1$, and (c) $t_x=2.5$ with $t_y=1$.
        Insets show the Fermi surface in the Brillouin zone for each case (assuming periodic boundary conditions) together with the propagation direction $\langle \hat{\boldsymbol{v}}_F\rangle$ obtained by evaluating the average Fermi vector direction Eq.~\eqref{eq:fermi_direction} from the analytical expression of the Fermi surface.
        The same vectors $\langle \hat{\boldsymbol{v}}_F\rangle$ are represented in the main panels, where they show agreement with the information propagation direction  \eqref{eq:propagation_direction}.
        This demonstrates that the dominant propagation in the information per multiscale points in the average direction of the velocity of the chiral modes on the Fermi surface.
    }
    \label{fig:fig4_local}
\end{figure}

The information propagation direction is the direction along which enlarging the subsystem yields the greatest increase in information.
We note a correspondence with the Fermi velocity orientation, suggesting that this direction captures the critical modes responsible for long-range entanglement in gapless states.
Motivated by the Widom formula~\cite{gioev2006entanglement,swingle2010entanglement,ding2012entanglement}, which relates entanglement to the momentum orientations on the Fermi surface relative to the subsystem boundary, we express an average Fermi velocity direction specialized to rectangular subsystems~\footnote{
    For a subsystem $\mathcal{C}$ we can write the Widom formula as $S(L)=\tfrac{1}{(2\pi)^{d-1}}\tfrac{L^{d-1}\log L}{12}\int_{\partial \mathcal{C}}\int_{\partial \Gamma} |\hat n \cdot \hat v_F| \mathrm{d}S_x\mathrm{d}S_p$ where $\partial \mathcal{C}$ is the subsystem boundary and $\hat n$ its normal, $\partial \Gamma$ is the Fermi surface and $\hat v_F$ its normal.
    $L$ is the linear dimension of $\mathcal{C}$.
    Assuming $\mathcal{C}$ is a rectangular subsystem, we can then write this compactly as an integral over space, where each boundary normal in space is weighted with an average Fermi vector orientation $\langle\hat v_F\rangle$ as
    $
    S(L)=\tfrac{1}{(2\pi)^{d-1}}\tfrac{L^{d-1}\log L}{6}\int_{\partial \mathcal{C}} |\hat n|\cdot\langle\hat v_F\rangle \mathrm{d}S_x
    $
    where we defined the average Fermi vector orientation as $\langle\hat v_F\rangle=\int_{\partial \Gamma} |\hat v_F| \mathrm{d}S_p$ where in both equations $|\cdot|$ denotes the element-wise absolute value.
    $\langle \hat v_F \rangle$ is the Fermi-surface average of the unit vector representing the local angle of the Fermi velocity with respect to the surface.
} as
\begin{align}
\langle\hat{\boldsymbol{v}}_F\rangle\propto\int_{\mathbf{k}\in\mathrm{FS}} \begin{pmatrix}
    |\hat{\boldsymbol{v}}^x_F(\mathbf{k})| \\
    |\hat{\boldsymbol{v}}^y_F(\mathbf{k})|
\end{pmatrix} \; \mathrm{d}s, \label{eq:fermi_direction}
\end{align}
where the omitted proportionality factor is such that $\langle\hat{\boldsymbol{v}}_F\rangle$ is normalized.
Here $|\cdot|$ denotes the absolute value, $\hat{\boldsymbol{v}}_F^\mu(\mathbf{k})$ is the $\mu\in\{x,y\}$ component of the normalized Fermi velocity and $\mathrm{d}s$ is the length element along the Fermi surface $\mathrm{FS}=\{\mathbf k\,|\,t_x\cos k_x+t_y\cos k_y=0\}$ for the tight-binding model~\eqref{eq:anderson_hamiltonian} with $W=0$ and at half-filling.
The information gain is largest when the subsystem multiscale is increased along $\langle\hat{\boldsymbol{v}}_F\rangle$.
To illustrate this, we analyze the scale dependence of the inclusion-exclusion information $i_{\tvec{0}{0}}^{\tvec{\ell_x}{\ell_y}}$ of a single subsystem, located at the bottom left corner of the system.
We consider three values of $t_x$ with fixed $t_y=1$ on a system of $N_x\times N_y=25\times40$ sites, which allows us to change the Fermi surface.
In the isotropic case $t_x=1$, the Fermi surface is diamond shaped and $\langle \hat{\boldsymbol{v}}_F\rangle=(1\ 1)/\sqrt{2}$ [see inset of Fig.~\ref{fig:fig4_local}(a)].
The corresponding information gain is maximal along this direction [Fig.~\ref{fig:fig4_local}(a)].
For $t_x=1.1$, the deformation of the Fermi surface tilts $\langle \hat{\boldsymbol{v}}_F \rangle$, and the information propagation direction tilts accordingly [Fig.~\ref{fig:fig4_local}(b)].
For stronger anisotropy, $t_x=2.5$, the information propagation direction is mostly along the $x$-axis, again consistent with the direction of $\langle \hat{\boldsymbol{v}}_F \rangle$ [Fig.~\ref{fig:fig4_local}(c)].
These observations illustrate that the inclusion-exclusion information provides a natural way to identify dominant propagation directions in critical states of higher-dimensional systems.
A systematic analysis of these critical directions based on a renormalization approach to scale invariance in the 2D information lattice is left for future work.

\subsection{Gapped bulk with a critical edge in the chiral $p_x+ip_y$ superconductor} \label{sec:edge_mode}

In the previous section, we considered a 2D free-fermion system whose Fermi surface gives rise to gapless excitations throughout the bulk.
In contrast, we now consider a gapped 2D system whose bulk excitations are fully gapped while its boundary supports a single chiral edge mode.
An example of such a system is the ground state $\ket{\Psi_\mathrm{TSC}}$ of the spinless chiral $p_x+i p_y$ topological superconductor~\cite{read2000paired,ivanov2000non, volovik1999fermion, potter2010multichannel}, given on a square lattice by the Bogoliubov-de-Gennes Hamiltonian~\footnote{
    The Hamiltonian in Eq.~\eqref{eq:BdG} is obtained by discretizing the continuum $p_x+i p_y$ Hamiltonian of Read and Green given by Eq.~(1) in~\cite{read2000paired}, where we instead assumed the momentum space pairing $\Delta_\mathbf{k}=i\Delta(k_x+ik_y)$ with $\Delta$ constant, via the substitution $k_\mu \to \sin k_\mu$ and $k_\mu^2 \to 2(1-\cos k_\mu)$ in both components $\mu\in\{x,y\}$.
    With this convention, the topological regime $\mu>0$ of the continuum corresponds to $|\mu|<4t$ on the lattice~\cite{asahi2012topological,sato2009topological,potter2010multichannel}.
}
\begin{align}
\label{eq:BdG}
    \hat{H}_\mathrm{TSC} =& -\mu \sum_i \hat{c}_i^\dagger \hat{c}_i - t \sum_{i, \hat n\in\{\hat x,\hat y\}} (\hat{c}_i^\dagger \hat{c}_{i+\hat n} + \text{H.c.}) \nonumber \\
    &+ \frac{\Delta}{2} \sum_{i} (\hat{c}_i \hat{c}_{i+\hat x} + \text{H.c.}) + i\frac{\Delta}{2} \sum_{i} (\hat{c}_i \hat{c}_{i+\hat y} - \text{H.c.}).
\end{align}
Here, $\mu$ is the chemical potential, $t$ is the hopping amplitude, and $\Delta$ the pairing strength.
For $|\mu| > 4t$, the ground state is topologically trivial and short-range correlated with Chern number $C=0$.
For $|\mu| < 4t$, the ground state is in a chiral topological phase supporting a single chiral Majorana edge mode whose direction of propagation (chirality) reverses as $\mu$ changes sign~\cite{asahi2012topological,sato2009topological,shen2011topological} (here Majorana edge mode refers to the collective edge excitation which, unlike Majorana zero modes, need not occur in pairs~\cite{beenakker2020search}).

\begin{figure}
    \centering
    \includegraphics[width=\linewidth]{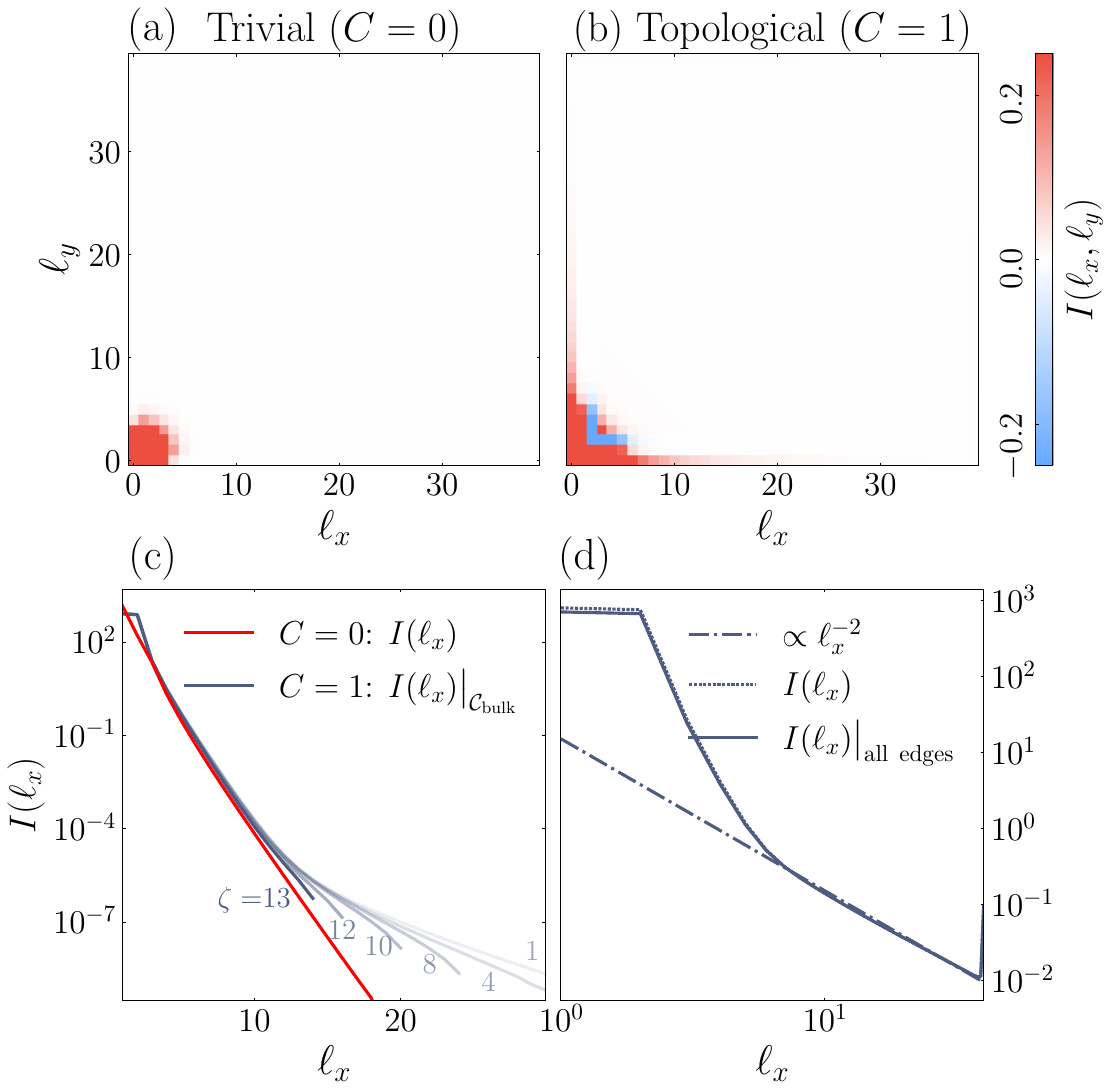}
    \caption{
       Information per multiscale $I(\ell_x,\ell_y)$ in the ground state of the $p_x+ip_y$ superconductor on a 40$\times$40 lattice in (a) the trivial phase ($\mu=6t$, Chern number $C=0$) and (b) the topological phase ($\mu=2t$, Chern number $C=1$).
        In the trivial ground state the information per multiscale is highly localized at short scales, while in the topological ground state it extends in the direction of 1D chain subsystems ($\ell_x=0$ or $\ell_y=0$).
        (c) The quasi-1D information per scale $I(\ell_x)$ in the trivial ground state (red line) decays exponentially with scale.
        (d) The quasi-1D information per scale $I(\ell_x)$ in the topological ground state (dotted line) decays algebraically with scale as $\ell_x^{-2}$ (dash-dotted line).
        By separating $I(\ell_x)$ into bulk and edge contributions, $I(\ell_x)|_{\mathcal{C}_\mathrm{bulk}}+I(\ell_x)|_\mathrm{all\ edges}$, we observe that the largest contribution at large scales come from the information on the edges (solid line).
        The remaining bulk contribution $I(\ell_x)|_{\mathcal{C}_\mathrm{bulk}}$ in the topological ground state is represented in (c) (gray curves).
        It approaches the exponential decay of the trivial ground state as the distance $\zeta$ (whose value labels each gray curve) between the edges and the bulk subsystem $\mathcal{C}_\mathrm{bulk}$ is increased.
        $I(\ell_x)|_{\mathcal{C}_\mathrm{bulk}}$ is normalized so that it sums to $N_x\times N_y$ bits.
        }
    \label{fig:fig5}
\end{figure}

The quasi-1D information per scale [Eq.~\eqref{eq:quasi-1d-local-scale}] for the ground states in the trivial ($\mu=6t,C=0$) and topological ($\mu=2t,C=1$) phases is shown in Fig.~\ref{fig:fig5}(c) and (d), respectively.
We set $\Delta=t=1$.
In the trivial ground state, $I(\ell_x)$ decays exponentially, signaling localized behavior, while in the topological ground state $I(\ell_x)$ decays algebraically, signaling critical behavior.
The corresponding information per multiscale is shown in Fig.~\ref{fig:fig5}(a) and (b).
Both states exhibit sharply peaked information at small multiscales.
In the topological state, we note the presence of negative values at scales $\ell_x=2$ or $\ell_y=2$.
They indicate that the same information can be accessed in overlapping subsystems at lower scales but different locations, similar to what is encountered at the short scales of the cat state in Fig.~\ref{fig:ghz}.
Furthermore, the information contribution of horizontal ($\ell_y=0$) and vertical ($\ell_x=0$) 1D chain subsystems extends further across length scales.
Unlike in Sec.~\ref{sec:gapped_critical}, where chain subsystem contributions at these scales originated from both the bulk and the edge subsystems, here they come only from the edge ones.

\begin{figure}
    \centering
    \includegraphics[width=.499\linewidth]{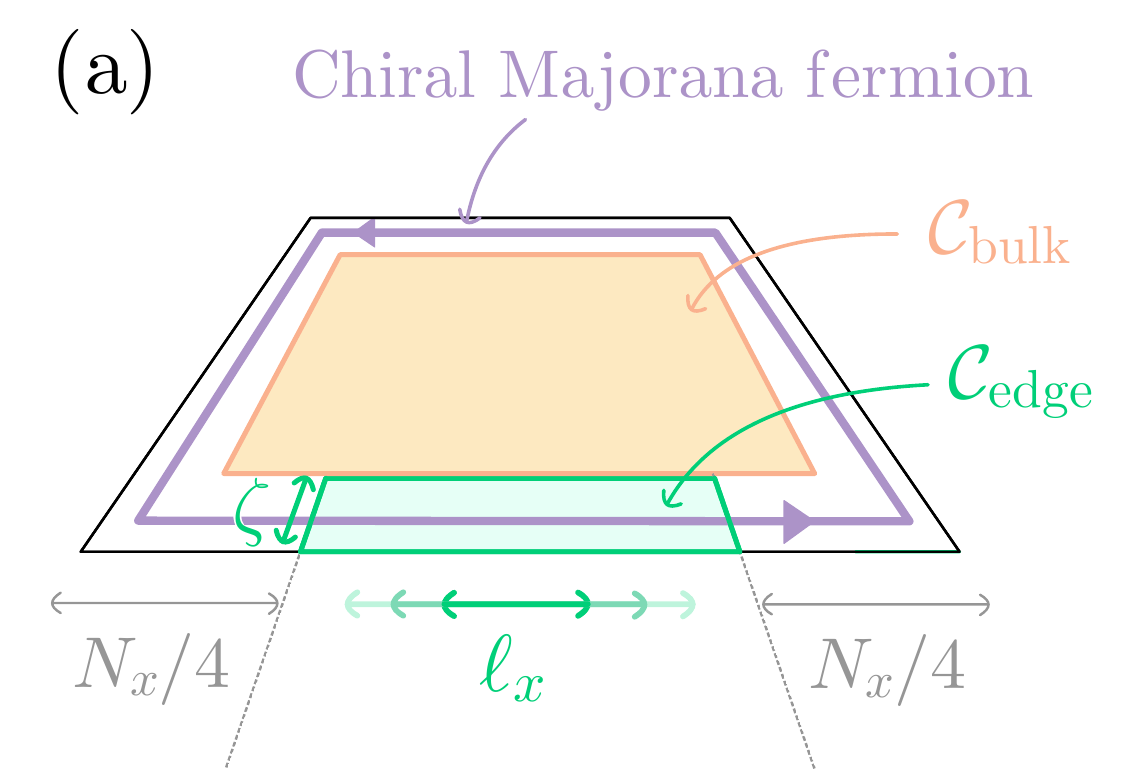}\includegraphics[width=.5\linewidth]{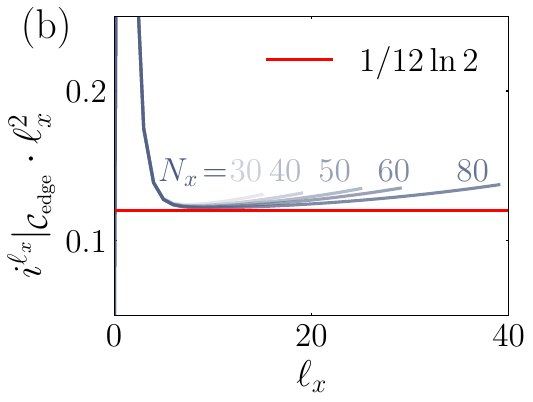}
    \caption{
        (a) The topological ground state $|\Psi_\mathrm{TSC}\rangle$ of the $p_x+ip_y$ superconductor supports a single chiral Majorana edge state at the boundary (represented in lilac).
        (b) The average value of the quasi-1D local information [Eq.~\eqref{eq:average-quasi-1D}] along the edge $i^{\ell_x}|_{\mathcal{C}_\mathrm{edge}}$ at each scale inside the subsystem $\mathcal{C}_\mathrm{edge}$ [shown in green in (a)] that covers the bottom edge.
        The subsystem $\mathcal{C}_\mathrm{edge}$ spans half the system horizontally ($N_x/2$) and has vertical scale $\zeta=3$.
        The scaling follows $i^{\ell_x}|_{\mathcal{C}_\mathrm{edge}} = \alpha \ell_x^{-2}$ as the system size $N_x$ is increased, with $\alpha=1/(12\ln{2})$ the universal scaling constant for a single chiral Majorana fermion.
    }
    \label{fig:central_charge}
\end{figure}

To demonstrate the spatial origin of these contributions, we consider further reduced quantities that retain the positional resolution.
For the topological ground state, we can isolate the local scaling behavior of the edges from that of the bulk.
We split the information per multiscale into two distinct local contributions $I(\boldsymbol{\ell})=I({\boldsymbol{\ell}})|_{\mathcal{C}_\mathrm{bulk}}+I({\boldsymbol{\ell}})|_\mathrm{all\ edges}$.
The first term in the sum is the bulk contribution at scale $\boldsymbol{\ell}$ given by $I({\boldsymbol{\ell}})|_{\smash{\mathcal{C}_\mathrm{bulk}}}$ following Eq.~\eqref{eq:informationonscaleinaregion_2d}, where $\mathcal{C}_{\mathrm{bulk}}$ denotes a large rectangular subsystem positioned in the middle of the system, taken to be at a distance $\zeta$ away from all edges [orange region in Fig.~\ref{fig:central_charge}(a)].
The second term is defined as the difference between the total information per multiscale and the bulk one $I(\boldsymbol{\ell})|_\mathrm{all\ edges}\!\!=I(\boldsymbol{\ell})-I({\boldsymbol{\ell}})|_{\mathcal{C}_\mathrm{bulk}}$ and hence describes the contributions coming from subsystems near the edges.
In Fig.~\ref{fig:fig5}(c), we show the corresponding quasi-1D information per scale $I(\ell_x)|_{\mathcal{C}_\mathrm{bulk}}$ [Eq.~\eqref{eq:quasi-1d-local-scale}] for increasing values of $\zeta$.
We find that as $\zeta$ increases and $\mathcal{C}_\mathrm{bulk}$ is deeper in the bulk, the closer $I(\ell_x)|_{\mathcal{C}_\mathrm{bulk}}$ approximates the exponentially decaying information distribution of the trivial ground state (red curve).
The scaling of the edge information is shown in Fig.~\ref{fig:fig5}(d) for $\zeta=13$, which asymptotically scales as $\propto\ell_x^{-2}$, and where the rapid convergence between $I(\ell_x)$ and $I(\ell_x)|_{\mathrm{all\ edges}}$ is due to the exponential decay of $I(\ell_x)|_{\mathcal{C}_\mathrm{bulk}}$ observed in Fig.~\ref{fig:fig5}(c).
Therefore, by considering the appropriate summation of inclusion-exclusion information inside a region, locally coexisting scaling behaviors in a system can be isolated from each other.

We examine the scaling of information on the edge more closely.
The edge theory of a two-dimensional symmetry-protected topological state is one-dimensional scale-invariant~\cite{chiu2016classification}.
In scale-invariant 1D states, the average local information value at each scale $i^\ell$ takes the form $i^\ell = \alpha \ell^{-2}$, where different values of $\alpha=\tfrac{c_L+c_R}{6\ln2}$ correspond to different universality classes, namely $c_{L/R}$ denotes the central charge of the modes of each chirality (left and right moving)~\cite{artiaco2025universal, swingle2010entanglement, calabrese2004entanglement}.
In our case, we have a single Majorana fermion with a chirality fixed by the sign of $\mu$, which we (arbitrarily) refer to as the right-mover; thus, $c_R=1/2$, $c_L=0$ and we expect $\alpha=1/(12\ln 2)$.
To isolate the 1D scaling along the edge, we consider the quasi-1D local information $i_{n_x}^{\ell_x}|_{{\mathcal{C}_\mathrm{edge}}}\!\!$ in Eq.~\eqref{eq:quasi-1d-local} inside a subsystem $\mathcal{C}_\mathrm{edge}$ positioned along the bottom edge.
$\mathcal{C}_\mathrm{edge}$ has position $(0\ \; N_x/4\!-\!1)$ and multiscale $(N_x/2\!-\!1\ \; \zeta)$, and is illustrated in green in Fig.~\ref{fig:central_charge}(a).
For generality we consider the average value of $i_{n_x}^{\ell_x}\big|_{\mathcal{C}_\mathrm{edge}}\!\!\!\!\!\!$~at each scale
\begin{equation}
    i^{\ell_x}\big|_{\smash{\mathcal{C}_\mathrm{edge}}} = \frac{1}{N} \sum_{n_x} i_{n_x}^{\ell_x}\big|_{\smash{\mathcal{C}_\mathrm{edge}}} \label{eq:average-quasi-1D}
\end{equation}
with $N=N_x/4-\ell_x$ the number of subsystems at scale $\ell_x$.
In Fig.~\ref{fig:central_charge}(b) we find that $i^{\ell_x}\big|_{\smash{\mathcal{C}_\mathrm{edge}}}$ approaches $\alpha\ell_x^{-2}$ with $\alpha=1/(12\ln 2)$.
The information lattice captures the universal scaling of the chiral channel in the two-dimensional setting.
More broadly, its spatially resolved viewpoint may provide a route to diagnosing richer edge features, such as finite-size effects, global boundary conditions, and localized edge defects.

\subsection{Topological order in the toric code} \label{sec:tc}

\begin{figure}
    \centering
    \rule{0pt}{0.1cm}
    \includegraphics[width=.55\linewidth]{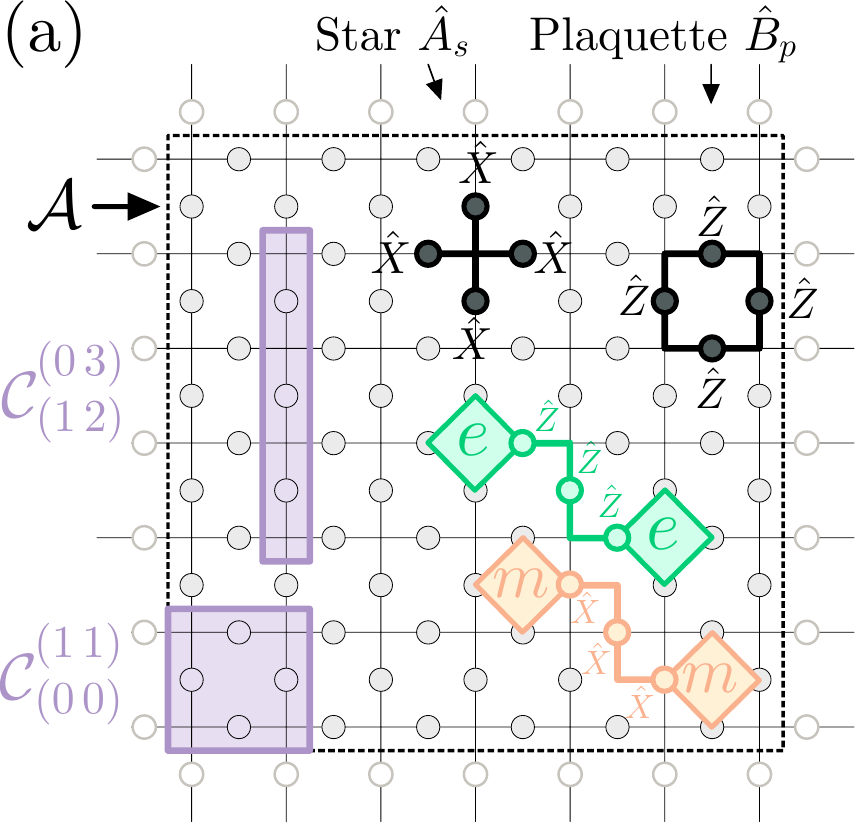}%
    \includegraphics[width=.45\linewidth]{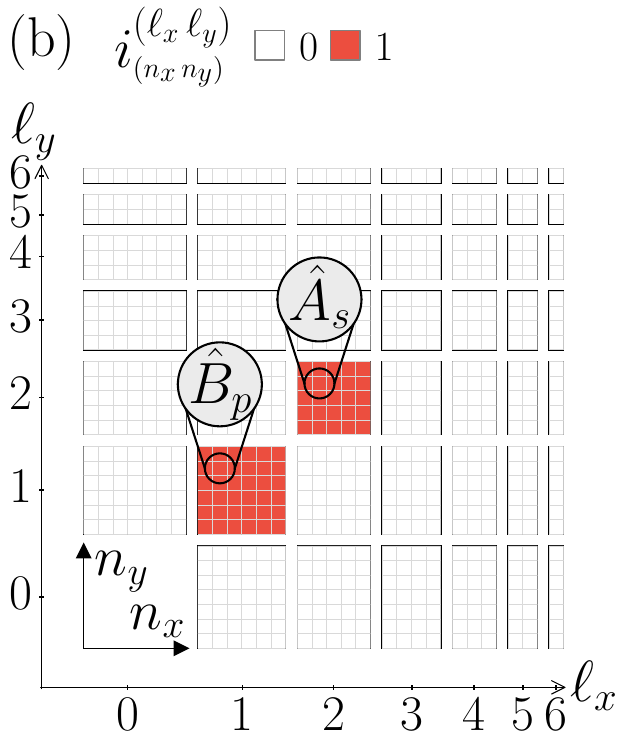}
    \caption{
        (a) In the toric code, star $\hat{A}_s$ and plaquette $\hat{B}_p$ operators (represented in black) act on spins (gray dots) located on the edges of a square lattice.
        The ground state satisfies the constraints $A_s=+1$ and $B_p=+1$ everywhere, while excited states correspond to pairs where these constraints are flipped in sign: a pair of $e$-anyons (green square) ($A_s=-1$) is created by a string $\hat{\gamma}_z$ (green line) and a pair of $m$-anyons (orange square) ($B_p=-1$) is created by a string $\hat{\gamma}_x$ (orange line).
        We consider a subsystem of $6\times 6$ plaquettes $\mathcal{A}$ (dashed line) of the toric code ground state on the infinite plane and its density matrix $\rho_\mathcal{A}$.
        Its subsystems are labeled $\mathcal{C}^{\tvec{\ell_x}{\ell_y}}_{\tvec{n_x}{n_y}}$ by groups of plaquettes (and their intersections), with $(n_x\ n_y)$ the coordinate of the bottom left corner and $(\ell_x\ \ell_y)$ their multiscale; two examples  $\mathcal{C}^{\tvec{1}{1}}_{\tvec{0}{0}}$ and $\mathcal{C}^{\tvec{0}{3}}_{\tvec{1}{2}}$ are shown in lilac.
        (b) The inclusion-exclusion information \eqref{eq:rectangular_info} is evaluated for $\rho_\mathcal{A}$ on all subsystems $\mathcal{C}^{\tvec{\ell_x}{\ell_y}}_{\tvec{n_x}{n_y}}$.
        The information of $\rho_\mathcal{A}$ is located in all the star and plaquette subsystems, each contributing 1 bit of information.
        No new information is present at larger scales.
    }
    \label{fig:toric}
\end{figure}

Finally, we consider the toric code, which is defined on a square lattice with a spin-$\tfrac12$ degree of freedom on every edge, as illustrated in Fig.~\ref{fig:toric}(a).
Its Hamiltonian is
\begin{align}
    \hat{H}_\mathrm{TC} &= -\sum_{s} \hat{A}_{s}\;-\;\sum_{p} \hat{B}_{p},
    \label{eq:TC_H}
    \\[2pt]
    \hat{A}_{s} &=\!\!\!\!\! \prod_{i\in \mathrm{star}(s) } \!\!\!\!\! \hat{X}_{i},
    \hspace{.5cm}
    \hat{B}_{p} = \!\!\!\!\!\!\!\! \prod_{i \in \mathrm{plaquette}(p) } \!\!\!\!\!\!\!\! \hat{Z}_{i},
    \label{eq:TC_stabs}
\end{align}
where $\hat{X}_i$ and $\hat{Z}_i$ denote the $x$ and $z$ Pauli operators on site $i$, respectively.
The star $\hat{A}_{s}$ and plaquette $\hat{B}_{p}$ operators are illustrated in Fig.~\ref{fig:toric}(a).
Since all star $\hat{A}_s$ and plaquette $\hat{B}_p$ operators commute, the ground-state manifold of the toric code is the simultaneous $+1$ eigenspace of all $\hat{A}_s$ and $\hat{B}_p$.
We write the corresponding eigenvalues as $A_s=+1$ and $B_p=+1$, referring to them as the star and plaquette \textit{constraints} that define the ground-state sector~\cite{kitaev2003fault}.
Products of Pauli operators along paths define string operators.
A string operator $\hat{\gamma}_z=\prod_i \hat{Z}_i$ supported on a path of edges in the direct lattice [green line in Fig.~\ref{fig:toric}(a)] anticommutes with the star operators at its endpoints.
Therefore, applying $\hat{\gamma}_z$ to the ground state creates two excitations with $A_s=-1$ at the endpoints of the string.
These endpoint excitations are the $e$ anyons, and we refer to such string operators as $e$ strings.
Similarly, a string operator $\hat{\gamma}_x=\prod_i \hat{X}_i$ supported on a path of edges on the dual lattice [orange line in Fig.~\ref{fig:toric}(a)] flips the two plaquette eigenvalues at its endpoints to $B_p=-1$.
These excitations are the $m$ anyons, and we refer to the string as an $m$ string.
A string with no endpoints, that is, a closed $e$ or $m$ string (or \textit{loop}), creates no excitations and leaves the ground state unchanged.
Therefore, the ground state is the equal-weight superposition of all possible loop configurations (i.e., the states obtained by acting with all loop operators on a reference state)~\cite{levin2005string,hamma2005bipartite}.
This loop picture reflects the topological order of the toric code: contractible loops leave the ground state unchanged, while non-contractible loops distinguish different topological sectors on manifolds with nontrivial topology.
The loop structure also manifests in the area law of the von Neumann entropy as $S(\mathcal{A})=|\partial \mathcal{A}|-\gamma$, where the offset $\gamma=1$ bit is the topological entanglement entropy~\cite{kitaev2006topological, levin2006detecting} and $|\partial \mathcal{A}|$ is the length of the boundary for a subsystem $\mathcal{A}$ assumed to be larger than the correlation length.
The topological term $\gamma$ arises from a single global $\mathbb Z_2$ constraint on the boundary configurations: loops must cross the boundary of $\mathcal{A}$ an even number of times, halving the number of allowed boundary configurations compared to a trivial gapped state~\cite{levin2006detecting}.

To characterize the toric code ground state on the information lattice, we first consider the infinite plane without boundaries, where the ground state is unique.
An explicit expression for the ground state is
\begin{align}
    \ket{\Psi_\mathrm{TC,inf}}=\prod_{p}\frac{(\hat 1+\hat{B}_p)}{\sqrt2}\ket{+}^{\otimes N}
    \label{eq:toric_wf}
\end{align}
where $\ket{+}$ denotes spin up in the basis of $\hat X$ and $N\rightarrow\infty$ is the number of qubits.
The product over the plaquette terms generates the superposition of all contractible loop configurations.
We examine the reduced density matrix $\rho_\mathcal{A}$ of $|\Psi_\mathrm{TC,inf}\rangle$ on a subsystem $\mathcal{A}$ which consists of $N_x\times N_y=6\times 6$ plaquettes, so that $\mathcal{A}$ has plaquette-type boundaries, as indicated by a dashed contour in Fig.~\ref{fig:toric}(a).
We decompose the total information contained in $\rho_\mathcal{A}$ across the subsystems $\mathcal{C}_{\boldsymbol{n}}^{\boldsymbol{\ell}}$ of $\mathcal{A}$ using the inclusion-exclusion information~\eqref{eq:inclusion_exclusion}.
For the decomposition, we take the subsystems $\mathcal{C}_{\boldsymbol{n}}^{\boldsymbol{\ell}}$ as the unions and intersections of plaquettes.
For example, the subsystem $\mathcal{C}_{\tvec{0}{0}}^{\tvec{1}{1}}$ is the single plaquette subsystem at the bottom left corner of $\mathcal{A}$, shown in lilac in Fig.~\ref{fig:toric}(a).
The subsystem $\mathcal{C}_{\tvec{1}{2}}^{\tvec{0}{3}}$, also indicated in lilac in Fig.~\ref{fig:toric}(a), consists of three vertical edges adjacent to plaquettes (it is the intersection between the two plaquette columns $\mathcal{C}_{\tvec{0}{2}}^{\tvec{1}{3}}$ and $\mathcal{C}_{\tvec{1}{2}}^{\tvec{1}{3}}$).
Since these subsystems are rectangular, the inclusion-exclusion information takes the form \eqref{eq:rectangular_info}.
One may also choose a finer or coarser resolution of subsystems (that is, decomposing the information into units smaller or larger than a plaquette).
In the present case we use plaquette subsystems, which is suitable for plaquette-type boundaries.
If $\mathcal A$ instead had star-type boundaries, an equivalent construction would be carried out on the dual lattice using star subsystems.

The information lattice of $\rho_\mathcal{A}$ is shown in Fig.~\ref{fig:toric}(b).
The single-site subsystems at scales $\boldsymbol{\ell}=(1\ 0)$ (horizontal edges) and $\boldsymbol{\ell}=(0\ 1)$ (vertical edges) contain no information about the state: their reduced density matrices are maximally mixed.
More generally, all subsystems that are strictly one-dimensional, i.e., those that contain only horizontal or vertical open strings such as $\boldsymbol{\ell}=(\ell\ 0)$ or $(0\ \ell)$ for any $\ell>1$, also contribute no information.
Instead, all the information of $\rho_\mathcal{A}$ is located at scales $\boldsymbol{\ell}=(1\ 1)$ and $\boldsymbol{\ell}=(2\ 2)$ with genuine two-dimensional extent, which correspond to the subsystems that contain single plaquettes and single stars~\footnote{
    In our resolution of subsystems, a single star does not correspond to a dedicated star-shaped subsystem; instead it appears at the center of a $2\times2$ plaquette subsystem, i.e., at scale $\boldsymbol{\ell}=(2\ 2)$.
    This reflects only the coarse graining of our subsystem decomposition: with a resolution of subsystems finer than the plaquette coarse graining, the information associated with individual stars can be assigned uniquely to star-shaped subsystems (not shown here).
}, respectively.
Each of these two-dimensional subsystems contributes exactly 1 bit of information, i.e., a single yes/no question, inferred from its density matrix: this corresponds to the definite $\mathbb{Z}_2$ value of the associated plaquette or star constraint~\cite{fattal2004entanglement,hamma2005bipartite}.
This reflects the loop picture of the toric code ground state and its $\mathbb Z_2$ topological order.
We also see that the state is \textit{local} as there is no information at larger scales: this means that $\rho_\mathcal{A}$ does not contain more information about the distribution of measurement outcomes on $\mathcal{A}$ than the reduced density matrices on the star and plaquette subsystems.
Therefore, the information in the toric code in the planar geometry is exclusively encoded in the local star and plaquette constraints.
This is different in nontrivial topologies (e.g., on the torus) where long-range information is contained in non-contractible loops.

The information lattice for excited states is identical to that of the ground state [Fig.~\ref{fig:toric}(b)], regardless of the shape and length of the string that connects the anyons.
All subsystems other than those containing a single star or plaquette still contribute 0 bits, which reflects the fact that all possible paths connecting the two anyons are physically indistinguishable.
The only information carried by the excited state is the location of the anyons, encoded in the negative eigenvalues of the corresponding star or plaquette operators.

\begin{figure}
    \centering
    \rule{0pt}{0.1cm}
    \includegraphics[width=.55\linewidth]{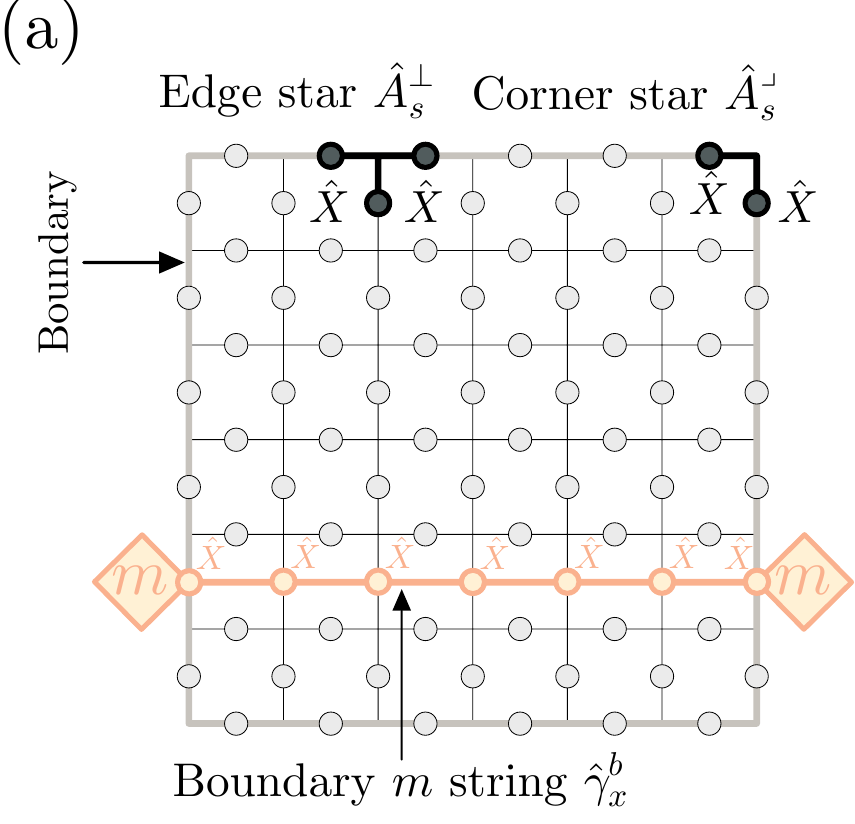}%
    \includegraphics[width=.45\linewidth]{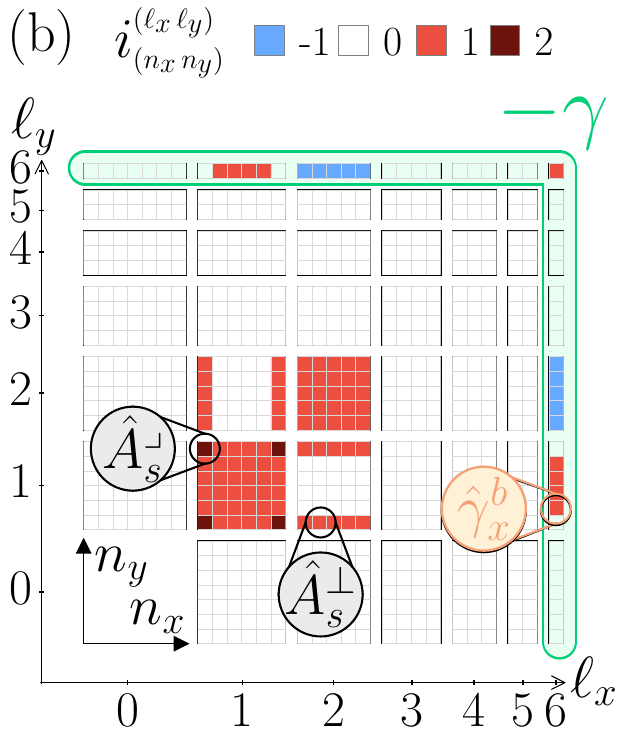}
    \caption{
        (a) Toric code on a $6\times6$ finite system with plaquette-type open boundaries.
        Because the usual star operators $\hat A_s$ are not defined at the boundary, they are replaced by edge stars $\hat A_s^\perp$ and corner stars $\hat A_s^\lrcorner$ (shown in black).
        In the ground state of Eq.~\eqref{eq:TC_obc}, their eigenvalues are fixed at $+1$, which defines the set of boundary constraints.
        An $m$ string denoted $\hat\gamma_x^b$ connecting two boundary qubits (orange) acts trivially on this ground state.
        Its endpoints correspond to $m$ anyons absorbed by the boundary, illustrating anyon condensation.
        (b) The ground state of Eq.~\eqref{eq:TC_obc} is represented on the information lattice.
        Boundary subsystems (edge and corner stars) each contribute 1 bit of information.
        At scales $\boldsymbol{\ell}=(1,N_y)$ and $(N_x,1)$ (plaquette columns or rows stretching from one boundary to the other), the boundary $m$-string $\hat\gamma_x^b$ contributes 1 bit, but this contribution is redundant: at the next scales $\boldsymbol{\ell}=(2,N_y)$ and $(N_x,2)$ a corresponding negative value appears, as two such strings together form a loop already generated by products of star operators.
        A global redundancy appears only when the full boundary is included (green contour).
        The sum of the inclusion-exclusion information values over the full boundary equals $-1$ bit, which signals the redundancy between the boundary operators in Eq.~\eqref{eq:star_redundancy}.
        For an arbitrary topological entanglement entropy, this value is $-\gamma$.
    }
    \label{fig:toric_obc}
\end{figure}

Next, we consider the ground state of the toric code on a finite system with open boundaries, representing an interface between the topologically ordered phase and a trivial gapped phase (the vacuum).
We take a system of $N_x\times N_y=6\times6$ plaquettes, as illustrated in Fig.~\ref{fig:toric_obc}(a).
With plaquette boundary conditions, the star operators $\hat A_s$ are absent on the boundary.
We denote edge and corner stars by $\hat A_s^\perp$ and $\hat A_s^\lrcorner$, respectively, as shown in Fig.~\ref{fig:toric_obc}(a).
If the Hamiltonian does not act on these boundary stars, the ground state is highly degenerate because the edge degrees of freedom remain unconstrained~\cite{cheipesh2019exact}.
We lift this degeneracy by adding small boundary terms, so the Hamiltonian becomes
\begin{align}
    \hat{H}_\mathrm{TC, open} &= \hat H_\mathrm{TC} - \frac{\epsilon}{2} \left(\sum_{s} \hat{A}_s^\perp +\sum_{s} \hat{A}_s^\lrcorner\right)
    \label{eq:TC_obc} \\[2pt]
    \hat{A}_s^\perp &=\!\!\!\!\! \prod_{i\in \mathrm{edge\ star}(s) } \!\!\!\!\! \hat{X}_{i},
    \hspace{.5cm}
    \hat{A}_s^\lrcorner = \!\!\!\!\!\!\!\! \prod_{i \in \mathrm{corner\ star}(s) } \!\!\!\!\!\!\!\! \hat{X}_{i}.
\end{align}
Here $\hat H_\mathrm{TC}$ [defined in \eqref{eq:TC_H}] acts on all stars and plaquettes in the finite system, and $\epsilon>0$ is the energy cost of flipping an edge or a corner star.
Since $\hat A_s^\perp$ and $\hat A_s^\lrcorner$ commute with all star and plaquette operators, their eigenvalues are $+1$ in the ground state, which defines the set of boundary constraints.
The ground state $\ket{\Psi_\mathrm{TC,open}}$ is unique and has the same form as Eq.~\eqref{eq:toric_wf}, with the product running over all plaquettes in the finite system.
Unlike in the infinite system, where only closed strings leave the state in the ground-state sector, the presence of boundaries allows some open strings to do so as well.
For a plaquette-type boundary, an $m$ string can be applied from one boundary to another without producing any excitations [for example, the orange line in Fig.~\ref{fig:toric_obc}(a)], while for a star-type boundary the same holds for $e$ strings.
In this sense, an anyon can be created or absorbed by the boundary at zero energy cost.
This is the phenomenon of anyon condensation (or string condensation) at the boundary~\cite{bravyi1998quantum,beigi2011quantum,kitaev2012models}.

The information lattice for $\rho=\ket{\Psi_\mathrm{TC,open}}\!\!\bra{\Psi_\mathrm{TC,open}}$ is shown in Fig.~\ref{fig:toric_obc}(b).
As in the infinite case of Fig.~\ref{fig:toric}(b), the stars and plaquettes each contribute 1 bit at their respective scales.
There are now additional contributions from the four corner stars at scales $\boldsymbol{\ell}=(1\ 1)$ and from the edge stars at $\boldsymbol{\ell}=(1\ 2)$ and $(2\ 1)$, indicated with circular insets in Fig.~\ref{fig:toric_obc}(b).
Just like the star and plaquette constraints, each $\mathbb{Z}_2$ boundary constraint contributes 1 bit.
However, the total information contributed by all subsystems with $\lVert\boldsymbol{\ell}\rVert_\infty \le 2$ is 61 bits (where $||\cdot||_\infty=\max\{\ell_x,\ell_y\}$ denotes the infinity-norm), while the full state contains only $I(\rho)=60$ bits.
The overcounted 1 bit at $\lVert\boldsymbol{\ell}\rVert_\infty \le 2$ is removed at the largest scale: summing the inclusion-exclusion values over all subsystems whose horizontal or vertical extent spans the full system [the vertices inside the green contour in Fig.~\ref{fig:toric_obc}(b)] yields $-1$ bit.
In fact, for an arbitrary topological entanglement entropy $\gamma$, the total large-scale contribution reduces to
\begin{align}
    \sum_{\substack{(\ell_x=N_x)\\\vee(\ell_y=N_y)}}\sum_{\boldsymbol{n}}i_{\boldsymbol{n}}^{\boldsymbol{\ell}}=-\gamma.
    \label{eq:long_range_summation}
\end{align}
Eq.~\eqref{eq:long_range_summation} signals that there is an overlap redundancy among the reduced density matrices of $\ket{\Psi_{\mathrm{TC,open}}}$ on the subsystems at the scales $\lVert\boldsymbol{\ell}\rVert_\infty \le 2$. 

According to the definition of overlap redundancy, Eq.~\eqref{eq:long_range_summation} implies that the information contribution associated with some subsystem satisfying $\lVert\boldsymbol{\ell}\rVert_\infty \le 2$  can also be inferred collectively through an alternative path $\mathcal{P}\subset\{\mathcal{C}_{\boldsymbol{n}}^{\boldsymbol{\ell}}\}$ of subsystems that do not contain that subsystem.
Moreover, because the negative contribution occurs only at the largest scale, this alternative path must extend across the entire system.
To make this concrete, consider a subsystem
$\mathcal{C}_{\boldsymbol{n}'}^{\tvec{2}{2}}$
supporting a star operator $\hat A_{s'}$.
The information on $\mathcal{C}_{\boldsymbol{n}'}^{\tvec{2}{2}}$ that is not already accessible from its proper subsystems is exactly the measurement statistics of $\hat A_{s'}$.
Equation~\eqref{eq:long_range_summation} tells us that these statistics can also be inferred from an alternative, system-spanning path $\mathcal{P}$ through
$\{\mathcal{C}_{\boldsymbol{n}}^{\boldsymbol{\ell}}\}$,
none of whose subsystems contains
$\mathcal{C}_{\boldsymbol{n}'}^{\tvec{2}{2}}$.
For the toric-code ground state, this prediction can be verified explicitly from the stabilizer structure.
The stabilizers satisfy
\begin{equation}
\prod_{s}\hat A_s \prod_s\hat A_s^\perp \prod_s\hat A_s^{\lrcorner} = \hat 1,
\label{eq:star_redundancy}
\end{equation}
so the constraint imposed by $\hat A_{s'}$ can indeed be inferred from all the remaining star-like constraints.
Their supports form such a system-spanning alternative path and provide access to the measurement statistics of $\hat A_{s'}$ without any subsystem along the path containing
$\mathcal{C}_{\boldsymbol{n}'}^{\tvec{2}{2}}$.
Thus, the stabilizer relation in Eq.~\eqref{eq:star_redundancy} explicitly realizes the overlap redundancy diagnosed by the information lattice.
This relation is the open-boundary counterpart of the familiar global stabilizer redundancy on the torus, where the product of all star operators and, separately, the product of all plaquette operators equals the identity.

We further analyze the individual long range values on the information lattice and interpret them in terms of the string condensation that occurs in the open boundary configuration.
The information value at the largest scale is $i_{\tvec{0}{0}}^{\tvec{N_x}{N_y}} = \gamma$.
This follows algebraically from the inclusion-exclusion information: at the largest scale $\boldsymbol{n}=(0\ 0)$, $\boldsymbol{\ell}=(N_x\ N_y)$, every term in Eq.~\eqref{eq:rectangular_info} is the information of a subsystem that touches the open boundary, except the last term $I^{\boldsymbol{\ell} \smminus \tvec{2}{2}}_{\boldsymbol{n} \smplus \tvec{1}{1}}$ (a centered subsystem reduced by two unit vectors along each axis).
For any subsystem that touches the gapped, trivial open boundary, the topological entanglement entropy is zero: any loop that straddles the subsystem boundary can be deformed into a string that condenses at the open boundary, so its topological contribution is lost~\footnote{
    For a single plaquette subsystem $\mathcal A$ in the bulk, the four stars $\hat A_1,\dots,\hat A_4$ straddling the boundary satisfy $\hat A_1\hat A_2\hat A_3\hat A_4=\hat 1_\mathcal A$ within $\mathcal{A}$, so the boundary eigenvalues obey $A_1A_2A_3A_4=+1$ and only $2^3$ boundary configurations remain.
    This redundant parity constraint reduces the von Neumann entropy by $1$ bit, described by the offset $\gamma=1$ bit~\cite{levin2006detecting}.
    If $\mathcal A$ touches a trivial (condensing) boundary, the relation instead involves an internal edge or corner star (e.g.\ $\hat A_1\hat A_2\hat A_3=\hat A_b$), so there is no constraint purely among boundary eigenvalues and the topological contribution disappears.
}.
Hence, the only term that carries $\gamma$ at the largest scale is $I^{\boldsymbol{\ell} \smminus \tvec{2}{2}}_{\boldsymbol{n} \smplus \tvec{1}{1}}$.
Meanwhile, the short-range (area-law) contributions are already fully accounted for at the smaller scales in the information lattice, so they cannot affect the value of $i_{\tvec{0}{0}}^{\tvec{N_x}{N_y}}$.

The negative values at scales $\boldsymbol{\ell}=(N_x\ 2)$ and $(2\ N_y)$ can be explained by an analogous reasoning using loops.
At scales $\boldsymbol{\ell}=(N_x\ 1)$ and $\boldsymbol{\ell}=(1\ N_y)$ (plaquette rows and columns respectively), the information of the boundary $m$ string $\hat \gamma_x^b$ shown in Fig.~\ref{fig:toric_obc}(a) contributes new information.
Indeed, within a subsystem of this scale, there are no star operators, and the operator $\hat \gamma_x^b$ is independent of all the others inside of that subsystem, so the reduced density matrix provides genuine new predictive power over measurement outcomes on this subsystem.
However, at the next scales $\boldsymbol{\ell}=(N_x\ 2)$ and $(2\ N_y)$, negative values indicate that some information was redundant: indeed, a pair of, e.g., horizontal string operators $\hat \gamma_x^b$ combined with edge and corner stars forms a loop, which is identical to a product of star operators, hence the correction at this scale.
This redundancy is only observed once the subsystem is large enough to contain the full loop.
The negative values at $\ell_x=2$ or $\ell_y=2$ therefore resolve the scale at which different long-range string operators are recognized as redundant through a closed loop relation. This makes the characteristic long-range loop structure of the topological state directly visible in the information lattice.

Altogether, the information lattice recovers the universal topological term $\gamma$ by exploiting string condensation at the boundaries.
This provides an alternative to the von Neumann entropy subtractions of bulk subsystems of Refs.~\cite{kitaev2006topological,levin2006detecting,kim2012perturbative}.
The same logic applies in the presence of domain walls: if the interior region of size $N_x\times N_y$ has topological entanglement entropy $\gamma_1$ and the exterior has $\gamma_2$, then both Eq.~\eqref{eq:long_range_summation} and the top-scale value $i_{\tvec{0}{0}}^{\tvec{N_x}{N_y}}=\gamma$ remain valid, with $\gamma$ replaced by $\gamma_1-\gamma_2$.

It is important to note that these topological contributions are assigned at the largest scales of the information lattice. As a result, the inclusion-exclusion information distinguishes the genuinely long-range topological contribution from spurious contributions to the topological entanglement entropy~\cite{zou2016spurious,kato2020entropic,kim2023universal}. A standard example is a one-dimensional cluster state embedded in an otherwise trivial two-dimensional state~\cite{zou2016spurious}, which can yield a nonzero topological entanglement entropy in conventional entropy subtractions despite the absence of topological order. In the information lattice, we find instead that this contribution is localized at the intermediate scales at which the cluster-state structure is resolved (not shown), and is therefore unambiguously separated from the largest-scale redundancy found above for the toric code.

\begin{figure}
    \centering
    \rule{0pt}{0.1cm}
    \includegraphics[width=.55\linewidth]{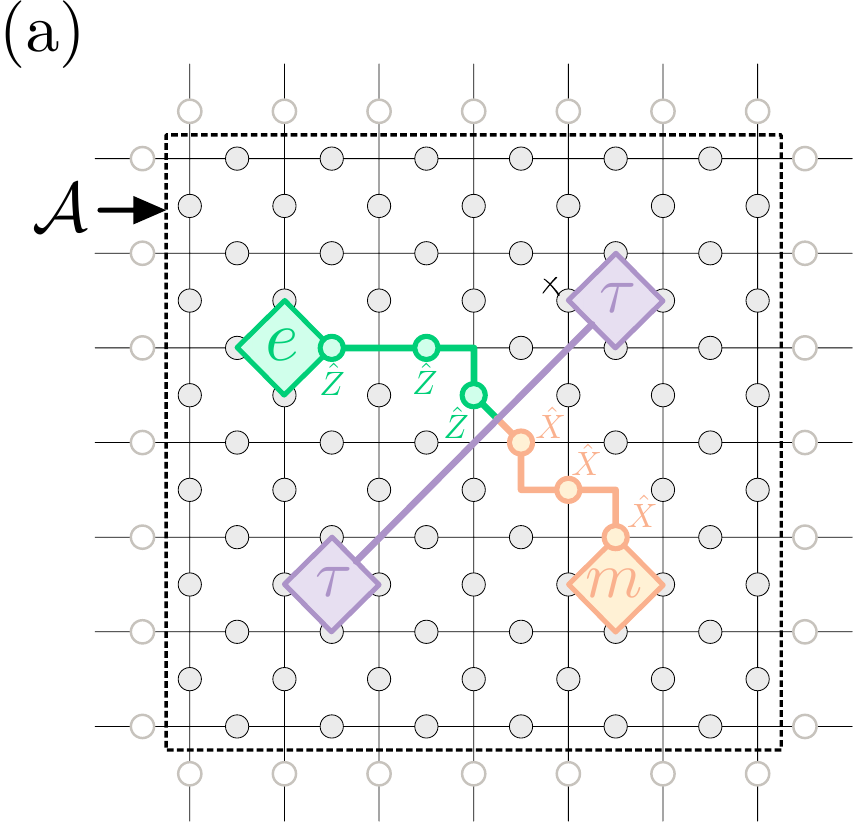}%
    \includegraphics[width=.45\linewidth]{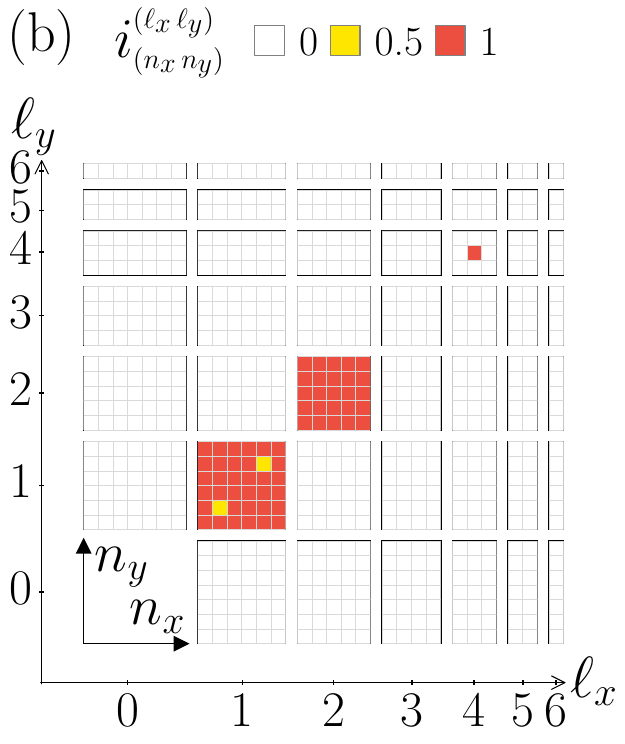}
    \caption{
        (a) In the toric code, a line defect (lilac line) is a lattice defect that turns $e$-anyons into $m$-anyons (as shown by the green-orange line).
        The endpoints of the line defect host twist defects $\tau$ (lilac squares) with non-Abelian fusion $\tau\otimes\tau = 1\oplus\varepsilon$, where $\varepsilon = e\otimes m$ is the fermionic excitation that flips both a star and an adjacent plaquette operator.
        We consider the toric code ground state on the infinite plane, where such a line defect is introduced at the indicated location on a $6\times6$ plaquette subsystem $\mathcal{A}$ on the plane.
        (b) The information lattice of $\rho_\mathcal{A}$ is similar to Fig.~\ref{fig:toric}(b), except that the subsystem enclosing a single twist contains $\tfrac12$ bit less information.
        This lost contribution is recovered in the subsystem that encloses both twists.
        Thus, the information lattice resolves the fusion channels of the non-Abelian defects across position and scale.
    }
    \label{fig:toric_twists}
\end{figure}

We finally discuss the signature of non-Abelian excitations on the information lattice.
In the toric code, bringing two anyons of the same type together merges the endpoints of the string that created them, turning the open string into a loop.
This means the anyons fuse to the vacuum (the ground state), expressed by the fusion rules $e\otimes e=m\otimes m=1$.
The fusion of an $e$ and $m$ anyon (i.e., the eigenvalues of adjacent star and plaquette are flipped to $-1$) results in a bound state denoted as $e\otimes m=\varepsilon$, which is fermionic~\cite{kitaev2003fault}.
The toric code anyons are Abelian because their fusion rules assign a single definite outcome to fusing any two anyons.
As a result, they leave no signature in the information lattice.
To see how this changes in the non-Abelian case, we consider the planar toric code with a line defect whose endpoints are non-Abelian.
A line defect [lilac line in Fig.~\ref{fig:toric_twists}(a)] is a lattice dislocation along which the roles of star and plaquette operators are interchanged: dragging an $e$ anyon across the line defect converts it into an $m$ anyon, and vice versa, as shown in Fig.~\ref{fig:toric_twists}(a)~\cite{bombin2010topological}.
The endpoints of this line defect are so-called twists $\tau$, which behave as Ising anyons and obey the non-Abelian fusion rule $\tau\otimes\tau = 1 \oplus \varepsilon$.
The two possible fusion outcomes $\{1,\varepsilon\}$ caused by the presence of a line defect forms a two-dimensional space, and thus a two-fold degeneracy of the ground state, even on the infinite plane.
This fusion space can be labeled by the $\pm1$ eigenvalues of a nonlocal fermionic parity operator $\hat P$, and braiding of the twists is described by a unitary acting on this fusion space~\cite{zheng2015demonstrating}.
To implement the defect in a Hamiltonian formulation, we modify the toric-code Hamiltonian~\eqref{eq:TC_H} by replacing the plaquette operators along the defect with mixed operators, e.g., $\hat B_p^{\mathrm{line}}=\hat Z_{i_1}\hat Z_{i_2}\hat X_{i_3}\hat X_{i_4}$ where the operator order follows the clockwise ordering of edges starting from the left edge of the plaquette.
To preserve commutativity, the star operators straddling the line defect are modified analogously, $\hat A_s^{\mathrm{line}}=\hat X_{i_1}\hat X_{i_2}\hat Z_{i_3}\hat Z_{i_4}$ assuming the same order.
Everywhere else, the Hamiltonian remains identical to that of the toric code~\footnote{
    Note that Ref.~\cite{bombin2010topological} works in another gauge, which is related to the one used here via a local unitary.
}.
For a single line defect, all pure states in the two-dimensional ground-state manifold give identical subsystem entropies.
In particular, for any subsystem $\mathcal{C}$ that encloses exactly one twist,  $S(\rho_\mathcal{C})=|\partial\mathcal{C}|-\log d_\tau$ with $d_\tau=\sqrt{2}$ the quantum dimension of the twist $\tau$, independent of which ground state is chosen~\cite{brown2013topological}.
Therefore, we consider any ground state of the infinite planar toric code with a line defect.

We take a region of $N_x\times N_y=6\times6$ plaquettes of the infinite plane with a single line at the position illustrated in Fig.~\ref{fig:toric_twists}(a).
In the information lattice shown in Fig.~\ref{fig:toric_twists}(b), we observe that subsystems enclosing a single twist show a reduction of information, which is recovered in the subsystem that encloses both twists.
This is similar to the information lattice for singlets in Fig.~\ref{fig:singlets}(a): a pair of twist defects $\tau$ possesses a two-dimensional fusion space $\{1,\varepsilon\}$, so the ground state can be a superposition of these fusion channels.
Note that, unlike a spin-$\tfrac12$ singlet which contributes 1 bit, each twist contributes only $\tfrac12$ bit.
This reflects the fact that a pair of twists carries the two-dimensional fusion space of two Majorana zero modes~\cite{bombin2010topological,teo2015theory}.
This nonlocal two-dimensional fusion space is a hallmark of non-Abelian anyons.
The information lattice not only identifies the presence of twist defects from their entanglement signatures but also provides a natural framework for tracking non-Abelian braiding operations across position and scale.
Specifically, it can be used to determine, at each stage of a braid, which twist pairings have a definite fusion outcome ($1$ or $\varepsilon$), although we do not analyze this systematically here.

In the toric code on a torus, the presence of non-contractible loops results in a fourfold degeneracy.
The information lattice analysis proceeds analogously, showing that part of the state information resides at the largest scales corresponding to these loops.
We leave the full analysis of the information lattice for systems with nontrivial topology for future work.

\section{Conclusion} \label{sec:conclusion}

We introduced a higher-dimensional information lattice to characterize many-body quantum states in spatial dimensions greater than one.
In periodic chains and in higher-dimensional geometries, overlapping subsystems may form loops.
This means that the same information may be shared nonlocally across spatially distinct yet overlapping subsystems, such that it cannot be assigned to any one of them alone.
This delocalization is itself part of the information structure of the state.
To resolve it, we defined the \textit{inclusion-exclusion local information}, which decomposes the total information of subsystem reduced density matrices into position- and scale-resolved contributions.
The inclusion-exclusion local information is assigned to the vertices on a higher-dimensional lattice, which are labeled by the position and scales of all subsystems.
This construction reveals how information is distributed in a quantum state: a positive value of the inclusion-exclusion local information indicates a net gain of information when considering a larger subsystem of the physical lattice relative to its smaller-scale subsystems.
Instead, a negative value signals that part of this information is already inferable from the joint distributions of measurement outcomes of its smaller-scale overlapping subsystems.
It therefore locates an overlap redundancy, namely a nonlocal dependence among the reduced density matrices of overlapping smaller-scale subsystems.
Such redundancies are a way in which higher-dimensional quantum states store nonlocal information, as illustrated by the topological contribution in the toric code.
In this way, the higher-dimensional information lattice turns the complexity of higher-dimensional geometries into a structured, position- and scale-resolved description of how information is organized in quantum many-body states.

We demonstrated through several examples that our decomposition isolates universal features of quantum states in higher dimensions.
For the two-dimensional Anderson model, we use quasi-one-dimensional information quantities derived from the inclusion-exclusion information to characterize different states.
In the localized ground state of the disordered phase, these quasi-1D information distributions decay exponentially with subsystem scale.
This allows us to define information-based intrinsic lengths along different directions, analogous to the intrinsic lengths defined for localized states in one dimension~\cite{artiaco2025universal}.
In the clean critical ground state quasi-1D information distributions instead decay algebraically with scale, and the inclusion-exclusion information defines a preferred direction of information propagation that aligns with the average Fermi velocity orientation, suggesting an information-based approach to Fermi-surface criticality.
We then considered a topological state with chiral edge modes, namely, the chiral $p_x+ip_y$ superconductor in its topological phase.
In the ground state, the inclusion-exclusion information isolates bulk and edge contributions: the bulk local information decays exponentially with subsystem size away from the open boundary, while the edge contribution exhibits the universal power-law scaling associated with a chiral Majorana edge mode.
We then considered the topologically ordered ground state of the toric code.
On the infinite plane without boundaries, the information lattice shows that all information is carried locally by star and plaquette constraints.
With open boundaries, the inclusion-exclusion information shows features associated with string condensation at the edges, and the topological entanglement entropy appears as a residual contribution at the largest scales of the lattice.
Finally, line defects with twist endpoints appear as local contributions on the information lattice.
In a non-Abelian braiding protocol, the information lattice points out the location and scale of the channels which produce definite fusion outcomes.
This provides an information-based diagnostic of non-Abelian defects and a framework for tracking their braiding across position and scale.

Several directions for future work remain.
On the static side, a systematic analysis of critical states using the information lattice could clarify how information propagation directions encode higher-dimensional universality classes, in particular, in the case of higher co-dimension gapless manifolds~\cite{pretko2017nodal} and interacting Fermi surfaces~\cite{shankar1994renormalization,ding2012entanglement,mcminis2013renyi,wang2014renyi,ogawa2012holographic}.
Our construction also lends itself to more complex instances of stabilizer states, as the precise location of all stabilizer generators is mapped on the lattice.
This opens the possibility to study long-range nonstabilizerness in higher dimensions~\cite{bilinskaya2025witnessing}.
A natural direction is to investigate whether our construction can resolve the local structure of domain walls between distinct topological orders with different anyon contents, even when the topological entanglement entropy is the same.
Our local description of information can help identify at what scale and location spurious contributions to the topological entanglement entropy are generated~\cite{zhang2012quasiparticle,haah2016invariant,kato2020entropic,kim2023universal}.
It is natural to apply the information lattice beyond pure ground states, for instance to mixed-state quantum phases, where one would like to characterize when local noise or dynamical processes preserve or destroy topological features~\cite{sang2024mixed,sang2025mixed}.
In coding-decoding problems, the study of information currents across scales may provide a scale-resolved understanding of decodability phase transitions and of the conditions under which information remains recoverable~\cite{sang2025mixed,gullans2020dynamical,sang2025stability}.
Applying our construction to systems on nontrivial topologies, where different winding sectors carry distinct information patterns, we expect the information lattice to be well suited for resolving how information is stored across these topological sectors.

On the dynamical side, a key challenge is to understand how information flows in higher-dimensional lattices and whether hydrodynamic regimes of information, similar to those identified in one dimension~\cite{klein2022time,artiaco2024efficient,bauer2025local}, can be established in other dimensions as well.
More generally, local currents in higher dimensions should be associated with effective structures whose dimensionality need not match the raw number of indices used to parametrize the information lattice.
Clarifying the local current structure is a prerequisite for extending Local Information Time Evolution (LITE).
~\cite{klein2022time,artiaco2024efficient}
Adapting LITE and general information-recovery methods, potentially in combination with sequential reconstruction schemes or cluster expansions~\cite{petz1988sufficiency,junge2018universal,vardhan2024petz,leifer2008quantum}, would open the door to information-lattice algorithms for efficient time evolution and information flow in higher dimensions.

\section*{Acknowledgments}

We thank David Aceituno Ch\'avez, Carlo W. J. Beenakker, Maria Hermanns, David J. Luitz, Vieri Mattei and Achim Rosch for useful discussions. This work received funding from the European Research Council (ERC) under the European Union’s Horizon 2020 research and innovation program (Grant Agreement No.~101001902) and the Knut and Alice Wallenberg Foundation (KAW) via the project Dynamic Quantum Matter (2019.0068). T.~K.~K.\ acknowledges funding from the Wenner-Gren Foundations. C.~A.\ acknowledges funding by the Deutsche Forschungsgemeinschaft (DFG, German Research Foundation) under Projektnummer 277101999 -- CRC 183 (project A01). The computations were enabled by resources provided by the National Academic Infrastructure for Supercomputing in Sweden (NAISS), partially funded by the Swedish Research Council through grant agreement no. 2022-06725. 

\section*{Data availability}

The data underlying the figures in this work are available on Zenodo~\cite{zenodo_data}.

\nocite{jakulin2003quantifying}
\bibliography{main}

@misc{zenodo_data,
  doi       = {10.5281/zenodo.18713641},
  url       = {https://doi.org/10.5281/zenodo.18713641},
  title     = {Data underlying the figures in ``Higher-Dimensional Information Lattice: Quantum State Characterization through Inclusion-Exclusion Local Information''},
  author    = {Fl{\'o}r, Ian Matthias and Artiaco, Claudia and Kvorning, Thomas Klein and Bardarson, Jens H.},
  year      = {2026},
  publisher = {Zenodo}
}

@article{winter2004,
	author = {Hayden, Patrick and Jozsa, Richard and Petz, D{\'e}nes and Winter, Andreas},
	date = {2004/04/01},
	doi = {10.1007/s00220-004-1049-z},
	id = {Hayden2004},
	isbn = {1432-0916},
	journal = {Commun. Math. Phys.},
	number = {2},
	pages = {359--374},
	title = {Structure of States Which Satisfy Strong Subadditivity of Quantum Entropy with Equality},
	url = {https://doi.org/10.1007/s00220-004-1049-z},
	volume = {246},
	year = {2004}
}

@article{fawzi2015quantum,
  title={Quantum conditional mutual information and approximate Markov chains},
  author={Fawzi, Omar and Renner, Renato},
  journal={Communications in Mathematical Physics},
  volume={340},
  number={2},
  pages={575--611},
  year={2015},
  publisher={Springer},
  doi={https://doi.org/10.1007/s00220-015-2466-x}
}

@article{zhang2024nonlocal,
  title = {Nonlocal growth of quantum conditional mutual information under decoherence},
  author = {Zhang, Yifan and Gopalakrishnan, Sarang},
  journal = {Phys. Rev. A},
  volume = {110},
  issue = {3},
  pages = {032426},
  numpages = {17},
  year = {2024},
  month = {Sep},
  publisher = {American Physical Society},
  doi = {10.1103/PhysRevA.110.032426}
}

@article{sang2025mixed,
  title = {Mixed-State Phases from Local Reversibility},
  author = {Sang, Shengqi and Lessa, Leonardo A. and Mong, Roger S. K. and Grover, Tarun and Wang, Chong and Hsieh, Timothy H.},
  journal = {Phys. Rev. X},
  volume = {16},
  issue = {4},
  pages = {041014},
  numpages = {17},
  year = {2026},
  month = {Oct},
  publisher = {American Physical Society},
  doi = {10.1103/1kr5-jlcb},
  url = {https://link.aps.org/doi/10.1103/1kr5-jlcb}
}

@article{lee2024universal,
  title = {Universal Spreading of Conditional Mutual Information in Noisy Random Circuits},
  author = {Lee, Su-un and Oh, Changhun and Wong, Yat and Chen, Senrui and Jiang, Liang},
  journal = {Phys. Rev. Lett.},
  volume = {133},
  issue = {20},
  pages = {200402},
  numpages = {6},
  year = {2024},
  month = {Nov},
  publisher = {American Physical Society},
  doi = {10.1103/PhysRevLett.133.200402},
  url = {https://link.aps.org/doi/10.1103/PhysRevLett.133.200402}
}

@article{zou2016spurious,
  title = {Spurious Long-Range Entanglement and Replica Correlation Length},
  author = {Zou, Liujun and Haah, Jeongwan},
  journal = {Phys. Rev. B},
  volume = {94},
  issue = {7},
  pages = {075151},
  year = {2016},
  month = {Aug},
  publisher = {American Physical Society},
  doi = {10.1103/PhysRevB.94.075151},
  url = {https://link.aps.org/doi/10.1103/PhysRevB.94.075151}
}

@article{zhang2025conditional,
  title = {Conditional Mutual Information and Information-Theoretic Phases of Decohered Gibbs States},
  author = {Zhang, Yifan F. and Gopalakrishnan, Sarang},
  journal = {Phys. Rev. Lett.},
  volume = {135},
  issue = {16},
  pages = {160401},
  numpages = {6},
  year = {2025},
  month = {Oct},
  publisher = {American Physical Society},
  doi = {10.1103/cjkh-1z6c},
  url = {https://link.aps.org/doi/10.1103/cjkh-1z6c}
}

@article{kuwahara2025clustering,
  title = {Clustering of Conditional Mutual Information and Quantum Markov Structure at Arbitrary Temperatures},
  author = {Kuwahara, Tomotaka},
  journal = {Phys. Rev. X},
  volume = {15},
  issue = {4},
  pages = {041010},
  numpages = {17},
  year = {2025},
  month = {Oct},
  publisher = {American Physical Society},
  doi = {10.1103/9hx7-pzxw},
  url = {https://link.aps.org/doi/10.1103/9hx7-pzxw}
}

@article{ryu2006holographic,
  title = {Holographic Derivation of Entanglement Entropy from the anti--de {S}itter Space/Conformal Field Theory Correspondence},
  author = {Ryu, Shinsei and Takayanagi, Tadashi},
  journal = {Phys. Rev. Lett.},
  volume = {96},
  issue = {18},
  pages = {181602},
  numpages = {4},
  year = {2006},
  month = {May},
  publisher = {American Physical Society},
  doi = {10.1103/PhysRevLett.96.181602},
  url = {https://link.aps.org/doi/10.1103/PhysRevLett.96.181602}
}

@article{klein2022time,
  title={Time-evolution of local information: thermalization dynamics of local observables},
  author={Klein Kvorning, Thomas and Herviou, Lo{\"\i}c and Bardarson, Jens H},
  journal={SciPost Phys.},
  volume={13},
  number={4},
  pages={080},
  year={2022},
  doi={10.21468/SciPostPhys.13.4.080}
}

@article{artiaco2024efficient,
  title={Efficient large-scale many-body quantum dynamics via local-information time evolution},
  author={Artiaco, Claudia and Fleckenstein, Christoph and Aceituno Ch{\'a}vez, David and Klein Kvorning, Thomas and Bardarson, Jens H},
  journal={PRX Quantum},
  volume={5},
  number={2},
  pages={020352},
  year={2024},
  publisher={APS},
  doi={10.1103/PRXQuantum.5.020352}
}

@article{artiaco2025universal,
  title={Universal Characterization of Quantum Many-Body States through Local Information},
  author={Artiaco, Claudia and Klein Kvorning, Thomas and Aceituno Ch{\'a}vez, David and Herviou, Lo{\"\i}c and Bardarson, Jens H},
  journal={Phys. Rev. Lett.},
  volume={134},
  number={19},
  pages={190401},
  year={2025},
  publisher={APS},
  doi={10.1103/PhysRevLett.134.190401}
}

@article{bauer2025local,
  title={Local information flow in quantum quench dynamics},
  author={Bauer, Nicolas P and Trauzettel, Bj{\"o}rn and Klein Kvorning, Thomas and Bardarson, Jens H and Artiaco, Claudia},
  journal={Phys. Rev. A},
  volume={112},
  number={2},
  pages={022221},
  year={2025},
  publisher={APS},
  doi={10.1103/v7gb-5gq8}
}

@article{nava2026information,
  title={Information transport and transport-induced entanglement in open fermion chains},
  author={Nava, Andrea and Artiaco, Claudia and Gefen, Yuval and Gornyi, Igor and Tsitsishvili, Mikheil and Zazunov, Alex and Egger, Reinhold},
  year={2026},
  journal={arXiv:2601.14153},
  url={https://arxiv.org/abs/2601.14153},
}

@article{sang2024mixed,
  title={Mixed-state quantum phases: Renormalization and quantum error correction},
  author={Sang, Shengqi and Zou, Yijian and Hsieh, Timothy H},
  journal={Phys. Rev. X},
  volume={14},
  number={3},
  pages={031044},
  year={2024},
  publisher={APS},
  doi={10.1103/PhysRevX.14.031044}
}

@article{peschel2003calculation,
  title={Calculation of reduced density matrices from correlation functions},
  author={Peschel, Ingo},
  journal={J. Phys. A},
  volume={36},
  number={14},
  pages={L205},
  year={2003},
  publisher={IOP Publishing},
  doi={10.1088/0305-4470/36/14/101}
}

@article{petz1988sufficiency,
  title={Sufficiency of channels over von {N}eumann algebras},
  author={Petz, D{\'e}nes},
  journal={Q. J. Math.},
  volume={39},
  number={1},
  pages={97--108},
  year={1988},
  publisher={Oxford University Press},
  doi={10.1093/qmath/39.1.97}
}

@article{hamma2005bipartite,
  title={Bipartite entanglement and entropic boundary law in lattice spin systems},
  author={Hamma, Alioscia and Ionicioiu, Radu and Zanardi, Paolo},
  journal={Phys. Rev. A},
  volume={71},
  number={2},
  pages={022315},
  year={2005},
  publisher={APS},
  doi={10.1103/PhysRevA.71.022315}
}

@article{kim2023universal,
  title = {Universal Lower Bound on Topological Entanglement Entropy},
  author = {Kim, Isaac H. and Levin, Michael and Lin, Ting-Chun and Ranard, Daniel and Shi, Bowen},
  journal = {Phys. Rev. Lett.},
  volume = {131},
  issue = {16},
  pages = {166601},
  numpages = {6},
  year = {2023},
  month = {Oct},
  publisher = {American Physical Society},
  doi = {10.1103/PhysRevLett.131.166601},
  url = {https://link.aps.org/doi/10.1103/PhysRevLett.131.166601}
}

@article{haah2016invariant,
  title={An invariant of topologically ordered states under local unitary transformations},
  author={Haah, Jeongwan},
  journal={Commun. Math. Phys.},
  volume={342},
  number={3},
  pages={771--801},
  year={2016},
  publisher={Springer},
  doi={https://doi.org/10.1007/s00220-016-2594-y}
}

@article{zhang2012quasiparticle,
  title = {Quasiparticle statistics and braiding from ground-state entanglement},
  author = {Zhang, Yi and Grover, Tarun and Turner, Ari and Oshikawa, Masaki and Vishwanath, Ashvin},
  journal = {Phys. Rev. B},
  volume = {85},
  issue = {23},
  pages = {235151},
  numpages = {15},
  year = {2012},
  month = {Jun},
  publisher = {American Physical Society},
  doi = {10.1103/PhysRevB.85.235151},
  url = {https://link.aps.org/doi/10.1103/PhysRevB.85.235151}
}

@article{kato2020entropic,
  title={An entropic invariant for 2D gapped quantum phases},
  author={Kato, Kohtaro and Naaijkens, Pieter},
  journal={J. Phys. A},
  volume={53},
  number={8},
  pages={085302},
  year={2020},
  publisher={IOP Publishing},
  doi={10.1088/1751-8121/ab63a5}
}

@article{wilczek1982quantum,
  title = {Quantum Mechanics of Fractional-Spin Particles},
  author = {Wilczek, Frank},
  journal = {Phys. Rev. Lett.},
  volume = {49},
  issue = {14},
  pages = {957--959},
  numpages = {0},
  year = {1982},
  month = {Oct},
  publisher = {American Physical Society},
  doi = {10.1103/PhysRevLett.49.957},
  url = {https://link.aps.org/doi/10.1103/PhysRevLett.49.957}
}

@article{arovas1984fractional,
  title = {Fractional Statistics and the Quantum {H}all Effect},
  author = {Arovas, Daniel and Schrieffer, J. R. and Wilczek, Frank},
  journal = {Phys. Rev. Lett.},
  volume = {53},
  issue = {7},
  pages = {722--723},
  numpages = {0},
  year = {1984},
  month = {Aug},
  publisher = {American Physical Society},
  doi = {10.1103/PhysRevLett.53.722},
  url = {https://link.aps.org/doi/10.1103/PhysRevLett.53.722}
}

@article{nayak2008nonabelian,
  title = {Non-Abelian anyons and topological quantum computation},
  author = {Nayak, Chetan and Simon, Steven H. and Stern, Ady and Freedman, Michael and Das Sarma, Sankar},
  journal = {Rev. Mod. Phys.},
  volume = {80},
  issue = {3},
  pages = {1083--1159},
  numpages = {0},
  year = {2008},
  month = {Sep},
  publisher = {American Physical Society},
  doi = {10.1103/RevModPhys.80.1083},
  url = {https://link.aps.org/doi/10.1103/RevModPhys.80.1083}
}

@article{wen1990topological,
  title={Topological orders in rigid states},
  author={Wen, Xiao-Gang},
  journal={Int. J. Mod. Phys. B},
  volume={4},
  number={02},
  pages={239--271},
  year={1990},
  publisher={World Scientific},
  doi={https://doi.org/10.1142/S0217979290000139}
}

@inproceedings{junge2018universal,
  title={Universal recovery maps and approximate sufficiency of quantum relative entropy},
  author={Junge, Marius and Renner, Renato and Sutter, David and Wilde, Mark M and Winter, Andreas},
  booktitle={Annales Henri Poincar{\'e}},
  volume={19},
  pages={2955--2978},
  year={2018},
  organization={Springer},
  doi={10.1007/s00023-018-0716-0}
}

@book{wilde2013quantum,
  title={Quantum information theory},
  author={Wilde, Mark},
  year={2013},
  publisher={Cambridge university press},
  doi={10.1017/CBO9781139525343}
}

@article{gioev2006entanglement,
  title={Entanglement entropy of fermions in any dimension and the {W}idom conjecture},
  author={Gioev, Dimitri and Klich, Israel},
  journal={Phys. Rev. Lett.},
  volume={96},
  number={10},
  pages={100503},
  year={2006},
  publisher={APS},
  doi={10.1103/PhysRevLett.96.100503}
}

@article{wolf2006violation,
  title = {Violation of the Entropic Area Law for Fermions},
  author = {Wolf, Michael M.},
  journal = {Phys. Rev. Lett.},
  volume = {96},
  issue = {1},
  pages = {010404},
  numpages = {4},
  year = {2006},
  month = {Jan},
  publisher = {American Physical Society},
  doi = {10.1103/PhysRevLett.96.010404},
  url = {https://link.aps.org/doi/10.1103/PhysRevLett.96.010404}
}

@article{ogawa2012holographic,
  title={Holographic {F}ermi surfaces and entanglement entropy},
  author={Ogawa, Noriaki and Takayanagi, Tadashi and Ugajin, Tomonori},
  journal={J. High Energy Phys.},
  volume={2012},
  number={1},
  pages={1--21},
  year={2012},
  publisher={Springer},
  doi={https://doi.org/10.1007/JHEP01(2012)125}
}

@article{mcminis2013renyi,
  title = {Renyi entropy of the interacting Fermi liquid},
  author = {McMinis, Jeremy and Tubman, Norm M.},
  journal = {Phys. Rev. B},
  volume = {87},
  issue = {8},
  pages = {081108},
  numpages = {4},
  year = {2013},
  month = {Feb},
  publisher = {American Physical Society},
  doi = {10.1103/PhysRevB.87.081108},
  url = {https://link.aps.org/doi/10.1103/PhysRevB.87.081108}
}

@article{wang2014renyi,
  title = {Renyi Entanglement Entropy of Interacting Fermions Calculated Using the Continuous-Time Quantum {M}onte {C}arlo Method},
  author = {Wang, Lei and Troyer, Matthias},
  journal = {Phys. Rev. Lett.},
  volume = {113},
  issue = {11},
  pages = {110401},
  numpages = {5},
  year = {2014},
  month = {Sep},
  publisher = {American Physical Society},
  doi = {10.1103/PhysRevLett.113.110401},
  url = {https://link.aps.org/doi/10.1103/PhysRevLett.113.110401}
}

@article{swingle2010entanglement,
  title={Entanglement entropy and the {F}ermi surface},
  author={Swingle, Brian},
  journal={Phys. Rev. Lett.},
  volume={105},
  number={5},
  pages={050502},
  year={2010},
  publisher={APS},
  doi={10.1103/PhysRevLett.105.050502}
}

@article{fattal2004entanglement,
    author = "Fattal, David and Cubitt, Toby S. and Yamamoto, Yoshihisa and Bravyi, Sergey and Chuang, Isaac L.",
    title = "{Entanglement in the stabilizer formalism}",
    year = {2004},
    journal = {arXiv:0406168},
    url={https://arxiv.org/abs/quant-ph/0406168},
}

@article{kitaev2006topological,
  title={Topological entanglement entropy},
  author={Kitaev, Alexei and Preskill, John},
  journal={Phys. Rev. Lett.},
  volume={96},
  number={11},
  pages={110404},
  year={2006},
  publisher={APS},
  doi={10.1103/PhysRevLett.96.110404}
}

@article{kitaev2012models,
  title={Models for gapped boundaries and domain walls},
  author={Kitaev, Alexei and Kong, Liang},
  journal={Commun. Math. Phys.},
  volume={313},
  number={2},
  pages={351--373},
  year={2012},
  publisher={Springer},
  doi={https://doi.org/10.1007/s00220-012-1500-5}
}

@article{bravyi1998quantum,
  title={Quantum codes on a lattice with boundary},
  author={Bravyi, Sergey B and Kitaev, A Yu},
  year={1998},
  journal={arXiv:quant-ph/9811052},
  url={https://arxiv.org/abs/quant-ph/9811052}
}

@article{beigi2011quantum,
  title={The quantum double model with boundary: condensations and symmetries},
  author={Beigi, Salman and Shor, Peter W and Whalen, Daniel},
  journal={Commun. Math. Phys.},
  volume={306},
  number={3},
  pages={663--694},
  year={2011},
  publisher={Springer},
  doi={https://doi.org/10.1007/s00220-011-1294-x}
}

@article{zheng2015demonstrating,
  title = {Demonstrating non-Abelian statistics of {M}ajorana fermions using twist defects},
  author = {Zheng, Huaixiu and Dua, Arpit and Jiang, Liang},
  journal = {Phys. Rev. B},
  volume = {92},
  issue = {24},
  pages = {245139},
  numpages = {10},
  year = {2015},
  month = {Dec},
  publisher = {American Physical Society},
  doi = {10.1103/PhysRevB.92.245139},
  url = {https://link.aps.org/doi/10.1103/PhysRevB.92.245139}
}

@article{teo2015theory,
  title={Theory of twist liquids: Gauging an anyonic symmetry},
  author={Teo, Jeffrey CY and Hughes, Taylor L and Fradkin, Eduardo},
  journal={Annals of Physics},
  volume={360},
  pages={349--445},
  year={2015},
  publisher={Elsevier},
  doi={https://doi.org/10.1016/j.aop.2015.05.012}
}

@article{levin2005string,
  title={String-net condensation: A physical mechanism for topological phases},
  author={Levin, Michael A and Wen, Xiao-Gang},
  journal={Phys. Rev. B},
  volume={71},
  number={4},
  pages={045110},
  year={2005},
  publisher={APS},
  doi={10.1103/PhysRevB.71.045110}
}

@article{brown2013topological,
  title={Topological entanglement entropy with a twist},
  author={Brown, Benjamin J and Bartlett, Stephen D and Doherty, Andrew C and Barrett, Sean D},
  journal={Phys. Rev. Lett.},
  volume={111},
  number={22},
  pages={220402},
  year={2013},
  publisher={APS},
  doi={10.1103/PhysRevLett.111.220402}
}

@article{kim2012perturbative,
  title={Perturbative analysis of topological entanglement entropy from conditional independence},
  author={Kim, Isaac H},
  journal={Phys. Rev. B},
  volume={86},
  number={24},
  pages={245116},
  year={2012},
  publisher={APS},
  doi={10.1103/PhysRevB.86.245116}
}

@article{bombin2010topological,
  title={Topological order with a twist: Ising anyons from an abelian model},
  author={Bomb{\'\i}n, H{\'e}ctor},
  journal={Phys. Rev. Lett.},
  volume={105},
  number={3},
  pages={030403},
  year={2010},
  publisher={APS},
  doi={10.1103/PhysRevLett.105.030403}
}

@article{kitaev2003fault,
  title={Fault-tolerant quantum computation by anyons},
  author={Kitaev, A Yu},
  journal={Ann. Phys.},
  volume={303},
  number={1},
  pages={2--30},
  year={2003},
  publisher={Elsevier},
  doi={10.1016/S0003-4916(02)00018-0}
}

@article{leifer2008quantum,
  title={Quantum graphical models and belief propagation},
  author={Leifer, Matthew S and Poulin, David},
  journal={Ann. Phys.},
  volume={323},
  number={8},
  pages={1899--1946},
  year={2008},
  publisher={Elsevier},
  doi={10.1016/j.aop.2007.10.001}
}

@article{casini2007universal,
  title={Universal terms for the entanglement entropy in 2+1 dimensions},
  author={Casini, Horacio and Huerta, Marina},
  journal={Nucl. Phys. B},
  volume={764},
  number={3},
  pages={183--201},
  year={2007},
  publisher={Elsevier},
  doi={10.1016/j.nuclphysb.2006.12.012}
}

@article{anderson1958absence,
  title={Absence of diffusion in certain random lattices},
  author={Anderson, Philip W},
  journal={Phys. Rev.},
  volume={109},
  number={5},
  pages={1492},
  year={1958},
  publisher={APS},
  doi={10.1103/PhysRev.109.1492}
}

@article{abrahams1979scaling,
  title={Scaling theory of localization: Absence of quantum diffusion in two dimensions},
  author={Abrahams, Elihu and Anderson, Philip W and Licciardello, Donald C and Ramakrishnan, Tiruppattur V},
  journal={Phys. Rev. Lett.},
  volume={42},
  number={10},
  pages={673},
  year={1979},
  publisher={APS},
  doi={10.1103/PhysRevLett.42.673}
}

@book{abrahams201050,
  title={50 years of Anderson Localization},
  author={Abrahams, Elihu},
  volume={24},
  year={2010},
  publisher={World Scientific},
  doi={10.1142/7663}
}

@article{cheipesh2019exact,
  title={Exact description of the boundary theory of the Kitaev toric code with open boundary conditions},
  author={Cheipesh, Yevheniia and Cevolani, Lorenzo and Kehrein, Stefan},
  journal={Phys. Rev. B},
  volume={99},
  number={2},
  pages={024422},
  year={2019},
  publisher={APS},
  doi={10.1103/PhysRevB.99.024422}
}

@book{joliffe2002pca,
  title={Principal Component Analysis},
  author={Jolliffe, I. T.},
  year= {2002},
  edition={2},
  series={Springer Series in Statistics},
  publisher={Springer},
  address= {New York},
  note={Section~1.1, ``Definition and Derivation of Principal Components'', p.~5},
  doi={10.1007/b98835}
}

@article{read2000paired,
  title={Paired states of fermions in two dimensions with breaking of parity and time-reversal symmetries and the fractional quantum {H}all effect},
  author={Read, Nicholas and Green, Dmitry},
  journal={Phys. Rev. B},
  volume={61},
  number={15},
  pages={10267},
  year={2000},
  publisher={APS},
  doi={10.1103/PhysRevB.61.10267}
}

@book{stanley2011enumerative,
  author={Richard P. Stanley},
  title={Enumerative Combinatorics, Volume 1},
  series={Cambridge Studies in Advanced Mathematics},
  edition={2nd},
  publisher={Cambridge University Press},
  year={2011},
  note={{Ch}.2, ``Sieve Methods", Sec.2.1  ``Inclusion--Exclusion"},
  doi={10.1007/978-1-4615-9763-6}
}

@incollection{rota1964foundations,
  title={{On the foundations of combinatorial theory: I. Theory of M{\"o}bius functions}},
  author={Rota, Gian-Carlo},
  booktitle={Classic Papers in Combinatorics},
  pages={332--360},
  year={1964},
  publisher={Springer},
  doi={10.1007/978-0-8176-4842-8_25}
}

@article{ting1962amount,
  title={On the amount of information},
  author={Ting, Hu Kuo},
  journal={Theory Probab. Appl.},
  volume={7},
  number={4},
  pages={439--447},
  year={1962},
  publisher={SIAM},
  doi={10.1137/1107041}
}

@article{chen2014entanglement,
  title={Entanglement contour},
  author={Chen, Yangang and Vidal, Guifre},
  journal={J. Stat. Mech.},
  volume={2014},
  number={10},
  pages={P10011},
  year={2014},
  publisher={IOP Publishing},
  doi={10.1088/1742-5468/2014/10/P10011}
}

@article{mcgill1954multivariate,
  title={Multivariate information transmission},
  author={McGill, William},
  journal={Trans. IRE Prof. Group Inf. Theory},
  volume={4},
  number={4},
  pages={93--111},
  year={1954},
  publisher={IEEE},
  doi={10.1007/BF02289159}
}

@article{yeung1991new,
  title={A new outlook on Shannon's information measures},
  author={Yeung, Raymond W},
  journal={IEEE Trans. Inf. Theory},
  volume={37},
  number={3},
  pages={466--474},
  year={1991},
  publisher={IEEE},
  doi={10.1109/18.79902}
}

@article{watanabe1960information,
  title={Information theoretical analysis of multivariate correlation},
  author={Watanabe, Satosi},
  journal={IBM J. Res. Dev.},
  volume={4},
  number={1},
  pages={66--82},
  year={1960},
  publisher={IBM},
  doi={10.1147/rd.41.0066}
}

@article{williams2010nonnegative,
  title={Nonnegative decomposition of multivariate information},
  author={Williams, Paul L and Beer, Randall D},
  year={2010},
  journal={arXiv:1004.2515},
  url={https://arxiv.org/abs/1004.2515}
}

@article{liu2014entanglement,
  title = {Entanglement Tsunami: Universal Scaling in Holographic Thermalization},
  author = {Liu, Hong and Suh, S. Josephine},
  journal = {Phys. Rev. Lett.},
  volume = {112},
  issue = {1},
  pages = {011601},
  numpages = {5},
  year = {2014},
  month = {Jan},
  publisher = {American Physical Society},
  doi = {10.1103/PhysRevLett.112.011601},
  url = {https://link.aps.org/doi/10.1103/PhysRevLett.112.011601}
}

@article{calabrese2016introduction,
  title={Introduction to ‘quantum integrability in out of equilibrium systems’},
  author={Calabrese, Pasquale and Essler, Fabian HL and Mussardo, Giuseppe},
  journal={J. Stat. Mech.},
  volume={2016},
  number={6},
  pages={064001},
  year={2016},
  publisher={IOP Publishing},
  doi={10.1088/1742-5468/2016/06/064001}
}

@Article{calabrese2020entanglement,
	title={{Entanglement spreading in non-equilibrium integrable systems}},
	author={Pasquale Calabrese},
	journal={SciPost Phys. Lect. Notes},
	pages={20},
	year={2020},
	publisher={SciPost},
	doi={10.21468/SciPostPhysLectNotes.20},
	url={https://scipost.org/10.21468/SciPostPhysLectNotes.20},
}

@article{bennett2000quantum,
  title={Quantum information and computation},
  author={Bennett, Charles H and DiVincenzo, David P},
  journal={Nature},
  volume={404},
  number={6775},
  pages={247--255},
  year={2000},
  publisher={Nature Publishing Group UK London},
  doi={https://doi.org/10.1038/35005001}
}

@article{terhal2015quantum,
  title = {Quantum error correction for quantum memories},
  author = {Terhal, Barbara M.},
  journal = {Rev. Mod. Phys.},
  volume = {87},
  issue = {2},
  pages = {307--346},
  numpages = {40},
  year = {2015},
  month = {Apr},
  publisher = {American Physical Society},
  doi = {10.1103/RevModPhys.87.307},
  url = {https://link.aps.org/doi/10.1103/RevModPhys.87.307}
}

@article{roy2021link,
  title={Link representation of the entanglement entropies for all bipartitions},
  author={Roy, Sudipto Singha and Santalla, Silvia N and Sierra, Germ{\'a}n and Rodr{\'\i}guez-Laguna, Javier},
  journal={J. Phys. A},
  volume={54},
  number={30},
  pages={305301},
  year={2021},
  publisher={IOP Publishing},
  doi= {10.1088/1751-8121/ac0a30},
}

@article{miao2025convergence,
  title={Convergence and quantum advantage of Trotterized {MERA} for strongly-correlated systems},
  author={Miao, Qiang and Barthel, Thomas},
  journal={Quantum},
  volume={9},
  pages={1631},
  year={2025},
  publisher={Verein zur F{\"o}rderung des Open Access Publizierens in den Quantenwissenschaften},
  doi={https://doi.org/10.22331/q-2025-02-11-1631}
}

@article{wong2025entanglement,
  title={Entanglement renormalization circuits for $2 d $ {G}aussian Fermion States},
  author={Wong, Sing Lam and Potter, Andrew C},
  year={2025},
  journal={arXiv:2506.04200},
  url={https://arxiv.org/abs/2506.04200}
}

@article{rolph2022local,
  title={Local measures of entanglement in black holes and {CFT}s},
  author={Rolph, Andrew},
  journal={SciPost Phys.},
  volume={12},
  number={3},
  pages={079},
  year={2022},
  doi={10.21468/SciPostPhys.12.3.079}
}

@article{zhang20252d,
  title={{2D} Entanglement from a Single {M}otzkin Spaghetto},
  author={Zhang, Zhao},
  year={2025},
  journal={arXiv:2506.02103},
  url={https://arxiv.org/abs/2506.02103}
}

@article{wen2020entanglement,
  title={Entanglement contour and modular flow from subset entanglement entropies},
  author={Wen, Qiang},
  journal={J. High Energy Phys.},
  volume={2020},
  number={5},
  pages={1--22},
  year={2020},
  publisher={Springer},
  doi={https://doi.org/10.1007/JHEP05(2020)018}
}

@article{rottoli2025entanglement,
  title = {Entanglement {H}amiltonians and the quasiparticle picture},
  author = {Rottoli, Federico and Rylands, Colin and Calabrese, Pasquale},
  journal = {Phys. Rev. B},
  volume = {111},
  issue = {14},
  pages = {L140302},
  numpages = {7},
  year = {2025},
  month = {Apr},
  publisher = {American Physical Society},
  doi = {10.1103/PhysRevB.111.L140302},
  url = {https://link.aps.org/doi/10.1103/PhysRevB.111.L140302}
}

@inproceedings{bell2003co,
  author={Anthony J. Bell},
  title= {The Co-Information Lattice},
  booktitle={Proceedings of the 4th International Symposium on Independent Component Analysis and Blind Signal Separation (ICA2003)},
  address={Nara, Japan},
  month=apr,
  year={2003},
  note={\href{https://api.semanticscholar.org/CorpusID:5031248}{Available online}}
}

@article{jakulin2003quantifying,
  title={Quantifying and visualizing attribute interactions},
  author={Jakulin, Aleks and Bratko, Ivan},
  year={2003},
  journal={arXiv:cs/0308002},
  url={https://arxiv.org/abs/cs/0308002}
}

@article{lieb1973proof,
  title={Proof of the strong subadditivity of quantum-mechanical entropy},
  author={Lieb, Elliott H and Ruskai, Mary Beth},
  journal={J. Math. Phys.},
  volume={14},
  number={12},
  year={1973},
  pages={1938},
  doi={10.1063/1.1666274}
}

@article{rosas2019quantifying,
  title={Quantifying high-order interdependencies via multivariate extensions of the mutual information},
  author={Rosas, Fernando E and Mediano, Pedro AM and Gastpar, Michael and Jensen, Henrik J},
  journal={Phys. Rev. E},
  volume={100},
  number={3},
  pages={032305},
  year={2019},
  publisher={APS},
  doi={10.1103/PhysRevE.100.032305}
}

@article{ding2012entanglement,
  title = {Entanglement Entropy of {F}ermi Liquids via Multidimensional Bosonization},
  author = {Ding, Wenxin and Seidel, Alexander and Yang, Kun},
  journal = {Phys. Rev. X},
  volume = {2},
  issue = {1},
  pages = {011012},
  numpages = {18},
  year = {2012},
  month = {Mar},
  publisher = {American Physical Society},
  doi = {10.1103/PhysRevX.2.011012},
  url = {https://link.aps.org/doi/10.1103/PhysRevX.2.011012}
}

@article{bilinskaya2025witnessing,
  title = {Witnessing Short- and Long-Range Nonstabilizerness via the Information Lattice},
  author = {Bilinskaya, Yuliya and Martínez, Miguel F. and Ghosh, Soumi and Klein Kvorning, Thomas and Artiaco, Claudia and Bardarson, Jens H.},
  year = {2025},
  journal = {arXiv:2510.26696},
  url = {https://arxiv.org/abs/2510.26696}
}

@article{artiaco2025out,
  title={Out-of-Equilibrium Dynamics in a {U(1)} Lattice Gauge Theory via Local Information Flows: Scattering and String Breaking},
  author={Artiaco, Claudia and Barata, Jo{\~a}o and Rico, Enrique},
  year={2025},
  journal={arXiv:2510.16101},
  url={https://arxiv.org/abs/2510.16101}
}

@article{barata2025hadronic,
  title={{Hadronic scattering in (1+1)D SU(2) lattice gauge theory from tensor networks}},
  author={Barata, Jo{\~a}o and Hormaza, Juan and Kang, Zhong-Bo and Qian, Wenyang},
  year={2025},
  journal={arXiv:2511.00154},
  url={https://arxiv.org/abs/2511.00154}
}

@article{white2018quantum,
  title = {Quantum dynamics of thermalizing systems},
  author = {White, Christopher David and Zaletel, Michael and Mong, Roger S. K. and Refael, Gil},
  journal = {Phys. Rev. B},
  volume = {97},
  issue = {3},
  pages = {035127},
  numpages = {14},
  year = {2018},
  month = {Jan},
  publisher = {American Physical Society},
  doi = {10.1103/PhysRevB.97.035127},
  url = {https://link.aps.org/doi/10.1103/PhysRevB.97.035127}
}

@article{ye2020emergent,
  title = {Emergent Hydrodynamics in Nonequilibrium Quantum Systems},
  author = {Ye, Bingtian and Machado, Francisco and White, Christopher David and Mong, Roger S. K. and Yao, Norman Y.},
  journal = {Phys. Rev. Lett.},
  volume = {125},
  issue = {3},
  pages = {030601},
  numpages = {6},
  year = {2020},
  month = {Jul},
  publisher = {American Physical Society},
  doi = {10.1103/PhysRevLett.125.030601},
  url = {https://link.aps.org/doi/10.1103/PhysRevLett.125.030601}
}

@article{ye2022universal,
  title = {Universal {Kardar-Parisi-Zhang} Dynamics in Integrable Quantum Systems},
  author = {Ye, Bingtian and Machado, Francisco and Kemp, Jack and Hutson, Ross B. and Yao, Norman Y.},
  journal = {Phys. Rev. Lett.},
  volume = {129},
  issue = {23},
  pages = {230602},
  numpages = {7},
  year = {2022},
  month = {Nov},
  publisher = {American Physical Society},
  doi = {10.1103/PhysRevLett.129.230602},
  url = {https://link.aps.org/doi/10.1103/PhysRevLett.129.230602}
}

@article{kuo2024energy,
  title = {Energy diffusion in weakly interacting chains with fermionic dissipation assisted operator evolution},
  author = {Kuo, En-Jui and Ware, Brayden and Lunts, Peter and Hafezi, Mohammad and White, Christopher David},
  journal = {Phys. Rev. B},
  volume = {110},
  issue = {7},
  pages = {075149},
  numpages = {17},
  year = {2024},
  month = {Aug},
  publisher = {American Physical Society},
  doi = {10.1103/PhysRevB.110.075149},
  url = {https://link.aps.org/doi/10.1103/PhysRevB.110.075149}
}

@article{yoo2023open,
  title = {Open-system spin transport and operator weight dissipation in spin chains},
  author = {Yoo, Yongchan and White, Christopher David and Swingle, Brian},
  journal = {Phys. Rev. B},
  volume = {107},
  issue = {11},
  pages = {115118},
  numpages = {17},
  year = {2023},
  month = {Mar},
  publisher = {American Physical Society},
  doi = {10.1103/PhysRevB.107.115118},
  url = {https://link.aps.org/doi/10.1103/PhysRevB.107.115118}
}

@article{yi2024comparing,
  title = {Comparing numerical methods for hydrodynamics in a one-dimensional lattice spin model},
  author = {Yi-Thomas, Stuart and Ware, Brayden and Sau, Jay D. and White, Christopher David},
  journal = {Phys. Rev. B},
  volume = {110},
  issue = {13},
  pages = {134308},
  numpages = {16},
  year = {2024},
  month = {Oct},
  publisher = {American Physical Society},
  doi = {10.1103/PhysRevB.110.134308},
  url = {https://link.aps.org/doi/10.1103/PhysRevB.110.134308}
}

@article{peng2023exploiting,
  title={Exploiting disorder to probe spin and energy hydrodynamics},
  author={Peng, Pai and Ye, Bingtian and Yao, Norman Y and Cappellaro, Paola},
  journal={Nat. Phys.},
  volume={19},
  number={7},
  pages={1027},
  year={2023},
  publisher={Nature Publishing Group UK London},
  doi={10.1038/s41567-023-02024-4}
}

@article{yang2025beyond,
  title = {Beyond fragmented dopant dynamics in quantum spin lattices: Robust localization and non-{G}aussian diffusion},
  author = {Yang, Mingru and Anand, Sajant and Nielsen, Kristian Knakkergaard},
  journal = {Phys. Rev. B},
  volume = {112},
  issue = {16},
  pages = {165129},
  numpages = {20},
  year = {2025},
  month = {Oct},
  publisher = {American Physical Society},
  doi = {10.1103/clhr-4h26},
  url = {https://link.aps.org/doi/10.1103/clhr-4h26}
}

@article{rakovszky2022dissipation,
  title = {Dissipation-assisted operator evolution method for capturing hydrodynamic transport},
  author = {Rakovszky, Tibor and von Keyserlingk, C. W. and Pollmann, Frank},
  journal = {Phys. Rev. B},
  volume = {105},
  issue = {7},
  pages = {075131},
  numpages = {10},
  year = {2022},
  month = {Feb},
  publisher = {American Physical Society},
  doi = {10.1103/PhysRevB.105.075131},
  url = {https://link.aps.org/doi/10.1103/PhysRevB.105.075131}
}

@article{shankar1994renormalization,
  title = {Renormalization-group approach to interacting fermions},
  author = {Shankar, R.},
  journal = {Rev. Mod. Phys.},
  volume = {66},
  issue = {1},
  pages = {129--192},
  numpages = {0},
  year = {1994},
  month = {Jan},
  publisher = {American Physical Society},
  doi = {10.1103/RevModPhys.66.129}
}

@article{horodecki2009quantum,
  title={Quantum entanglement},
  author={Horodecki, Ryszard and Horodecki, Pawe{\l} and Horodecki, Micha{\l} and Horodecki, Karol},
  journal={Rev. Mod. Phys.},
  volume={81},
  number={2},
  pages={865--942},
  year={2009},
  publisher={APS},
  doi={10.1103/RevModPhys.81.865}
}

@article{beenakker2020search,
  title={Search for non-Abelian {M}ajorana braiding statistics in superconductors},
  author={Beenakker, Carlo},
  journal={SciPost Phys. Lect. Notes},
  volume = {15},
  pages={},
  year={2020},
  doi={10.21468/SciPostPhysLectNotes.15}
}

@article{potter2010multichannel,
  title = {Multichannel Generalization of {K}itaev's {M}ajorana End States and a Practical Route to Realize Them in Thin Films},
  author = {Potter, Andrew C. and Lee, Patrick A.},
  journal = {Phys. Rev. Lett.},
  volume = {105},
  issue = {22},
  pages = {227003},
  numpages = {4},
  year = {2010},
  month = {Nov},
  publisher = {American Physical Society},
  doi = {10.1103/PhysRevLett.105.227003},
  url = {https://link.aps.org/doi/10.1103/PhysRevLett.105.227003}
}

@article{ivanov2000non,
  title = {Non-Abelian Statistics of Half-Quantum Vortices in $\mathit{p}$-Wave Superconductors},
  author = {Ivanov, D. A.},
  journal = {Phys. Rev. Lett.},
  volume = {86},
  issue = {2},
  pages = {268--271},
  numpages = {0},
  year = {2001},
  month = {Jan},
  publisher = {American Physical Society},
  doi = {10.1103/PhysRevLett.86.268},
  url = {https://link.aps.org/doi/10.1103/PhysRevLett.86.268}
}

@article{volovik1999fermion,
  title={Fermion zero modes on vortices in chiral superconductors},
  author={Volovik, GE},
  journal={JETP Lett.},
  volume={70},
  number={9},
  pages={609--614},
  year={1999},
  publisher={Springer},
  doi={https://doi.org/10.1134/1.568223}
}

@article{asahi2012topological,
  title = {Topological indices, defects, and {M}ajorana fermions in chiral superconductors},
  author = {Asahi, Daichi and Nagaosa, Naoto},
  journal = {Phys. Rev. B},
  volume = {86},
  issue = {10},
  pages = {100504},
  numpages = {5},
  year = {2012},
  month = {Sep},
  publisher = {American Physical Society},
  doi = {10.1103/PhysRevB.86.100504},
  url = {https://link.aps.org/doi/10.1103/PhysRevB.86.100504}
}

@article{sato2009topological,
  title = {Topological properties of spin-triplet superconductors and {F}ermi surface topology in the normal state},
  author = {Sato, Masatoshi},
  journal = {Phys. Rev. B},
  volume = {79},
  issue = {21},
  pages = {214526},
  numpages = {6},
  year = {2009},
  month = {Jun},
  publisher = {American Physical Society},
  doi = {10.1103/PhysRevB.79.214526},
  url = {https://link.aps.org/doi/10.1103/PhysRevB.79.214526}
}

@article{pretko2017nodal,
  title = {Nodal-line entanglement entropy: Generalized {W}idom formula from entanglement {H}amiltonians},
  author = {Pretko, Michael},
  journal = {Phys. Rev. B},
  volume = {95},
  issue = {23},
  pages = {235111},
  numpages = {11},
  year = {2017},
  month = {Jun},
  publisher = {American Physical Society},
  doi = {10.1103/PhysRevB.95.235111},
  url = {https://link.aps.org/doi/10.1103/PhysRevB.95.235111}
}

@article{metlitski2009entanglement,
  title = {Entanglement entropy in the {$O(N)$} model},
  author = {Metlitski, Max A. and Fuertes, Carlos A. and Sachdev, Subir},
  journal = {Phys. Rev. B},
  volume = {80},
  issue = {11},
  pages = {115122},
  numpages = {20},
  year = {2009},
  month = {Sep},
  publisher = {American Physical Society},
  doi = {10.1103/PhysRevB.80.115122},
  url = {https://link.aps.org/doi/10.1103/PhysRevB.80.115122}
}

@article{melko2013entanglement,
  title = {Entanglement at a Two-Dimensional Quantum Critical Point: A Numerical Linked-Cluster Expansion Study},
  author = {Kallin, Ann B. and Hyatt, Katharine and Singh, Rajiv R. P. and Melko, Roger G.},
  journal = {Phys. Rev. Lett.},
  volume = {110},
  issue = {13},
  pages = {135702},
  numpages = {5},
  year = {2013},
  month = {Mar},
  publisher = {American Physical Society},
  doi = {10.1103/PhysRevLett.110.135702},
  url = {https://link.aps.org/doi/10.1103/PhysRevLett.110.135702}
}

@article{sang2025stability,
  title = {Stability of Mixed-State Quantum Phases via Finite {M}arkov Length},
  author = {Sang, Shengqi and Hsieh, Timothy H.},
  journal = {Phys. Rev. Lett.},
  volume = {134},
  issue = {7},
  pages = {070403},
  numpages = {7},
  year = {2025},
  month = {Feb},
  publisher = {American Physical Society},
  doi = {10.1103/PhysRevLett.134.070403},
  url = {https://link.aps.org/doi/10.1103/PhysRevLett.134.070403}
}

@article{gullans2020dynamical,
  title = {Dynamical Purification Phase Transition Induced by Quantum Measurements},
  author = {Gullans, Michael J. and Huse, David A.},
  journal = {Phys. Rev. X},
  volume = {10},
  issue = {4},
  pages = {041020},
  numpages = {28},
  year = {2020},
  month = {Oct},
  publisher = {American Physical Society},
  doi = {10.1103/PhysRevX.10.041020},
  url = {https://link.aps.org/doi/10.1103/PhysRevX.10.041020}
}

@article{fradkin2006entanglement,
  title = {Entanglement Entropy of {2D} Conformal Quantum Critical Points: Hearing the Shape of a Quantum Drum},
  author = {Fradkin, Eduardo and Moore, Joel E.},
  journal = {Phys. Rev. Lett.},
  volume = {97},
  issue = {5},
  pages = {050404},
  numpages = {4},
  year = {2006},
  month = {Aug},
  publisher = {American Physical Society},
  doi = {10.1103/PhysRevLett.97.050404},
  url = {https://link.aps.org/doi/10.1103/PhysRevLett.97.050404}
}

@article{vardhan2024petz,
  title={Petz recovery from subsystems in conformal field theory},
  author={Vardhan, Shreya and Wei, Annie Y and Zou, Yijian},
  journal={J. High Energy Phys.},
  volume={2024},
  number={3},
  pages={1--67},
  year={2024},
  publisher={Springer},
  doi={https://doi.org/10.1007/JHEP03(2024)016}
}

@inproceedings{shen2011topological,
  title={Topological insulator and the Dirac equation},
  author={Shen, Shun-Qing and Shan, Wen-Yu and Lu, Hai-Zhou},
  booktitle={Spin},
  volume={1},
  number={01},
  pages={33--44},
  year={2011},
  organization={World Scientific},
  doi={https://doi.org/10.1142/S2010324711000057}
}

@article{ostlund1995thermodynamic,
  title = {Thermodynamic Limit of Density Matrix Renormalization},
  author = {\"Ostlund, Stellan and Rommer, Stefan},
  journal = {Phys. Rev. Lett.},
  volume = {75},
  issue = {19},
  pages = {3537--3540},
  numpages = {0},
  year = {1995},
  month = {Nov},
  publisher = {American Physical Society},
  doi = {10.1103/PhysRevLett.75.3537},
  url = {https://link.aps.org/doi/10.1103/PhysRevLett.75.3537}
}

@article{chiu2016classification,
  title = {Classification of topological quantum matter with symmetries},
  author = {Chiu, Ching-Kai and Teo, Jeffrey C. Y. and Schnyder, Andreas P. and Ryu, Shinsei},
  journal = {Rev. Mod. Phys.},
  volume = {88},
  issue = {3},
  pages = {035005},
  numpages = {63},
  year = {2016},
  month = {Aug},
  publisher = {American Physical Society},
  doi = {10.1103/RevModPhys.88.035005},
  url = {https://link.aps.org/doi/10.1103/RevModPhys.88.035005}
}

@article{srednicki1993entropy,
  title = {Entropy and area},
  author = {Srednicki, Mark},
  journal = {Phys. Rev. Lett.},
  volume = {71},
  issue = {5},
  pages = {666--669},
  numpages = {0},
  year = {1993},
  month = {Aug},
  publisher = {American Physical Society},
  doi = {10.1103/PhysRevLett.71.666},
  url = {https://link.aps.org/doi/10.1103/PhysRevLett.71.666}
}

@article{levin2006detecting, 
year = {2006}, 
title = {Detecting Topological Order in a Ground State Wave Function}, 
author = {Levin, Michael and Wen, Xiao-Gang}, 
journal = {Phys. Rev. Lett.}, 
doi = {10.1103/physrevlett.96.110405}, 
pmid = {16605803}, 
pages = {110405}, 
number = {11}, 
volume = {96} 
}

@article{Verstraete2008,
author = { F.   Verstraete  and  V.   Murg  and  J.I.   Cirac },
title = {Matrix product states, projected entangled pair states, and variational renormalization group methods for quantum spin systems},
journal = {Adv. Phys.},
volume = {57},
number = {2},
pages = {143-224},
year  = {2008},
publisher = {Taylor & Francis},
doi = {10.1080/14789940801912366},
}

@article{Cirac2021,
  title = {Matrix product states and projected entangled pair states: Concepts, symmetries, theorems},
  author = {Cirac, J. Ignacio and P\'erez-Garc\'{\i}a, David and Schuch, Norbert and Verstraete, Frank},
  journal = {Rev. Mod. Phys.},
  volume = {93},
  issue = {4},
  pages = {045003},
  numpages = {65},
  year = {2021},
  month = {Dec},
  publisher = {American Physical Society},
  doi = {10.1103/RevModPhys.93.045003},
  url = {https://link.aps.org/doi/10.1103/RevModPhys.93.045003}
}

@article{white1992density,
  title = {Density matrix formulation for quantum renormalization groups},
  author = {White, Steven R.},
  journal = {Phys. Rev. Lett.},
  volume = {69},
  issue = {19},
  pages = {2863--2866},
  numpages = {0},
  year = {1992},
  month = {Nov},
  publisher = {American Physical Society},
  doi = {10.1103/PhysRevLett.69.2863},
  url = {https://link.aps.org/doi/10.1103/PhysRevLett.69.2863}
}

@article{nahum2017quantum,
  title = {Quantum Entanglement Growth under Random Unitary Dynamics},
  author = {Nahum, Adam and Ruhman, Jonathan and Vijay, Sagar and Haah, Jeongwan},
  journal = {Phys. Rev. X},
  volume = {7},
  issue = {3},
  pages = {031016},
  numpages = {30},
  year = {2017},
  month = {Jul},
  publisher = {American Physical Society},
  doi = {10.1103/PhysRevX.7.031016},
  url = {https://link.aps.org/doi/10.1103/PhysRevX.7.031016}
}

@article{calabrese2005evolution,
doi = {10.1088/1742-5468/2005/04/P04010},
url = {https://dx.doi.org/10.1088/1742-5468/2005/04/P04010},
year = {2005},
month = {apr},
publisher = {},
volume = {2005},
number = {04},
pages = {P04010},
author = {Pasquale Calabrese and John Cardy},
title = {Evolution of entanglement entropy in one-dimensional systems},
journal = {J. Stat. Mech.}
}

@article{mezei2018membrane,
  title = {Membrane theory of entanglement dynamics from holography},
  author = {Mezei, M\'ark},
  journal = {Phys. Rev. D},
  volume = {98},
  issue = {10},
  pages = {106025},
  numpages = {9},
  year = {2018},
  month = {Nov},
  publisher = {American Physical Society},
  doi = {10.1103/PhysRevD.98.106025},
  url = {https://link.aps.org/doi/10.1103/PhysRevD.98.106025}
}

@article{coser2017contour,
  title={A contour for the entanglement entropies in harmonic lattices},
  author={Coser, Andrea and De Nobili, Cristiano and Tonni, Erik},
  journal={J. Phys. A},
  volume={50},
  number={31},
  pages={314001},
  year={2017},
  publisher={IOP Publishing},
  doi={10.1088/1751-8121/aa7902}
}

@article{surace2019simulating,
  title = {Simulating the out-of-equilibrium dynamics of local observables by trading entanglement for mixture},
  author = {Surace, J. and Piani, M. and Tagliacozzo, L.},
  journal = {Phys. Rev. B},
  volume = {99},
  issue = {23},
  pages = {235115},
  numpages = {11},
  year = {2019},
  month = {Jun},
  publisher = {American Physical Society},
  doi = {10.1103/PhysRevB.99.235115},
  url = {https://link.aps.org/doi/10.1103/PhysRevB.99.235115}
}

@article{pastori2019disentangling,
  title = {Disentangling sources of quantum entanglement in quench dynamics},
  author = {Pastori, Lorenzo and Heyl, Markus and Budich, Jan Carl},
  journal = {Phys. Rev. Res.},
  volume = {1},
  issue = {1},
  pages = {012007},
  numpages = {5},
  year = {2019},
  month = {Aug},
  publisher = {American Physical Society},
  doi = {10.1103/PhysRevResearch.1.012007},
  url = {https://link.aps.org/doi/10.1103/PhysRevResearch.1.012007}
}

@article{rams2020breaking,
  title = {Breaking the Entanglement Barrier: Tensor Network Simulation of Quantum Transport},
  author = {Rams, Marek M. and Zwolak, Michael},
  journal = {Phys. Rev. Lett.},
  volume = {124},
  issue = {13},
  pages = {137701},
  numpages = {6},
  year = {2020},
  month = {Mar},
  publisher = {American Physical Society},
  doi = {10.1103/PhysRevLett.124.137701},
  url = {https://link.aps.org/doi/10.1103/PhysRevLett.124.137701}
}

@article{frias2024converting,
  title = {Converting Long-Range Entanglement into Mixture: Tensor-Network Approach to Local Equilibration},
  author = {Fr\'{\i}as-P\'erez, Miguel and Tagliacozzo, Luca and Ba\~nuls, Mari Carmen},
  journal = {Phys. Rev. Lett.},
  volume = {132},
  issue = {10},
  pages = {100402},
  numpages = {6},
  year = {2024},
  month = {Mar},
  publisher = {American Physical Society},
  doi = {10.1103/PhysRevLett.132.100402},
  url = {https://link.aps.org/doi/10.1103/PhysRevLett.132.100402}
}

@article{wurtz2018cluster,
title = {Cluster truncated {W}igner approximation in strongly interacting systems},
journal = {Ann. Phys.},
volume = {395},
pages = {341-365},
year = {2018},
issn = {0003-4916},
doi = {https://doi.org/10.1016/j.aop.2018.06.001},
url = {https://www.sciencedirect.com/science/article/pii/S0003491618301647},
author = {Jonathan Wurtz and Anatoli Polkovnikov and Dries Sels}
}

@article{harkins2025nanoscale,
    author = {Kieren Harkins  and Christoph Fleckenstein  and Noella D’Souza  and Paul M. Schindler  and David Marchiori  and Claudia Artiaco  and Quentin Reynard-Feytis  and Ushoshi Basumallick  and William Beatrez  and Arjun Pillai  and Matthias Hagn  and Aniruddha Nayak  and Samantha Breuer  and Xudong Lv  and Maxwell McAllister  and Paul Reshetikhin  and Emanuel Druga  and Marin Bukov  and Ashok Ajoy },
    title = {Nanoscale engineering and dynamic stabilization of mesoscopic spin textures},
    journal = {Sci. Adv.},
    volume = {11},
    number = {13},
    pages = {eadn9021},
    year = {2025},
    doi = {10.1126/sciadv.adn9021},
    URL = {https://www.science.org/doi/abs/10.1126/sciadv.adn9021}
}

@article{evers2008anderson, 
year = {2008}, 
title = {{Anderson transitions}}, 
author = {Evers, Ferdinand and Mirlin, Alexander D}, 
journal = {Rev. Mod. Phys.}, 
doi = {10.1103/revmodphys.80.1355}, 
pages = {1355 -- 1417}, 
number = {4}, 
volume = {80}, 
}

@article{eisert2010colloquium,
  title = {Colloquium: Area laws for the entanglement entropy},
  author = {Eisert, J. and Cramer, M. and Plenio, M. B.},
  journal = {Rev. Mod. Phys.},
  volume = {82},
  issue = {1},
  pages = {277--306},
  numpages = {0},
  year = {2010},
  month = {Feb},
  publisher = {American Physical Society},
  doi = {10.1103/RevModPhys.82.277},
  url = {https://link.aps.org/doi/10.1103/RevModPhys.82.277}
}

@article{vidal2003entanglement,
  title = {Entanglement in Quantum Critical Phenomena},
  author = {Vidal, G. and Latorre, J. I. and Rico, E. and Kitaev, A.},
  journal = {Phys. Rev. Lett.},
  volume = {90},
  issue = {22},
  pages = {227902},
  numpages = {4},
  year = {2003},
  month = {Jun},
  publisher = {American Physical Society},
  doi = {10.1103/PhysRevLett.90.227902},
  url = {https://link.aps.org/doi/10.1103/PhysRevLett.90.227902}
}

@article{calabrese2004entanglement,
doi = {10.1088/1742-5468/2004/06/P06002},
url = {https://dx.doi.org/10.1088/1742-5468/2004/06/P06002},
year = {2004},
month = {jun},
publisher = {},
volume = {2004},
number = {06},
pages = {P06002},
author = {Pasquale Calabrese and  John Cardy},
title = {Entanglement entropy and quantum field theory},
journal = {J. Stat. Mech.}
}

@article{laflorencie2016quantum,
  title={Quantum entanglement in condensed matter systems},
  author={Laflorencie, Nicolas},
  journal={Phys. Rep.},
  volume={646},
  pages={1--59},
  year={2016},
  publisher={Elsevier},
  doi={10.1016/j.physrep.2016.06.008}
}

@article{keyserlingk2022operator,
  title = {Operator backflow and the classical simulation of quantum transport},
  author = {von Keyserlingk, Curt and Pollmann, Frank and Rakovszky, Tibor},
  journal = {Phys. Rev. B},
  volume = {105},
  issue = {24},
  pages = {245101},
  numpages = {16},
  year = {2022},
  month = {Jun},
  publisher = {American Physical Society},
  doi = {10.1103/PhysRevB.105.245101},
  url = {https://link.aps.org/doi/10.1103/PhysRevB.105.245101}
}

@article{lloyd2024ballistic,
  title = {Ballistic to diffusive crossover in a weakly interacting {F}ermi gas},
  author = {Lloyd, Jerome and Rakovszky, Tibor and Pollmann, Frank and von Keyserlingk, Curt},
  journal = {Phys. Rev. B},
  volume = {109},
  issue = {20},
  pages = {205108},
  numpages = {14},
  year = {2024},
  month = {May},
  publisher = {American Physical Society},
  doi = {10.1103/PhysRevB.109.205108},
  url = {https://link.aps.org/doi/10.1103/PhysRevB.109.205108}
}

@article{doherty2005detecting,
  title = {Detecting multipartite entanglement},
  author = {Doherty, Andrew C. and Parrilo, Pablo A. and Spedalieri, Federico M.},
  journal = {Phys. Rev. A},
  volume = {71},
  issue = {3},
  pages = {032333},
  numpages = {9},
  year = {2005},
  month = {Mar},
  publisher = {American Physical Society},
  doi = {10.1103/PhysRevA.71.032333},
  url = {https://link.aps.org/doi/10.1103/PhysRevA.71.032333}
}

@article{szalay2015multipartite,
  title = {Multipartite entanglement measures},
  author = {Szalay, Szil\'ard},
  journal = {Phys. Rev. A},
  volume = {92},
  issue = {4},
  pages = {042329},
  numpages = {34},
  year = {2015},
  month = {Oct},
  publisher = {American Physical Society},
  doi = {10.1103/PhysRevA.92.042329},
  url = {https://link.aps.org/doi/10.1103/PhysRevA.92.042329}
}

@incollection{walter2016multipartite,
  title={Multipartite entanglement},
  author={Walter, Michael and Gross, David and Eisert, Jens},
  journal={Quantum Inf.: From Found. Quantum Technol. Appl.},
  pages={293--330},
  year={2016},
  publisher={Wiley Online Library},
  doi={https://doi.org/10.1002/9783527805785.ch14}
}

@article{pezze2009entanglement,
  title = {Entanglement, Nonlinear Dynamics, and the {H}eisenberg Limit},
  author = {Pezz\'e, Luca and Smerzi, Augusto},
  journal = {Phys. Rev. Lett.},
  volume = {102},
  issue = {10},
  pages = {100401},
  numpages = {4},
  year = {2009},
  month = {Mar},
  publisher = {American Physical Society},
  doi = {10.1103/PhysRevLett.102.100401},
  url = {https://link.aps.org/doi/10.1103/PhysRevLett.102.100401}
}

@article{hyllus2012fisher,
  title = {Fisher information and multiparticle entanglement},
  author = {Hyllus, Philipp and Laskowski, Wies\l{}aw and Krischek, Roland and Schwemmer, Christian and Wieczorek, Witlef and Weinfurter, Harald and Pezz\'e, Luca and Smerzi, Augusto},
  journal = {Phys. Rev. A},
  volume = {85},
  issue = {2},
  pages = {022321},
  numpages = {10},
  year = {2012},
  month = {Feb},
  publisher = {American Physical Society},
  doi = {10.1103/PhysRevA.85.022321},
  url = {https://link.aps.org/doi/10.1103/PhysRevA.85.022321}
}

@article{toth2012multipartite,
  title = {Multipartite entanglement and high-precision metrology},
  author = {T\'oth, G\'eza},
  journal = {Phys. Rev. A},
  volume = {85},
  issue = {2},
  pages = {022322},
  numpages = {8},
  year = {2012},
  month = {Feb},
  publisher = {American Physical Society},
  doi = {10.1103/PhysRevA.85.022322},
  url = {https://link.aps.org/doi/10.1103/PhysRevA.85.022322}
}

@article{strobel2014fisher,
author = {Helmut Strobel  and Wolfgang Muessel  and Daniel Linnemann  and Tilman Zibold  and David B. Hume  and Luca Pezzè  and Augusto Smerzi  and Markus K. Oberthaler },
title = {Fisher information and entanglement of non-{G}aussian spin states},
journal = {Science},
volume = {345},
number = {6195},
pages = {424-427},
year = {2014},
doi = {10.1126/science.1250147},
URL = {https://www.science.org/doi/abs/10.1126/science.1250147}
}

@article{hauke2016measuring,
  title={Measuring multipartite entanglement through dynamic susceptibilities},
  author={Hauke, Philipp and Heyl, Markus and Tagliacozzo, Luca and Zoller, Peter},
  journal={Nat. Phys.},
  volume={12},
  number={8},
  pages={778--782},
  year={2016},
  publisher={Nature Publishing Group UK London},
  doi={https://doi.org/10.1038/nphys3700}
}

@article{singha2020entanglement,
  title = {Entanglement as geometry and flow},
  author = {Singha Roy, Sudipto and Santalla, Silvia N. and Rodr\'{\i}guez-Laguna, Javier and Sierra, Germ\'an},
  journal = {Phys. Rev. B},
  volume = {101},
  issue = {19},
  pages = {195134},
  numpages = {12},
  year = {2020},
  month = {May},
  publisher = {American Physical Society},
  doi = {10.1103/PhysRevB.101.195134},
  url = {https://link.aps.org/doi/10.1103/PhysRevB.101.195134}
}

@article{santalla2023entanglement,
  title = {Entanglement links and the quasiparticle picture},
  author = {Santalla, Silvia N. and Ram\'{\i}rez, Giovanni and Roy, Sudipto Singha and Sierra, Germ\'an and Rodr\'{\i}guez-Laguna, Javier},
  journal = {Phys. Rev. B},
  volume = {107},
  issue = {12},
  pages = {L121114},
  numpages = {6},
  year = {2023},
  month = {Mar},
  publisher = {American Physical Society},
  doi = {10.1103/PhysRevB.107.L121114},
  url = {https://link.aps.org/doi/10.1103/PhysRevB.107.L121114}
}

@article{wolf2008area,
  title = {Area Laws in Quantum Systems: Mutual Information and Correlations},
  author = {Wolf, Michael M. and Verstraete, Frank and Hastings, Matthew B. and Cirac, J. Ignacio},
  journal = {Phys. Rev. Lett.},
  volume = {100},
  issue = {7},
  pages = {070502},
  numpages = {4},
  year = {2008},
  month = {Feb},
  publisher = {American Physical Society},
  doi = {10.1103/PhysRevLett.100.070502},
  url = {https://link.aps.org/doi/10.1103/PhysRevLett.100.070502}
}

@article{vanraamsdonk2010building,
author = {Van Raamsdonk, Mark},
title = {Building up spacetime with quantum entanglement},
journal = {Gen. Relativ. Gravit.},
volume = {42},
number = {10},
pages = {2323-2329},
year = {2010},
doi = {10.1007/s10714-010-1034-0},
URL = {https://doi.org/10.1007/s10714-010-1034-0}
}

@article{swingle2012entanglement,
  title = {Entanglement renormalization and holography},
  author = {Swingle, Brian},
  journal = {Phys. Rev. D},
  volume = {86},
  issue = {6},
  pages = {065007},
  numpages = {8},
  year = {2012},
  month = {Sep},
  publisher = {American Physical Society},
  doi = {10.1103/PhysRevD.86.065007},
  url = {https://link.aps.org/doi/10.1103/PhysRevD.86.065007}
}

@article{schollwock2011density,
title = {The density-matrix renormalization group in the age of matrix product states},
journal = {Ann. Phys. (Amsterdam)},
volume = {326},
number = {1},
pages = {96-192},
year = {2011},
issn = {0003-4916},
doi = {https://doi.org/10.1016/j.aop.2010.09.012},
url = {https://www.sciencedirect.com/science/article/pii/S0003491610001752},
author = {Ulrich Schollwöck}
}

@book{zeng2019quantum,
  title={Quantum information meets quantum matter},
  author={Zeng, Bei and Chen, Xie and Zhou, Duan-Lu and Wen, Xiao-Gang and others},
  year={2019},
  publisher={Springer}
}

@article{page1993average,
  title = {Average entropy of a subsystem},
  author = {Page, Don N.},
  journal = {Phys. Rev. Lett.},
  volume = {71},
  issue = {9},
  pages = {1291--1294},
  numpages = {0},
  year = {1993},
  month = {Aug},
  publisher = {American Physical Society},
  doi = {10.1103/PhysRevLett.71.1291},
  url = {https://link.aps.org/doi/10.1103/PhysRevLett.71.1291}
}

@article{vidmar2017entanglement,
  title = {Entanglement Entropy of Eigenstates of Quantum Chaotic Hamiltonians},
  author = {Vidmar, Lev and Rigol, Marcos},
  journal = {Phys. Rev. Lett.},
  volume = {119},
  issue = {22},
  pages = {220603},
  numpages = {6},
  year = {2017},
  month = {Nov},
  publisher = {American Physical Society},
  doi = {10.1103/PhysRevLett.119.220603},
  url = {https://link.aps.org/doi/10.1103/PhysRevLett.119.220603}
}

@article{bianchi2022volume,
  title = {Volume-Law Entanglement Entropy of Typical Pure Quantum States},
  author = {Bianchi, Eugenio and Hackl, Lucas and Kieburg, Mario and Rigol, Marcos and Vidmar, Lev},
  journal = {PRX Quantum},
  volume = {3},
  issue = {3},
  pages = {030201},
  numpages = {77},
  year = {2022},
  month = {Jul},
  publisher = {American Physical Society},
  doi = {10.1103/PRXQuantum.3.030201},
  url = {https://link.aps.org/doi/10.1103/PRXQuantum.3.030201}
}

\end{document}